\documentclass[smallcondensed]{svjour3}     
\usepackage{graphicx}
\usepackage{mathptmx}      
\makeatletter
\let\cl@chapter\undefined
\makeatletter

\usepackage{microtype}
\usepackage{xspace}
\usepackage{xcolor}
\usepackage{amsmath,amssymb}
\usepackage{hyperref}
\usepackage[ruled,vlined]{algorithm2e}
\SetKwInOut{Input}{input}
\SetKwInOut{Output}{output}
\SetKwProg{Fn}{function}{:}{end}
\SetFuncSty{textsf}
\usepackage[capitalise]{cleveref}
\usepackage{tabularx}
\usepackage{booktabs}
\usepackage{array}
\usepackage{subfig}
\usepackage{placeins}
\usepackage{paralist}
\usepackage{scratch3}
\usepackage{amsfonts}
\usepackage[%
  group-minimum-digits=4,%
  list-final-separator={, and },%
  add-integer-zero=false,%
  free-standing-units,%
  round-mode=figures,%
  round-precision=3,%
  binary-units,%
  detect-weight=true,%
  detect-inline-weight=math,%
]{siunitx}

\usepackage{fbox}
\newcommand{\summary}[2]{
        \vspace{1mm}
        \noindent
        \fbox{%
            \parbox{.97\linewidth}{%
                    \textbf{#1 Summary.}
                #2
            }%
        }%
}%

\usepackage{listings}
\usepackage[round]{natbib}
\newcommand{\RandomAlgorithmsWorkingProjects}{983}
\newcommand{\RandomAlgorithmsFullCoverageProjects}{696}

\newcommand{\RandomAlgorithmsTrivialProjects}{391}
\newcommand{\RandomAlgorithmsNonTrivialProjects}{592}

\newcommand{\RandomAlgorithmsAverageCoverageRANDOM}{92.65}

\newcommand{\RandomAlgorithmsAverageCoverageMOSA}{95.40}

\newcommand{\RandomAlgorithmsAverageCoverageMIO}{95.54}

\newcommand{\RandomAlgorithmsAverageExecutionTimeRANDOM}{2.81}

\newcommand{\RandomAlgorithmsAverageExecutionTimeMOSA}{3.07}

\newcommand{\RandomAlgorithmsAverageExecutionTimeMIO}{3.16}

\newcommand{\RandomAlgorithmsVDCoverageMOSARandom}{0.55}

\newcommand{\RandomAlgorithmsCovSigBetterMOSAvsRANDOM}{146}
\newcommand{\RandomAlgorithmsCovBetterMOSAvsRANDOM}{174}
\newcommand{\RandomAlgorithmsCovEqualMOSAvsRANDOM}{727}
\newcommand{\RandomAlgorithmsCovWorseMOSAvsRANDOM}{82}
\newcommand{\RandomAlgorithmsCovSigWorseMOSAvsRANDOM}{38}
\newcommand{\RandomAlgorithmsVDCoverageMIORandom}{0.56}

\newcommand{\RandomAlgorithmsCovSigBetterMIOvsRANDOM}{159}
\newcommand{\RandomAlgorithmsCovBetterMIOvsRANDOM}{198}
\newcommand{\RandomAlgorithmsCovEqualMIOvsRANDOM}{741}
\newcommand{\RandomAlgorithmsCovWorseMIOvsRANDOM}{44}
\newcommand{\RandomAlgorithmsCovSigWorseMIOvsRANDOM}{15}
\newcommand{\RandomAlgorithmsVDCoverageMIOMOSA}{0.51}

\newcommand{\RandomAlgorithmsCovSigBetterMIOvsMOSA}{40}
\newcommand{\RandomAlgorithmsCovBetterMIOvsMOSA}{135}
\newcommand{\RandomAlgorithmsCovEqualMIOvsMOSA}{807}
\newcommand{\RandomAlgorithmsCovWorseMIOvsMOSA}{41}
\newcommand{\RandomAlgorithmsCovSigWorseMIOvsMOSA}{6}

\newcommand{\TopRatedAlgorithmsTotalProjects}{1000}
\newcommand{\TopRatedAlgorithmsWorkingProjects}{947}
\newcommand{\TopRatedAlgorithmsFullCoverageProjects}{111}

\newcommand{\TopRatedAlgorithmsTrivialProjects}{29}
\newcommand{\TopRatedAlgorithmsNonTrivialProjects}{918}

\newcommand{\TopRatedAlgorithmsAverageCoverageRANDOM}{62.74}

\newcommand{\TopRatedAlgorithmsAverageCoverageMOSA}{68.97}

\newcommand{\TopRatedAlgorithmsAverageCoverageMIO}{69.17}

\newcommand{\TopRatedAlgorithmsAverageExecutionTimeRANDOM}{9.14}

\newcommand{\TopRatedAlgorithmsAverageExecutionTimeMOSA}{12.74}

\newcommand{\TopRatedAlgorithmsAverageExecutionTimeMIO}{15.55}

\newcommand{\TopRatedAlgorithmsVDCoverageMOSARandom}{0.53}

\newcommand{\TopRatedAlgorithmsCovSigBetterMOSAvsRANDOM}{293}
\newcommand{\TopRatedAlgorithmsCovBetterMOSAvsRANDOM}{376}
\newcommand{\TopRatedAlgorithmsCovEqualMOSAvsRANDOM}{165}
\newcommand{\TopRatedAlgorithmsCovWorseMOSAvsRANDOM}{406}
\newcommand{\TopRatedAlgorithmsCovSigWorseMOSAvsRANDOM}{251}
\newcommand{\TopRatedAlgorithmsVDCoverageMIORandom}{0.55}

\newcommand{\TopRatedAlgorithmsCovSigBetterMIOvsRANDOM}{306}
\newcommand{\TopRatedAlgorithmsCovBetterMIOvsRANDOM}{414}
\newcommand{\TopRatedAlgorithmsCovEqualMIOvsRANDOM}{157}
\newcommand{\TopRatedAlgorithmsCovWorseMIOvsRANDOM}{376}
\newcommand{\TopRatedAlgorithmsCovSigWorseMIOvsRANDOM}{223}
\newcommand{\TopRatedAlgorithmsVDCoverageMIOMOSA}{0.50}

\newcommand{\TopRatedAlgorithmsCovSigBetterMIOvsMOSA}{89}
\newcommand{\TopRatedAlgorithmsCovBetterMIOvsMOSA}{342}
\newcommand{\TopRatedAlgorithmsCovEqualMIOvsMOSA}{278}
\newcommand{\TopRatedAlgorithmsCovWorseMIOvsMOSA}{327}
\newcommand{\TopRatedAlgorithmsCovSigWorseMIOvsMOSA}{77}

\newcommand{\AlgTrivThresholdRandom}{2.05}
\newcommand{\AlgTrivThresholdTop}{1.50}

\newcommand{\RandomAlgorithmsAverageMinimizedEqualCovRANDOM}{39.12}

\newcommand{\RandomAlgorithmsAverageMinimizedEqualCovMOSA}{38.89}

\newcommand{\RandomAlgorithmsAverageMinimizedEqualCovMIO}{45.20}

\newcommand{\TopRatedAlgorithmsAverageMinimizedEqualCovRANDOM}{33.40}

\newcommand{\TopRatedAlgorithmsAverageMinimizedEqualCovMOSA}{34.10}

\newcommand{\TopRatedAlgorithmsAverageMinimizedEqualCovMIO}{55.65}

\newcommand{\TopRatedAlgorithmsAverageExecutionTimeEqualRANDOM}{6.36}
\newcommand{\TopRatedAlgorithmsAverageExecutionTimeEqualMOSA}{6.78}
\newcommand{\TopRatedAlgorithmsAverageExecutionTimeEqualMIO}{7.07}
\newcommand{\MutRandomGeneratedMutantsRandom}{1273516}

\newcommand{\MutRandomMutationScoreTcSkipRandom}{52.23}

\newcommand{\MutRandomMutationScorePerProjectRandom}{47.44}
\newcommand{\MutRandomGeneratedMutantsMOSA}{1250248}

\newcommand{\MutRandomMutationScoreTcSkipMOSA}{54.14}

\newcommand{\MutRandomMutationScorePerProjectMOSA}{46.70}
\newcommand{\MutRandomGeneratedMutantsMIO}{1253983}

\newcommand{\MutRandomMutationScoreTcSkipMIO}{55.67}

\newcommand{\MutRandomMutationScorePerProjectMIO}{50.28}

\newcommand\definetool[2]{\newcommand{#1}{{\textsc{#2}}\xspace}}
\definetool{\Scratch}{Scratch}
\definetool{\whisker}{Whisker}
\definetool{\servant}{Servant}
\definecolor{commentgray}{rgb}{0.31,0.31,0.31}
\definecolor{codegray}{rgb}{0.5,0.5,0.5}
\definecolor{codepurple}{rgb}{0.58,0,0.82}
\definecolor{backcolour}{rgb}{0.94,0.95,0.96}
\definecolor{darkgreen}{rgb}{0.11,0.520.21}

\setscratch{scale=.5}

\usepackage{tikz}
\newcommand{\blockleft}{\begin{mbox}\sf\begin{tikz}[baseline=(X.base)]\node[draw=black!30,fill=black!3,semithick,rectangle,inner sep=1pt, minimum size=1em, outer sep=0pt, rounded corners=1pt] (X)}%
\newcommand{\blockright}{;\end{tikz}\normalfont\end{mbox}}%

\lstdefinelanguage{JavaScript}{
  morekeywords=[1]{break, continue, delete, else, for, function, if, in,
    new, return, this, typeof, var, void, while, with},
  morekeywords=[2]{false, null, true, boolean, number, undefined,
    Array, Boolean, Date, Math, Number, String, Object},
  morekeywords=[3]{eval, parseInt, parseFloat, escape, unescape},
  sensitive,
  morecomment=[s]{/*}{*/},
  morecomment=[l]//,
  morecomment=[s]{/**}{*/}, 
  morestring=[b]',
  morestring=[b]"
}[keywords, comments, strings]
\lstalias[]{ES6}[ECMAScript2015]{JavaScript}
\lstdefinelanguage[ECMAScript2015]{JavaScript}[]{JavaScript}{
  morekeywords=[1]{await, async, case, catch, class, const, default, do,
    enum, export, extends, finally, from, implements, import, instanceof,
    let, static, super, switch, throw, try},
  morestring=[b]` 
}

\lstdefinestyle{code}{
  backgroundcolor=\color{backcolour},
  commentstyle=\color{commentgray},
  keywordstyle=\color{violet},
  numberstyle=\tiny\color{codegray},
  stringstyle=\color{darkgreen},
  basicstyle=\linespread{1.1}\ttfamily,
  breakatwhitespace=false,
  breaklines=true,
  captionpos=b,
  keepspaces=true,
  numbers=left,
  numbersep=5pt,
  showspaces=false,
  showstringspaces=false,
  showtabs=false,
  tabsize=2,
  frame=tlbr,
  framerule=0pt,
}
\newcommand{\event}[1]{\textsf{#1}}
\newcommand{\List}[1]{\ensuremath{\langle#1\rangle}}

\newcommand{\rand}{\textsf{Random1000}\xspace}
\newcommand{\topp}{\textsf{Top1000}\xspace}
\newcommand{\gitId}{\textsf{c1e68361}\xspace}
\newcommand{\gitIdVm}{\textsf{ed4055f2}\xspace}

\begin{document}

\title{Automated Test Generation for Scratch Programs
}


\author{Adina Deiner \and
		Patric Feldmeier \and
        Gordon Fraser \and
        Sebastian Schweikl \and
		Wengran Wang
        \thanks{Authors listed in alphabetical order.}}

\authorrunning{Deiner et al.}

\institute{University of Passau, Innstr. 33, 94032 Passau, Germany}

\date{Received: date / Accepted: date}

\maketitle

\begin{abstract}
	%
	%
	The importance of programming education has led to dedicated educational
programming environments, where users visually arrange block-based programming
constructs that typically control graphical, interactive game-like programs.
The \Scratch programming environment is particularly popular, with more than 90 million registered users at the time of this writing.
	%
	%
	While the block-based nature of \Scratch helps learners by preventing
syntactical mistakes, there nevertheless remains a need to provide feedback and
support in order to implement desired functionality. To support individual
learning and classroom settings, this feedback and support should ideally be
provided in an automated fashion, which requires tests to enable dynamic
program analysis.
	%
	%
	In prior work we introduced \whisker, a framework that enables automated
testing of \Scratch programs. However, creating these automated tests for
\Scratch programs is challenging. In this paper, we therefore investigate how
to automatically generate \whisker tests.
	%
	%
	Generating tests for \Scratch raises important challenges: First, game-like
programs are typically randomised, leading to flaky tests. Second, \Scratch
programs usually consist of animations and interactions with long delays,
inhibiting the application of classical test generation approaches. Thus, the
new application domain raises the question which test generation technique is
best suited to produce high coverage tests.
	%
	%
	We investigate these questions using an extension of the \whisker test
framework for automated test generation. Evaluation on common programming
exercises, a random sample of 1000 \Scratch user programs, and the 1000 most
popular \Scratch programs demonstrates that our approach enables \whisker to
reliably accelerate test executions, and even though many \Scratch programs are
small and easy to cover, there are many unique challenges for which advanced
search-based test generation using many-objective algorithms is needed in order
to achieve high coverage.
\keywords{Search-based testing \and Block-based programming \and Scratch.}
\end{abstract}

%

\section{Introduction}
\label{sec:introduction}

Computer programming nowadays is at the forefront of education: Not only is
programming considered an important skill that is included in general computer
science education, it also plays a central role in the teaching of
computational thinking~\citep{Lee2011}. The term ``computational thinking''
refers to the ability to think or solve problems based on computing methods,
and includes aspects such as abstraction, data representation, and logically
organising data. A core vehicle to teach these aspects is programming. Since
computational thinking is increasingly integrated into core curricula
at primary school level, even the youngest learners nowadays learn how to
create simple computer programs.

Teaching young learners programming requires dedicated programming languages
and programming environments. Common novice programming environments, such as
\Scratch~\citep{maloney2010scratch}, \textsc{Snap}~\citep{harvey2013snap}, and
\textsc{Alice}~\citep{cooper2000alice} engage young learners by allowing them
to build programming artefacts such as apps and games, which connects
computation with their real-world interests~\citep{seymour1980mindstorms}.
%
Novice programming environments typically have two distinguishing
features: First, to avoid the necessity to memorize and type textual
programming commands as well as the common syntactic overhead caused by braces
or indentation, programs are created visually by dragging and dropping
block-shaped commands from ``drawers'' containing all possible blocks. The blocks
have specific shapes and only matching blocks snap together, such that it is
only possible to produce syntactically valid programs. Second, the programs
typically control graphical sprites in a game-like, interactive environment.
Accordingly, many programming commands are high-level statements that control
the behaviour of these graphical sprites.

While these simplifications and application scenarios reduce complexity and
make programming accessible and engaging, the learning process is nevertheless
challenging from multiple points of view: Learners may struggle implementing
programs due to misconceptions \citep{sirkia2012exploring}, and even though
there are no syntax errors there still is an infinite number of possible ways
to assemble blocks in incorrect ways~\citep{fraedrich2020}. Teachers therefore
need to support their students, but to do so they need to comprehend each
individual learner's program, which can be a daunting task in the light of
large classrooms. Consequently, there is a need to support learners and
teachers with the help of automated tools.

A primary means to enable automated tools to inform learners and teachers is by
testing the programs. 
Given insights on the implemented behaviour, automated tools can identify
missing or incorrect functionality, they can suggest which parts of the program
to fix, how to fix them, or which steps to perform next in order to solve the
overall task. A common prerequisite, however, is that tests can be automated.
In the context of \Scratch programs, the \whisker
framework~\citep{TestingScratch} provides a means to automate testing. A
\whisker-test automatically sends user events such as key presses or mouse
clicks to the program, and observes the resulting behaviour. However, creating
\whisker tests is a challenging task. For example, the test suite for a simple
fruit catching game used in the original \whisker study (shown in
\cref{fig:scratch}) consists of 869 lines of JavaScript code. Some of the
complexity of creating such tests can be alleviated by providing suitable user
interfaces and more abstract means to specify the
tests~\citep{wang2021snapcheck}. However, at the end of the day the tests
nevertheless require non-trivial manual labour, which is problematic
considering the likely target audience of teachers who may not be adequately
educated software engineers.

In this paper we aim to mitigate this problem by relieving users of the task
of creating tests themselves. Given a \Scratch program, we aim to automatically
generate a set of \whisker tests that execute all parts of the code. The program under test could be
an example solution for a given task, such that the test suite can then be
executed against all student solutions. Alternatively, the program could be a
student solution for which a dynamic analysis is desired.
However, even though \Scratch programs tend to be small and playful, generating
tests for them automatically is nevertheless challenging. In an initial proof
of concept, we demonstrated the feasibility of using search techniques to
automatically generate sequences of interactions with \Scratch
programs~\citep{deiner2020search}, but also revealed multiple
obstacles that make test generation difficult: Unlike traditional code,
\Scratch program executions tend to take substantial time due to the frequent
use of motion- and sound-related animations and time-delays. Automated test
generation techniques, however, rely on frequently executing programs. The
original \whisker test execution framework~\citep{TestingScratch} reduced
non-determinism by controlling random number generators, but experience has
shown that the timing aspects of these programs and the unpredictability of the
scheduler when executing these highly concurrent programs nevertheless make
deterministic executions difficult. Finally, there are technical challenges
related to the questions of which interactions a program should receive, and
which algorithm to use in order to explore possible sequences of such
interactions.

In order to address these challenges, we extend our prior work on the \whisker
testing framework~\citep{TestingScratch} and \whisker test
generation~\citep{deiner2020search}. In detail, the contributions of this paper
are as follows:
\begin{itemize}
	\item We modify the execution model of the \Scratch
virtual machine that makes executions deterministic, even in the light of timing
and concurrency. Given this modification, tests are fully reproducible and can
be executed in a highly accelerated fashion.
	\item We propose a strategy that takes the
source code as well as the runtime state of a program during its execution into
account to determine which user events are
suitable for interacting with a program under test.
	\item We adapt random and search-based test generation approaches to the scenario of generating high-coverage test suites for \Scratch programs, which includes improvements to encoding, fitness function, and algorithms.
	\item We implement a full-fledged test generation framework that combines generated event sequences with regression assertions.
	\item We empirically study the ability of these test generation techniques  to cover and test the code of real \Scratch programs using two large datasets.
	%
\end{itemize}

Our experiments demonstrate that, even when accelerating test execution by a
factor of~10, test results are completely deterministic and devoid of any
flakiness using our approach. While we find that many users create programs
that are small and easy to cover fully, there are countless unique challenges
to test generation for \Scratch, ranging from extracting suitable events for
exercising a program to calculating appropriate reachability estimates to guide
test generation. Our extension of the \whisker framework implements techniques
that, collectively, allow many-objective search-based test generation to
achieve an average of 99.5\% coverage on common user-written programs, and on average 72\% coverage on popular projects. Consequently, \whisker represents an
important step towards enabling dynamic analysis of \Scratch programs and
countless resulting possible applications in the area of supporting programming
learners. To support researchers in developing these applications, and to develop new techniques to improve coverage further, \whisker and
its mature automated test generation framework are freely available as open
source software.

\section{Background}
\label{sec:background}

\subsection{\Scratch Programs}\label{sec:Scratch-Programs}

\newcommand{\inttype}{\ensuremath{\mathit{int}}\xspace}%
\newcommand{\floattype}{\ensuremath{\mathit{float}}\xspace}%
\newcommand{\stringtype}{\ensuremath{\mathit{string}}\xspace}%
\newcommand{\actortype}{\ensuremath{\mathit{actor}}\xspace}%
\newcommand{\listtype}{\ensuremath{\mathit{list}}\xspace}%

\newcommand{\abstdomain}{\ensuremath{D}\xspace}%
\newcommand{\conc}[1]{\ensuremath{[\![ #1 ]\!]}\xspace}%
\newcommand{\abst}[1]{\ensuremath{\langle\!\langle #1 \rangle\!\rangle}\xspace}%
\newcommand{\sem}{\conc}%

\newcommand{\scratchapp}{\ensuremath{\mathit{App}}\xspace}%
\newcommand{\ctrllocations}{\ensuremath{L}\xspace}%
\newcommand{\datalocations}{\ensuremath{X}\xspace}%
\newcommand{\dataloc}{\ensuremath{x}\xspace}%
\newcommand{\datavalues}{\ensuremath{V}\xspace}%
\newcommand{\datavalue}{\ensuremath{v}\xspace}%
\newcommand{\controlflows}{\ensuremath{G}\xspace}%
\newcommand{\scripts}{\ensuremath{S}\xspace}%
\newcommand{\locations}{\ctrllocations}%
\newcommand{\ops}{\ensuremath{Ops}\xspace}%
\newcommand{\location}{\ensuremath{l}\xspace}%
\newcommand{\process}{\ensuremath{p}\xspace}%
\newcommand{\processes}{\ensuremath{P}\xspace}%
\newcommand{\concrete}{\ensuremath{\gamma}\xspace}%
\newcommand{\concretes}{\ensuremath{C}\xspace}%
\newcommand{\program}{\ensuremath{{\mathrm{App}}}\xspace}%
\newcommand{\step}{\textsf{step}\xspace}
\newcommand{\executiontrace}{\ensuremath{\bar \concrete}\xspace}%

\newcommand{\pcomp}{\ensuremath{\mathsf{pstate}}\xspace}%
\newcommand{\pc}{\ensuremath{\mathsf{pc}}\xspace}%
\newcommand{\pid}{\ensuremath{\mathsf{pid}}\xspace}%
\newcommand{\pgroup}{\ensuremath{\mathsf{pgroup}}\xspace}%
\newcommand{\pwaitfor}{\ensuremath{\mathsf{pwaitfor}}\xspace}%
\newcommand{\ptime}{\ensuremath{\mathsf{ptime}}\xspace}%
\newcommand{\compwait}{\ensuremath{\mathsf{Wait}}\xspace}%
\newcommand{\comprun}{\ensuremath{\mathsf{Running}}\xspace}%
\newcommand{\compyield}{\ensuremath{\mathsf{Yield}}\xspace}%
\newcommand{\compdone}{\ensuremath{\mathsf{Done}}\xspace}%
\newcommand{\trace}{\ensuremath{\overline{\concrete}}\xspace}%
\newcommand{\traces}{\ensuremath{\concretes^\infty}\xspace}%
\newcommand{\actorset}{\ensuremath{A}\xspace}%
\newcommand{\actors}{\ensuremath{\mathcal{A}}\xspace}%
\newcommand{\actor}{\ensuremath{a}\xspace}%
\newcommand{\properties}{\ensuremath{\phi}\xspace}%
\newcommand{\batches}{\ensuremath{B}\xspace}%
\newcommand{\batch}{\ensuremath{b}\xspace}%

\newcommand{\inlineFigure}[2]{$\vcenter{\hbox{\includegraphics[scale=#1]{#2}}}$}%

\begin{figure}[tb]
	\centering
	\includegraphics[width=0.8\textwidth]{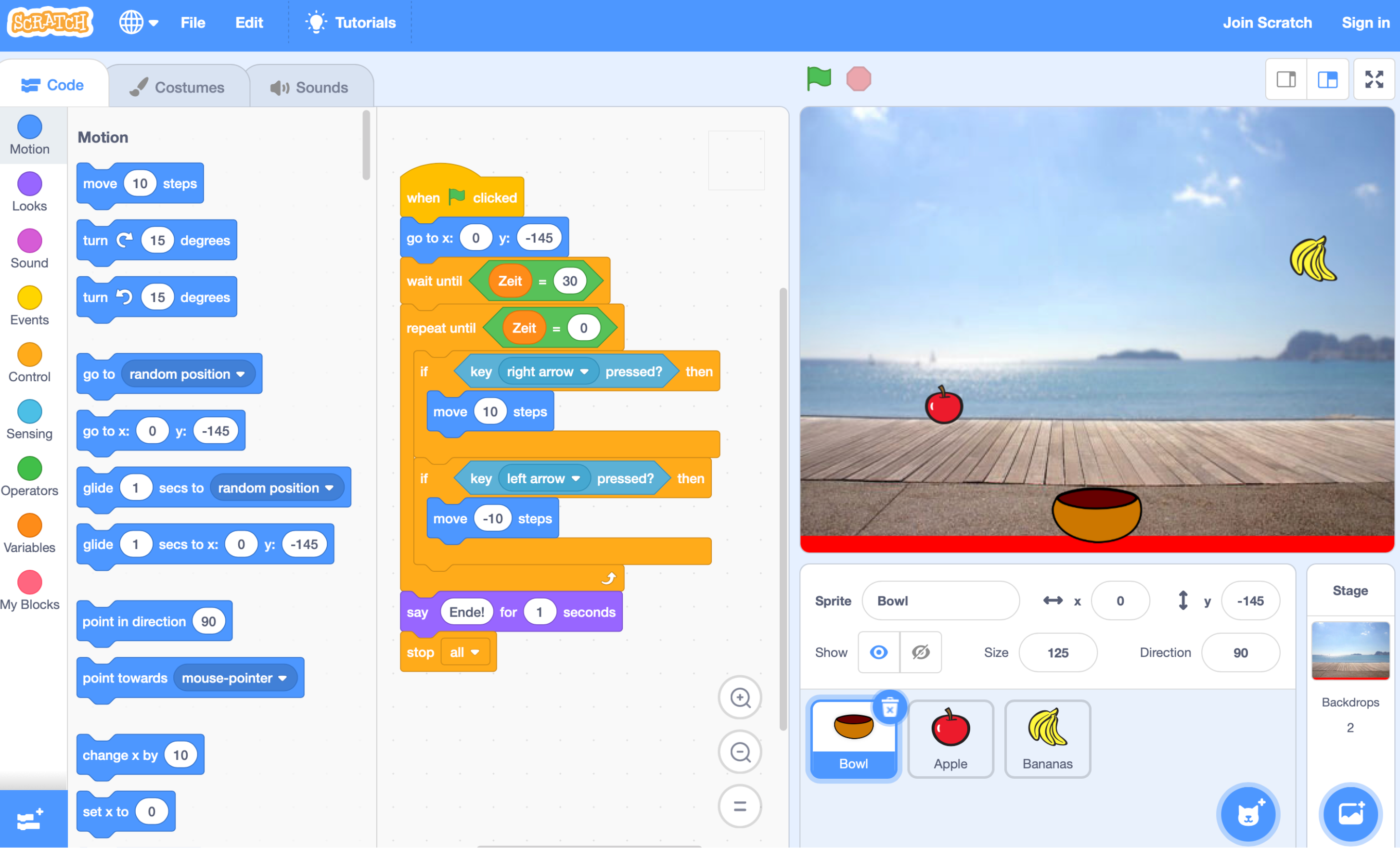}
	\caption{\label{fig:scratch}The \Scratch user interface}
\end{figure}

A \Scratch program 
 revolves around a \emph{stage} on which
graphical \emph{sprites} process user inputs and interact. The program in
\cref{fig:scratch} shows a \Scratch program with three sprites: The bowl, the
bananas and the apple; the stage
contains the background image.
Conceptually, we define a \Scratch program as a set of \emph{actors}, one of
which is the stage and the others are sprites. Actors are rendered on a
canvas; each actor is rendered on a separate layer~\citep{TestingScratch}.
An actor is composed of a set of scripts, a set of custom blocks, sound and
image resources.
The resources are used for example to provide different background images on
the stage, or to decorate sprites with different costumes. The currently chosen
costume is an example of an \emph{attribute} of an actor, and other attributes
include position, rotation, or size. Actors can also define \emph{variables}
which are untyped and can contain numeric or textual data.
%
%
Scripts consist of
individual \emph{blocks} stacked together. The \Scratch language consists of
different types of blocks that have different shapes, and programs are arranged
by combining blocks in ways that are permitted by their shapes.
\begin{itemize}
	\item Hat blocks 
\begin{scratch}
\blockinit{\hspace{1em}...\hspace{1em}}
\end{scratch}:
	Each script can have only one hat block. The hat block represents an event handler that triggers the execution of the corresponding script. If a script does not have a hat block, it is not connected to the program and will not be included in any regular program executions; however, the code can still be executed by the user double clicking on it.
	\item Stack blocks
\begin{scratch}
\blockmove{\hspace{1em}...\hspace{1em}}
\end{scratch},
\begin{scratch}
\blocklook{\hspace{1em}...\hspace{1em}}
\end{scratch}, ...:
	 Regular program statements, for example to control the appearance or motion of sprites, can be stacked on top of each other. The stacking represents the order of the control flow between the statements.
	\item C blocks
\begin{scratch}[scale=0.3]
\blockif{\hspace{1em}...\hspace{1em}}
{
\blockspace[0.4]
}
\end{scratch}:
These blocks are named after their shape and represent control flow (if, if-else, loops). The conditionally executed code is contained within the C-shape. 
	\item Reporter blocks \ovalmove{\hspace{1em}...\hspace*{1em}}: These blocks represent variables and expressions and can be used as parameters of other blocks.
	\item Boolean blocks  \booloperator{\hspace{1em}...\hspace{1em}}: These are special reporter blocks that represent Boolean values.
	%
	\item Cap blocks
\begin{scratch}
\blockstop{\hspace{1em}...\hspace{1em}}
\end{scratch}:
These are blocks after which no stack blocks can be attached as they either terminate the execution or it never proceeds beyond them (e.g., forever loops).
\item Custom blocks \inlineFigure{0.1}{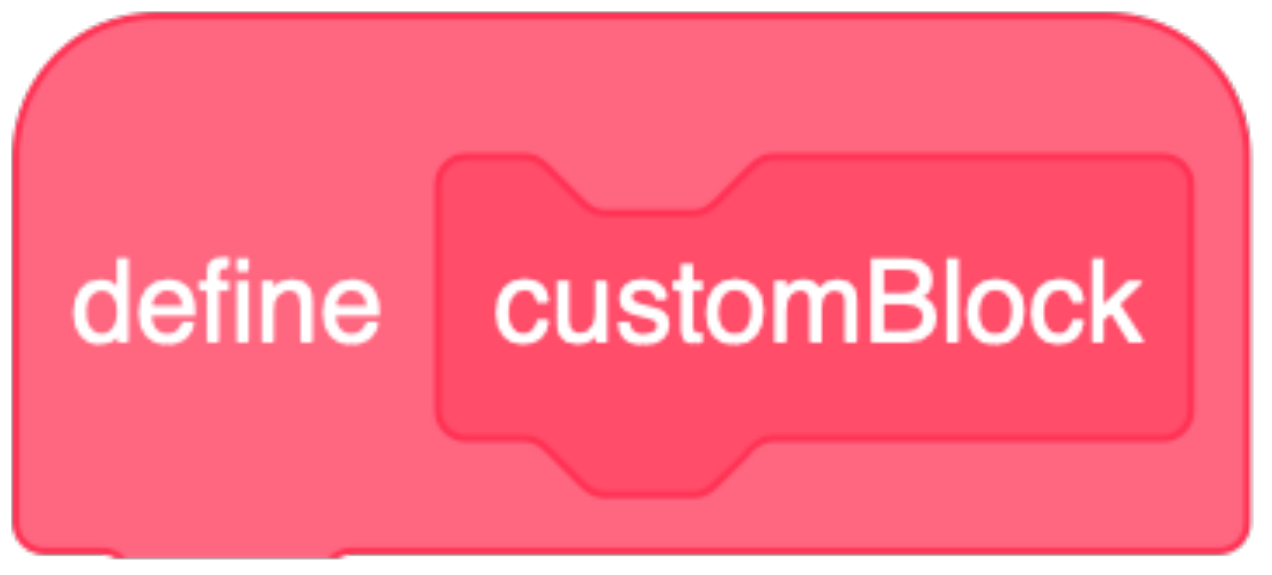} \inlineFigure{0.1}{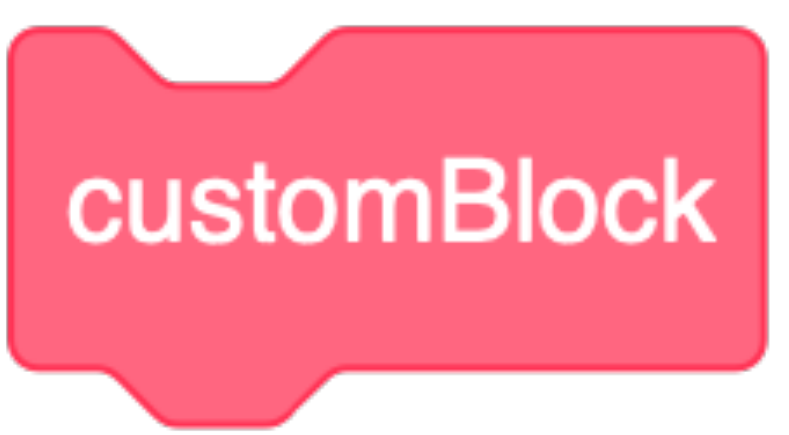}: These blocks are essentially macro scripts. An instance of a custom block triggers the execution of the corresponding macro script.
\end{itemize}


%

\Scratch programs are executed in the \Scratch virtual machine, and controlled by the user 
via mouse, keyboard, microphone, or other input devices. That
is, a program can react to mouse movement, mouse button presses, keyboard key
presses, sound levels, or entering answers to 
\begin{scratch}
  \blocksensing{ask \ovalnum{question?} and wait}
\end{scratch}
blocks. In addition, there is a global \event{Greenflag} event 
\begin{scratch}
  \blockinit{When \greenflag clicked}
\end{scratch}
which represents the user starting the
program through the green flag icon in the user interface (cf. \cref{fig:scratch}). The \Scratch
language also contains broadcast statements
\inlineFigure{0.12}{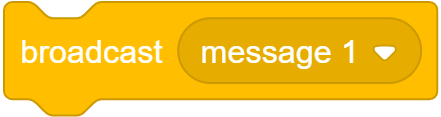}
, which trigger corresponding
message receiver hat-blocks
\begin{scratch}
\blockinit{When I receive \selectmenu{message 1}}
\end{scratch}%
. 
The execution of a script is initiated when the event corresponding to its hat
block occurs, resulting in a \emph{process}~$\process$.
Executing a \Scratch program therefore results
in the creation of a collection of concurrent processes~$\processes$,
and the state of each process is defined by the control location as well as the values of all variables and attributes of the actor.

\begin{figure}[tb]
\centering
\includegraphics[width=0.8\textwidth]{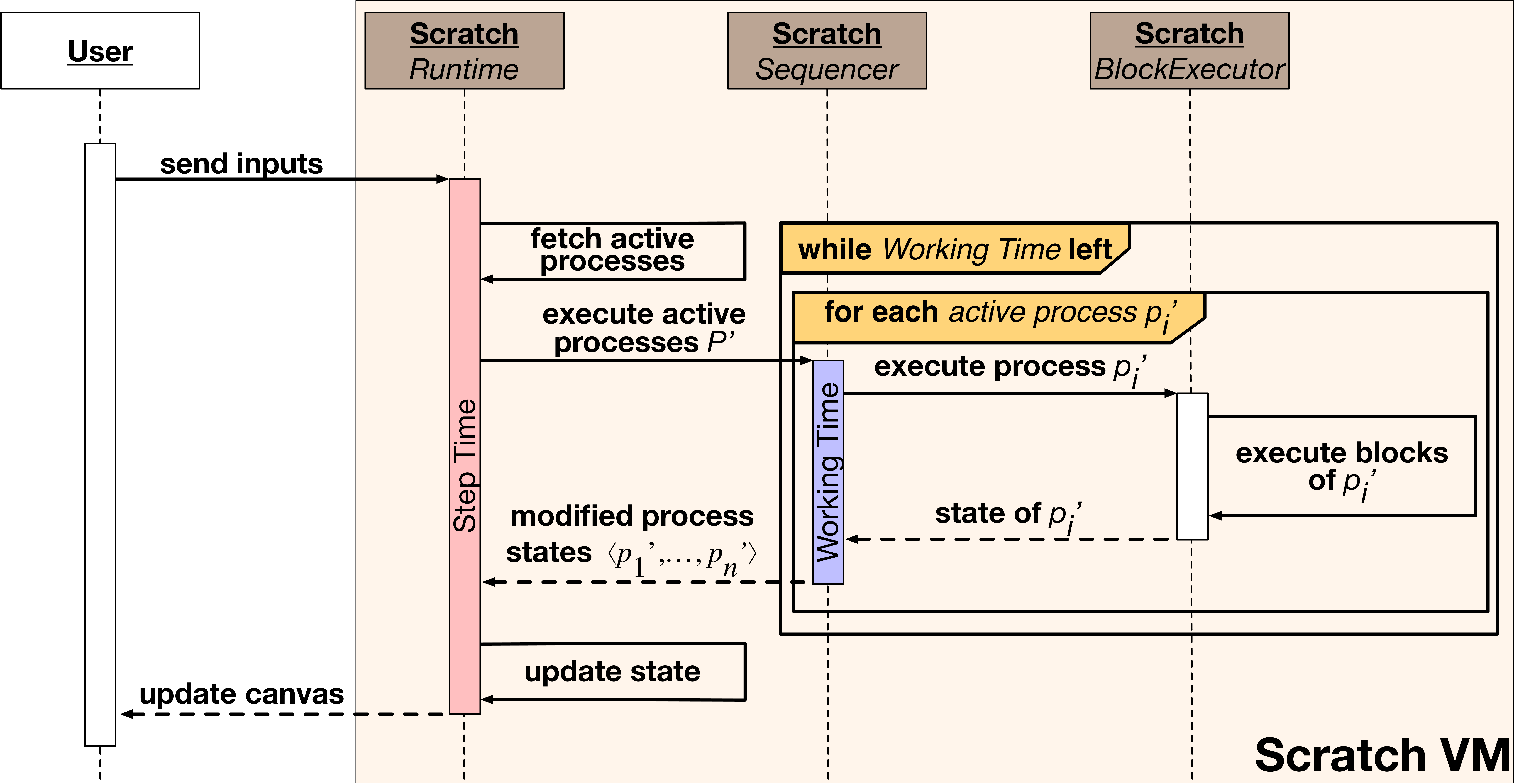}
\caption{Simplified scheduling function of the \Scratch VM.}\label{fig:Scratch-Scheduling}
\end{figure}

%
Execution is operationalised by the \emph{step} function of the virtual machine.
\Cref{fig:Scratch-Scheduling} shows a simplified version of one scheduling step performed by the \Scratch VM to update its internal state.
Each step has a predefined \emph{step time} duration and starts by determining which scripts are currently active and have to be executed. 
The collection of active scripts~$\processes' \subset \processes$ consists of processes triggered by recent user inputs and already active scripts from previous time steps. 
All active processes are then handed over for execution to the \emph{sequencer}. 
The sequencer mimics parallelism by sequentially executing all received processes in batches $\batches = \langle \process_1', \ldots, \process_n' \rangle $ until the \emph{working time}, which is by default set to two-thirds of the \emph{step time}, has elapsed.
In order to avoid non-deterministic behaviour, 
 the execution of a process batch is never interrupted, even if the \emph{working time} has depleted.
Whenever a single process~$\process'$ is scheduled for execution, the process is transferred to the \emph{block executor}.
Upon receiving a script to execute, the block executor processes each block of the given process~$\process'$ until all blocks of the script's process have been executed or specific blocks, forcing the process to halt,
are encountered.
These process halting blocks consist of: 
\begin{itemize}
\item 
\begin{scratch} 
\blockcontrol{wait \ovalnum{x} seconds}
\end{scratch} and \begin{scratch} 
\blockcontrol{wait until  \boolsensing{cond}}
\end{scratch}
blocks, which force the process to wait until a user-defined timeout $x$ has run out or some condition \boolsensing{cond} is met.
\item \begin{scratch} 
\blocklook{think/say \ovalnum{} for \ovalnum{x} seconds} \end{scratch} 
blocks, which create a think/speech bubble for the specified amount of time $x$ on top of the sprite containing the block.
\item \begin{scratch} 
\blockmove{glide \ovalnum{x} seconds to ...}
\end{scratch} blocks, which move the given sprite gradually within a time frame~$x$ to a specified location.
\item \begin{scratch} 
\blocksound{play sound \selectmenu{x} until done}
\end{scratch} blocks, which force the program execution to halt until the defined sound file $x$ has been played completely.
\item \inlineFigure{0.1}{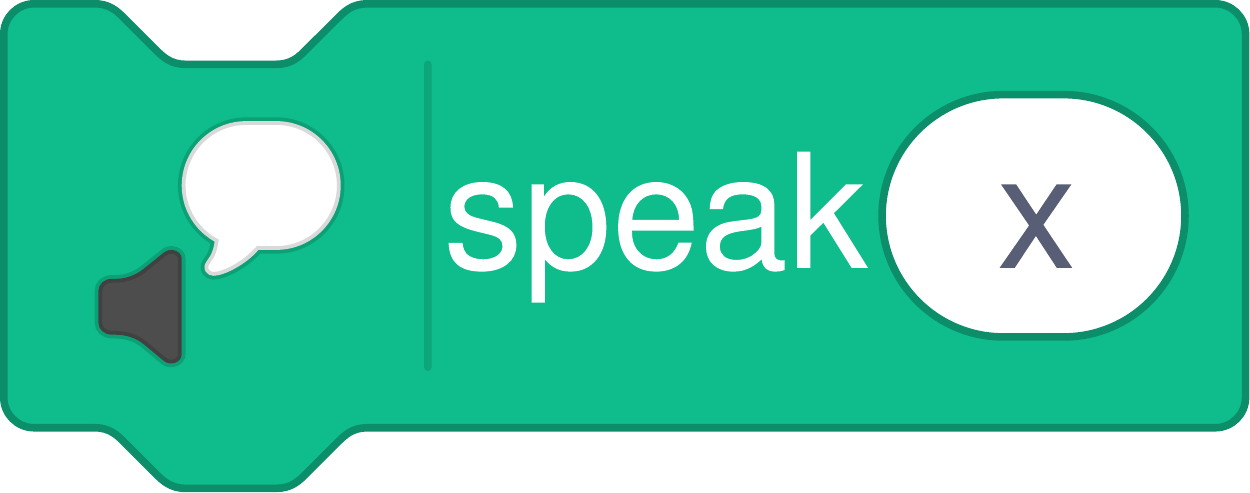} blocks, which work like a \begin{scratch} \blocksound{play sound \selectmenu{y} until done} \end{scratch} block by translating the given text argument $x$ into the sound file $y$.
\item The last block contained within 
\begin{scratch}[scale=0.4]\blockinfloop{forever}{\blockspace[0.4]}\end{scratch}, 
\begin{scratch}[scale=0.4]\blockrepeat{repeat \ovalnum{}}{\blockspace[0.4]}\end{scratch} and 
\begin{scratch}[scale=0.4]\blockrepeat{repeat until \boolempty[2em]}{\blockspace[0.4]}\end{scratch}
blocks, which forces the process to halt until the next process batch is executed.
\end{itemize}

Eventually, the state of the currently executed process~$p'$ is reported back to the sequencer.
As soon as the \emph{working time} has depleted and the full batch of processes has been executed, the collection of modified process states~$\langle \process_1', \ldots, \process_n' \rangle \in P'$ is handed back to the runtime environment.
Finally, the runtime environment updates the internal state 
of the \Scratch VM and notifies the user by redrawing the canvas to meet the state changes.


\subsection{The \whisker Testing Framework}

\label{sec:whisker}

Testing a program means executing the program, observing the program's
behaviour, and checking this behaviour against expectations.
\whisker~\citep{TestingScratch} automates this process for
\Scratch programs: Conceptually, a \whisker test consists of a \emph{test
harness}, which sends user events to the \Scratch program under test, and a set of \emph{\Scratch observers}, which encode
properties~$\properties$ that should be checked on the program under test.

\begin{figure}[tb]
	\centering
\includegraphics[width=0.8\textwidth]{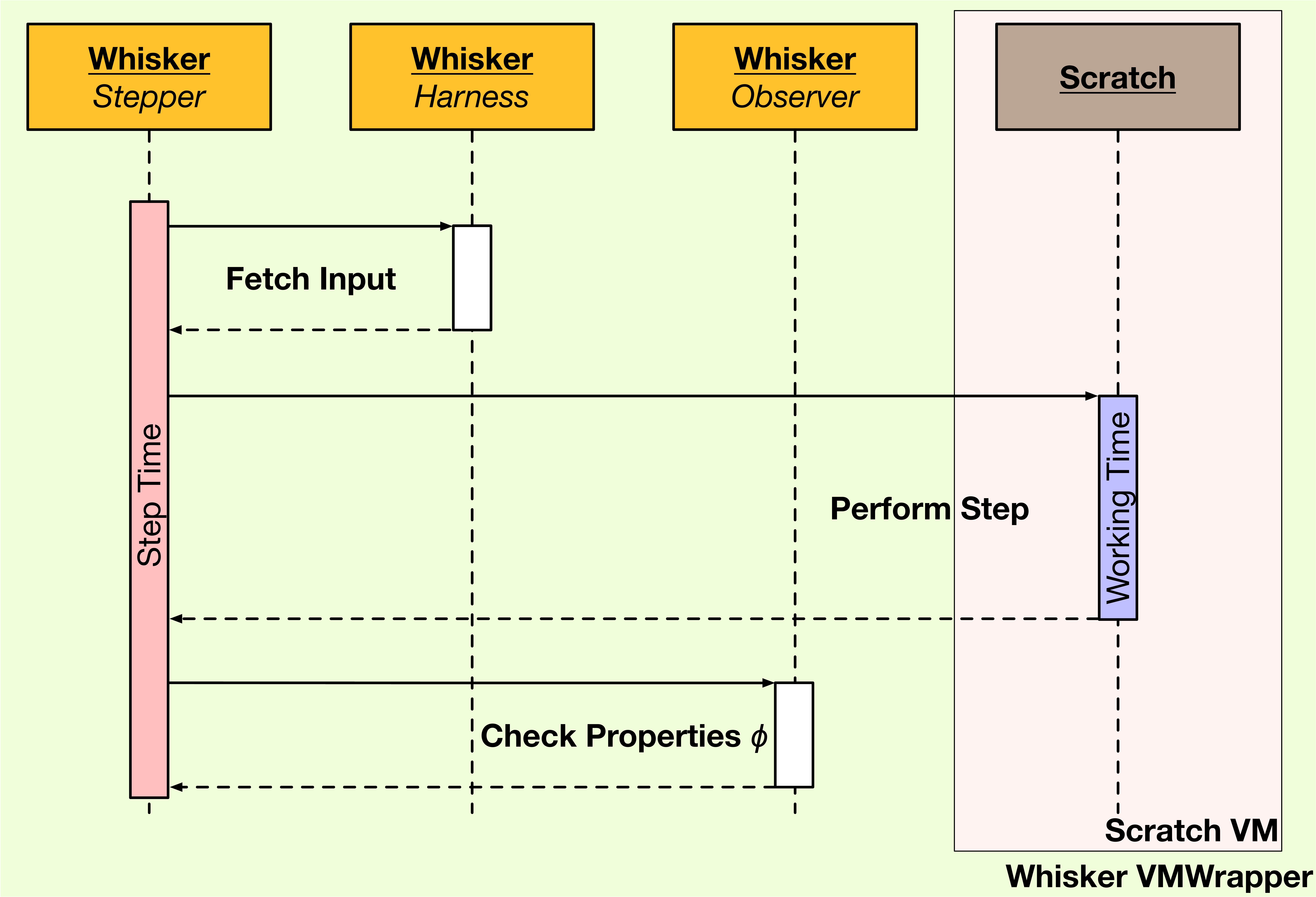}
\caption{Execution step of the \whisker VMWrapper selecting an input, sending that input to the \Scratch VM, and then matching the resulting state against the expected properties~$\properties$.}\label{fig:Whisker-Scheduling}
\end{figure}

As illustrated by \cref{fig:Whisker-Scheduling}, \whisker executes tests by wrapping the \Scratch VM's scheduling function and inheriting its \emph{step time} from the runtime environment. 
First, \whisker queries the test harness for an input to be sent to the \Scratch program under test.
Then it performs a \emph{step} by sending the obtained input in the form of an event to the \Scratch program.
The \Scratch VM then invokes its scheduling function, as shown in \cref{fig:Scratch-Scheduling}.
After the \emph{working time} has expired, the scheduling function stops and reports the new state 
 back to \whisker.
This state is handed over to the test observer, which checks if the actual state 
 matches the expected properties~$\properties$. If it does not, then a failing has been found and is reported in the form of an \emph{error witness}~\citep{ScratchErrorWitness}, which contains the whole input sequence leading to the violation of the given property.

 A static test harness provides inputs encoded in JavaScript to the program. Arbitrary events can be sent based on time intervals or when certain conditions hold.
 As an example, Listing~\ref{lst:example_test} shows a \whisker test for the project in \cref{fig:scratch}. The test consists of pressing the left cursor key for 10 steps, and then checking that the bowl-sprite has not moved to the right.

\begin{center}
\begin{minipage}[t]{0.8\textwidth}
\begin{lstlisting}[basicstyle=\small\ttfamily,language=JavaScript,label=lst:example_test,numbers=none,caption=Example \whisker test case for the game shown in \cref{fig:scratch}]
const test = async function(t) {
    let sprite = t.getSprite('Bowl');
    let oldX = sprite.x;
    t.keyPress('Left', 10);
    await t.runForSteps(10);
    t.assert.ok(oldX >= sprite.x);
    t.end();
}
\end{lstlisting}
\end{minipage}
\end{center}


%
\whisker also supports dynamic test harnesses, where the program is exercised with randomly generated sequences of inputs. Although these are often sufficient to fully cover simple programs, previous work~\citep{TestingScratch} has shown that more complex programs are not always fully covered. In addition, in cases where an instructor or a researcher needs to author tests for multiple students' solutions in one assignment, defining one set of inputs is almost never sufficient to cover all different student programs, where each student may have their own unique implementation for properties such as actor movement speed and game begin/end conditions \citep{wang2021execution}. Finally, specifying \Scratch observers may be easier for a given sequence of user inputs in contrast to specifying the expected outcome for arbitrary sequences inputs.
Therefore, the aim of this paper is to generate static test harnesses, i.e., test suites that reach all statements of a program under test.

\section{Accelerated and Deterministic Test Execution}

The nature of \Scratch programs causes two issues for automated testing: First,
the frequent use of animations and timed behaviour causes executions to
take long. Second, this time-dependent behaviour, the randomised nature of games, and various implementation aspects of \Scratch may lead to non-determinism. To enable
efficient and reliable testing of \Scratch programs we therefore modified the \Scratch VM to decrease the execution time and to make executions deterministic. Additionally, since some 
\Scratch blocks check for sound originating
from a device's microphone, we modified the \Scratch VM to allow
\whisker to generate virtual sound levels 
without 
requiring an actual physical microphone.

\label{sec:execution}
\newcommand{\pHalt}{\ensuremath{\mathsf{pstateHalt}}\xspace}%
\newcommand{\tempArg}{\ensuremath{\mathsf{x}}\xspace}%

\subsection{Accelerating Execution}
\label{section:acceleration}

To increase test
execution speed 
 we apply two essential modifications to the \Scratch VM: First, the rate at which inputs are sent to the \Scratch VM is increased; second, all blocks that halt the execution of a process to wait for time to pass (see \cref{sec:Scratch-Programs}) are instrumented to reduce their time-dependent arguments in proportion to the chosen acceleration factor.

\begin{figure}[tb]
\centering
\includegraphics[width=0.8\textwidth]{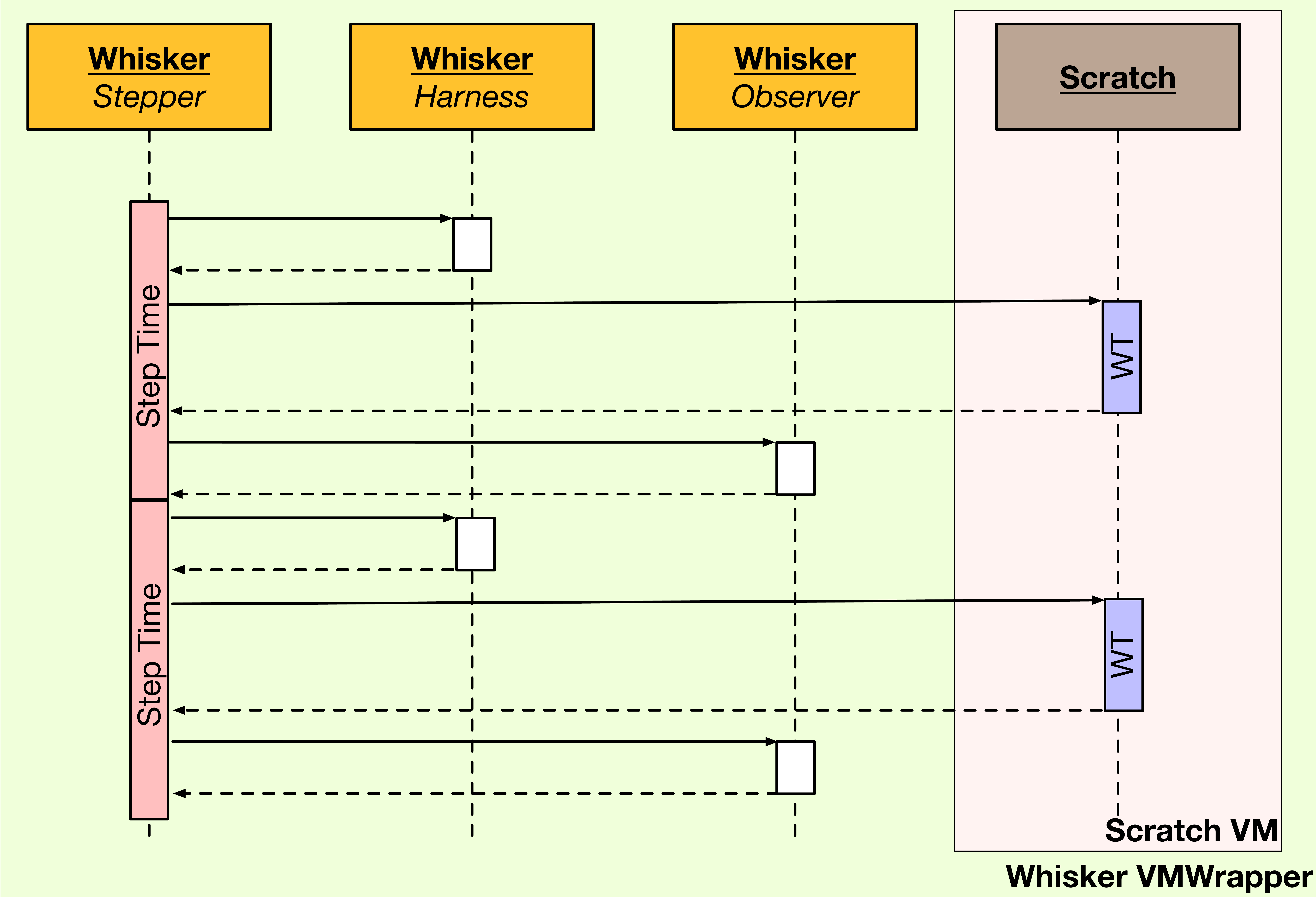}
\caption{Accelerating a \Scratch program by halving \whisker's \emph{step time} and the \Scratch VM's \emph{working time} (``\textsf{WT}'').}\label{fig:Acceleration}
\end{figure}

While the \Scratch VM tries to execute process batches until the \emph{working time}
has elapsed, due to the nature of \Scratch programs most processes sooner or
later hit some statement that causes waiting. Indeed, we have observed that
processes tend to spend more time waiting than executing within a step.
Therefore, to effectively increase the rate at which inputs are sent to the
\Scratch VM, we
reduce the \emph{step time} of the \whisker VMWrapper 
by the selected acceleration factor. 
Since the \emph{step time} is directly linked to the \Scratch VM's \emph{working time}, as explained in \cref{sec:Scratch-Programs}, reducing the \emph{step time} automatically leads to an increased update interval of the \Scratch VM.
Thus, a decrease in the \emph{step time} results in sending inputs more frequently to the 
program under test and furthermore increases the rate at which the \Scratch VM processes 
these inputs.
To illustrate the increased execution speed, \cref{fig:Acceleration} depicts an accelerated 
version of \whisker's scheduling function shown in \cref{fig:Whisker-Scheduling}.
In the accelerated scheduling function, an acceleration factor of two cuts the \emph{step} and 
\emph{working time} in half, resulting in twice as many steps within the same time frame.

However, 
%
 some blocks force a process to enter an execution halting process state 
until a timeout or some user-defined condition is met
(\cref{sec:Scratch-Programs}). Hence, even if the speed at which the \Scratch
VM updates its internal state is increased, these blocks would still force the
process they are contained in to halt for the given amount of time.
Therefore,
statements that contain a time-dependent argument such as \begin{scratch}\blockcontrol{wait \ovalnum{x} seconds}\end{scratch}, \begin{scratch}\blocklook{say \ovalnum{} for \ovalnum{x} seconds}\end{scratch},\begin{scratch}\blocklook{think \ovalnum{} for \ovalnum{x} seconds} \end{scratch}, and \begin{scratch} \blockmove{glide \ovalnum{x} seconds to}\end{scratch} are instrumented to reduce their time argument $\tempArg$ by the given acceleration factor. 
On the other hand, execution halting blocks that do not directly hold any time-dependent arguments like \begin{scratch} \blocksound{play sound \selectmenu{x} until done}\end{scratch} and \inlineFigure{0.1}{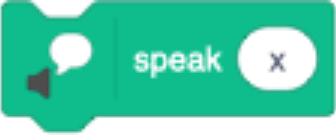} are accelerated by reducing the play duration of the given or translated sound file appropriately.
Lastly, statements that stop the execution of a process until a specific condition \boolsensing{cond?} is met, as the \begin{scratch} \blockcontrol{wait until \boolsensing{cond?}}\end{scratch} block, are not altered at all because these conditions \boolsensing{cond?} already emerge earlier due to accelerated program execution.

By applying both VM modifications to the original \Scratch VM, we construct a modified \Scratch VM used by \whisker, capable of effectively increasing test execution speed by a user-defined acceleration factor.

\subsection{Ensuring Determinism}\label{sec:ensureDeterminism}

\newcommand{\stepCount}{\ensuremath{\mathsf{sc}}\xspace}%
\newcommand{\reqSteps}{\ensuremath{\mathsf{s}}\xspace}%
\newcommand{\activeprocesses}{\ensuremath{\processes'}\xspace}%

Most of the projects created within the \Scratch programming environment represent simple games.
A very prominent characteristic of games is their use of random number generators, which results in non-deterministic behaviour.
Unfortunately, this frequently leads to flaky test suites \citep{EmpiricalStudyOnFlakyTests, FlakyTestsPython}, which is undesirable in any testing tool but especially problematic in \whisker's application scenario:
For example, a \whisker test related to a graded assignment could pass on a student's system but fail on the teacher's machine, leading to a potentially unfair grading process. 
In order to avoid flaky tests originating from randomised program behaviour, \whisker offers the option to seed the \Scratch VM's random number generator to a user-defined seed, by replacing the global \texttt{Math.random} function with a seeded one at runtime.
%

\begin{figure}[tb]
\centering
\includegraphics[width=0.8\textwidth]{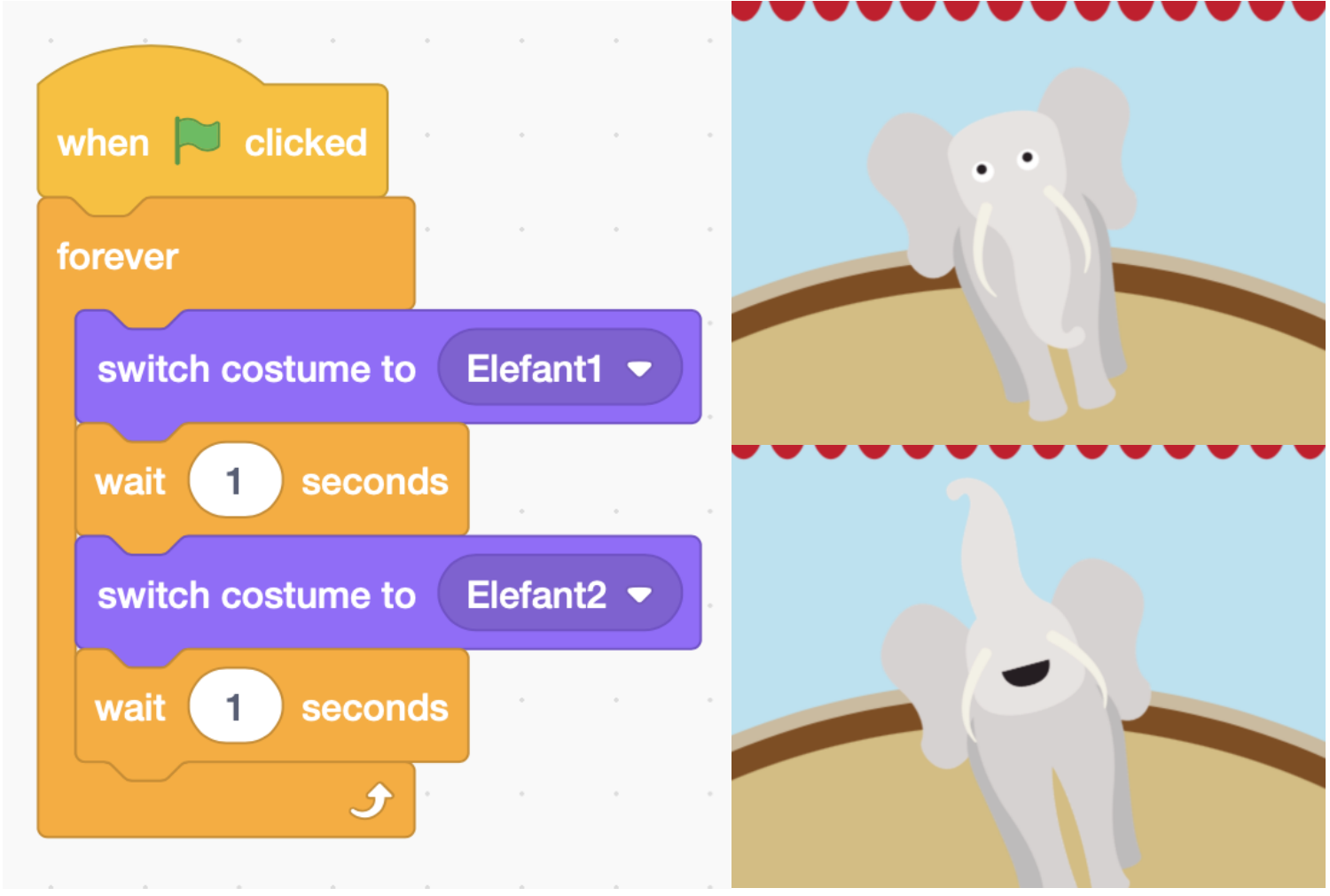}
\caption{\Scratch project containing an elephant that changes its visual appearance every second.}\label{fig:ElephantWait}
\end{figure}

However, random number generators do not pose the only source of non-deterministic behaviour.
In consequence of its many temporal dependencies, the \Scratch VM itself is also susceptible to non-determinism. 
This becomes especially apparent in programs containing blocks that take temporal values as an argument.
To exemplify this problem, consider \cref{fig:ElephantWait}, showcasing a program consisting of an elephant (originating from a study by \citet{geldreich2016programming}) that changes its costume (and thus its visual appearance) every second.
When executing this program for the same amount of time on different machines, the last selected costume might not always be the same.

\begin{figure}[tb]
  \centering
  \subfloat[Effect of code execution speed variances]{\includegraphics[width=\textwidth]{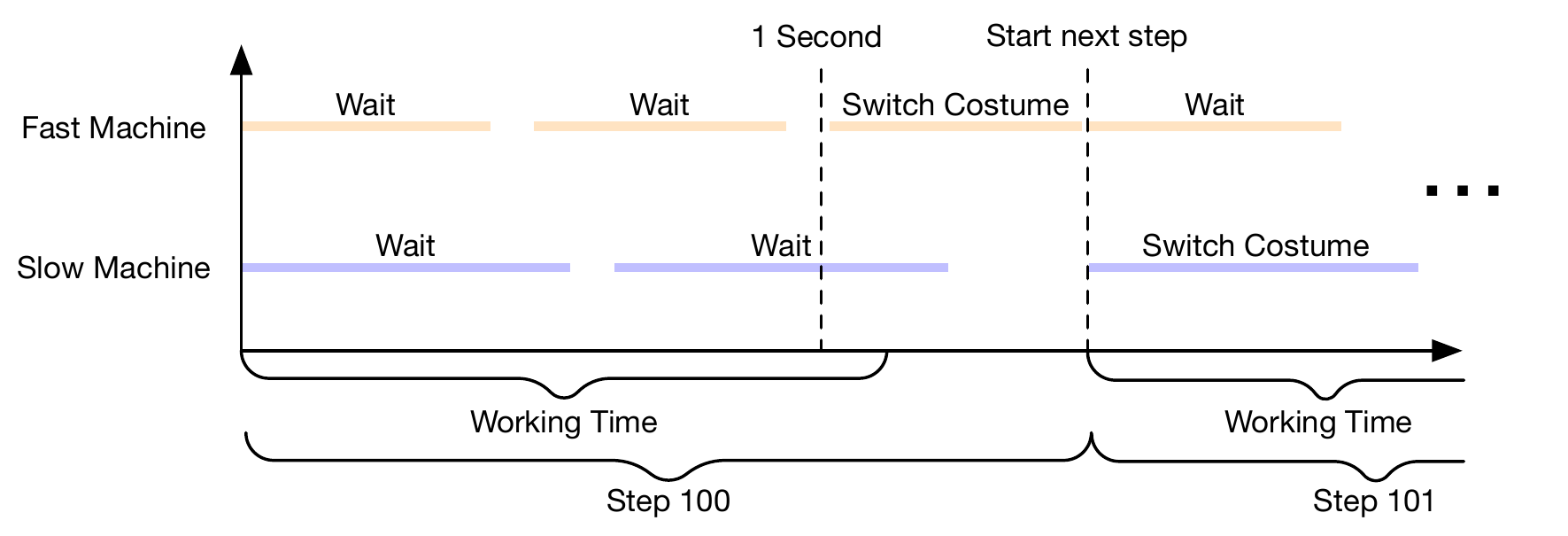}}\\
  \subfloat[State on fast machine]{\includegraphics[width=0.45\textwidth]{figures/ElephantFast}}
  \hfill
  \subfloat[State on slow machine]{\includegraphics[width=0.45\textwidth]{figures/ElephantSlow}}
  \caption{Effect of code execution speed variances on the Elephant project with each beam representing one process batch $\batch \in \batches$: Fast machines execute process batches more frequently within one step and thus have a higher chance of striking the end of the 1 second block earlier than machines having a low execution speed, resulting in diverging elephant states.}
  \label{fig:ExecutionSpeedVar}
\end{figure}

This non-deterministic behaviour originates from the way the sequencer repeatedly executes active processes. 
As depicted in \cref{fig:Scratch-Scheduling}, the sequencer obtains all currently active processes $\activeprocesses$ from the runtime environment and keeps executing batches of those processes until the allocated \emph{working time} has elapsed.
The root cause of non-deterministic behaviour resides in the sequencer processing each batch of processes $\batch \in \batches$ as a whole and its inability to interrupt the execution of a batch $\batch$ even if the allocated \emph{working time} has run out. 
Therefore, differences in code execution speed between systems of diverging performances will allow the sequencer to step through and execute a varying number of process batches in the course of one \emph{working time} interval.
Hence, fast machines have a higher chance of striking a specific time interval earlier than slow machines.
As depicted in~\cref{fig:ExecutionSpeedVar}, for the elephant example, this behaviour eventually leads to earlier costume changes in faster machines.



Accelerating execution may facilitate non-deterministic behaviour since the window of acceptable execution speed variances shrinks proportionally to the acceleration factor used. 
For example, an acceleration factor of \( 5 \) reduces the wait duration in the elephant project from one second down to \( 0.2 \) seconds.
With a wait duration of only \( 0.2 \) seconds, the probability of more performant machines repeatedly triggering a state change faster than less performant ones increases.
Moreover, as illustrated in \cref{fig:Acceleration}, higher acceleration factors decrease the \emph{working time} interval, further enforcing diverging program behaviour.

To eliminate non-deterministic behaviour originating from the scheduling function, we further modify the accelerated \Scratch VM established within \cref{section:acceleration}:
First, time-dependent arguments within specific \Scratch blocks are replaced with a discrete measure which is added to the \Scratch VM and based on the number of executed steps so far. Since the \Scratch VM implements different ways to treat time in different types of blocks, there are multiple different ways this change has to be implemented.
Second, in a similar way, time-dependent \whisker event parameters, defining how long inputs should be sent to the \Scratch VM, are modified to be based on the number of executed steps as well.

In order to replace the imprecise measurement of time, a step counter $\stepCount \in \mathbb{N}$ is added to the \Scratch VM's runtime environment.
The step counter is decoupled from the exact unit of time measured in seconds and responsible for counting the number of steps \whisker has executed so far.
Whenever a time-dependent \begin{scratch} \blockcontrol{wait \ovalnum{x} seconds} \end{scratch} block is encountered during the execution of a \Scratch program, the temporal argument $\tempArg \in \mathbb{Q}_{\geq 0}$ of the waiting block measured in seconds is translated into the corresponding number of steps $\reqSteps \in \mathbb{N}$.
Converting seconds into the appropriate number of steps is done by dividing the time with the fixed \emph{step time}, which already incorporates the chosen acceleration factor. 
For example, concerning the \begin{scratch} \blockcontrol{wait \ovalnum{1} seconds} \end{scratch}  blocks of the depicted elephant project and assuming a \emph{step time} of \( \SI{10}{\milli\second} \), the duration of all waiting blocks $\tempArg$ is transformed into 
\( \reqSteps = \SI{1}{\second} / \SI{10}{\milli\second} =  \num{100} \) steps.
At the time of entering the process halting block, the current step count $\stepCount_0$ is added to the calculated number of steps $\reqSteps$ to obtain the step count value $\stepCount_r = \stepCount_0 + \reqSteps$ at which execution of the halted process can resume.
Every time the process containing the time-dependent block is executed again, the current step count $\stepCount$ is checked, and execution is resumed iff $\stepCount > \stepCount_r$.

Very similar to the wait block is the \ovalsensing{timer} block, which measures the time spent since program execution started or the time passed since a \begin{scratch} \blocksensing{reset timer}\end{scratch} block was encountered.
Non-deterministic real-time measurements in timer blocks are replaced by a new variable, which is increased by a value of \emph{0.075} after every executed step and set back to zero after encountering a reset block.
The value of 0.075 was empirically derived by minimizing the difference to real-time measurements.
Although those changes may introduce slight and, in most cases, non-perceivable behavioral differences to the original \Scratch VM, they nonetheless ensure deterministic program behavior and are therefore preferred over real-time measurements, which will always lead to flaky behavior. 

Whereas wait and timer blocks are realised in \Scratch as custom timers, \begin{scratch} \blocklook{say/think \ovalnum{} for \ovalnum{x} seconds}\end{scratch}, \begin{scratch} \blockmove{glide \ovalnum{x} seconds to}\end{scratch} \begin{scratch} \blocksound{play sound \selectmenu{x} until done}\end{scratch} and \inlineFigure{0.1}{figures/blockSpeak} blocks are implemented differently and hence require special treatment.

Instead of maintaining a simple timer, \begin{scratch} \blocklook{say/think \ovalnum{} for \ovalnum{x} seconds}\end{scratch} blocks withhold a \emph{promise} until a timeout set via JavaScript's \lstinline|setTimeout()| function has elapsed.
The blocked process is kept in the $ \compyield $ 
process state and can only resume its execution if it attains the withhold promise from the yield forcing block. 
In addition to the already present non-determinism caused by the \Scratch VM, the \lstinline|setTimeout()| function is known to be very inaccurate, therefore enforcing non-deterministic behaviour even more\footnote{\url{https://developer.mozilla.org/en-US/docs/Web/API/WindowOrWorkerGlobalScope/setTimeout}}.
Hence, it does not suffice to simply replace the timeout duration $\tempArg$ with the corresponding number of steps $\reqSteps$.
To avoid the problematic JavaScript function, these blocks were modified to resemble a \begin{scratch}\blockcontrol{wait \ovalnum{x} seconds}\end{scratch} block by setting the blocked process into the $\compwait$ 
 process state instead of the $\compyield$ 
  process state.
Furthermore, in order to retain the functionality of \begin{scratch} \blocklook{say/think \ovalnum{} for \ovalnum{x} seconds}\end{scratch} blocks, a speech/think bubble is placed before, and removed after, the simulated wait above the corresponding $\actortype$.

Blocks using \begin{scratch} \blockmove{glide  \ovalnum{x} seconds to}\end{scratch} statements, on the other hand, repeatedly change an $\actortype$'s position on the canvas in relation to the elapsed time.
These blocks are instrumented by transforming the total gliding duration $\tempArg$ to the corresponding number of steps $\reqSteps$.
Additionally, whenever a specific glide block is entered for the first time, we calculate and store the glide terminating step count $\stepCount_r = \stepCount + \reqSteps$, after which the respective block reaches its destination.
Finally, by setting the the current step count $\stepCount$ in relation to the glide terminating step count $\stepCount_r$, the position of the sprite can precisely be determined in each execution step.

Lastly, \begin{scratch} \blocksound{play sound \selectmenu{x} until done}\end{scratch} and \inlineFigure{0.1}{figures/blockSpeak} blocks behave  similar since both halt execution until the given or translated file $x$ has been played. Because sound files specify their duration, the step count $\stepCount_r$ at which execution will resume can be calculated the same way as for the \begin{scratch} \blockcontrol{wait \ovalnum{x} seconds} \end{scratch} blocks by translating the duration into the corresponding number of steps $\reqSteps$.

Besides their translated sound file, \inlineFigure{0.1}{figures/blockSpeak} blocks contain another source of flaky behaviour by querying a remote server to produce a sound file for the text argument $x$.
Due to network uncertainties~\citep{EmpiricalStudyOnFlakyTests}, the translation of the text argument always takes different lengths of time, which in turn delays the calculation of $\stepCount_r$ and eventually leads to non-deterministic behaviour.
To eliminate network uncertainties, we modified the \Scratch VM to translate and cache all resulting sound files of \inlineFigure{0.1}{figures/blockSpeak} blocks during the generation of the block hosting sprite, i.e., before the execution of the \Scratch program.

\begin{figure}[tb]
\includegraphics[width=\textwidth]{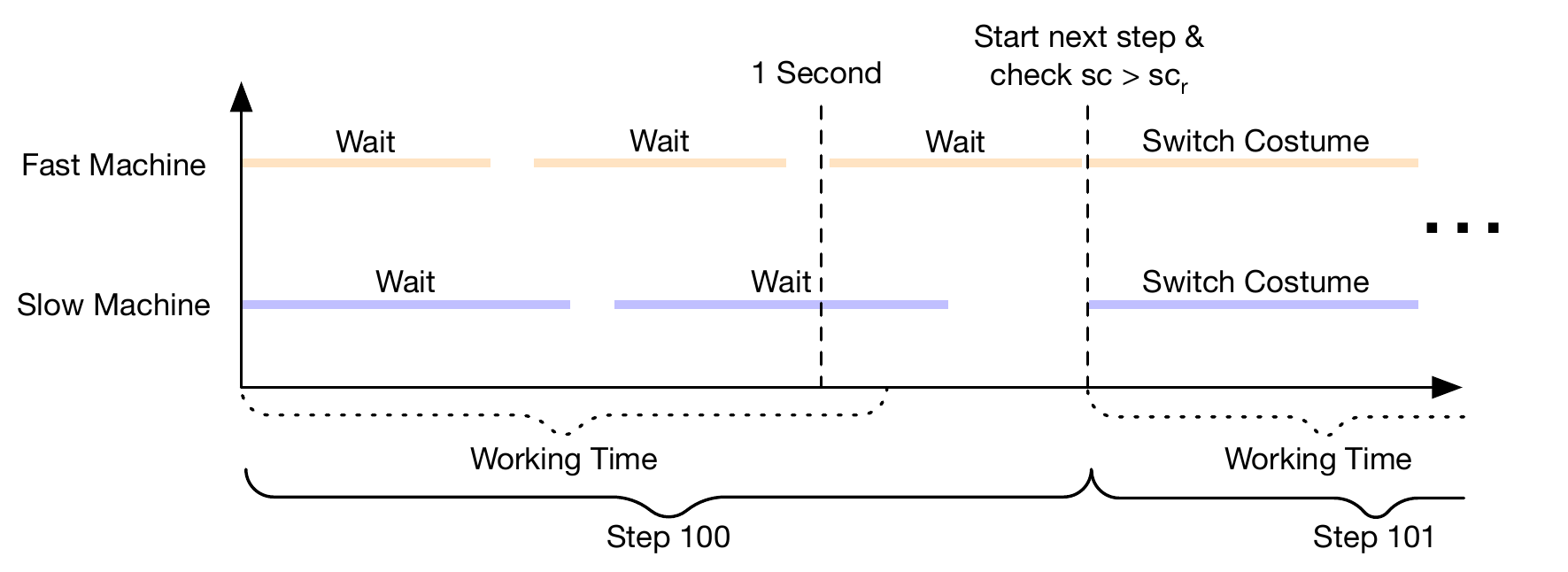}
\caption{Instead of waiting for exactly one second, the costume-changing blocks of the elephant now have to wait for exactly 100 steps until they are allowed to change their costume. The \emph{working time} interval is no longer needed.}
\label{fig:StepCounter}
\end{figure}

By abstracting the measure of time $\tempArg$ to logical execution steps $\reqSteps$, we can exactly define at which step count $\stepCount_r$ a halted process can be resumed.
As a consequence, locks originating from execution halting blocks can only be released between steps and no longer within a single step, which makes the \emph{working time} obsolete. 
Therefore, the \Scratch VM's \emph{sequencer} is further modified to only execute a single process batch instead of executing as many process batches as possible within the \emph{working time}.
Since the implemented acceleration technique instruments execution halting blocks and because a single process is executed until the end of its corresponding script or until reaching a process halting block, the removal of the \emph{working time} does not change the behaviour of the \Scratch VM.
Regarding the elephant project, as shown in \cref{fig:StepCounter}, instead of changing the costume as soon as one second has passed, the step counter ensures that the elephant will constantly change its costume every time precisely 100 steps have elapsed, no matter how fast the given machine executes process batches.  

Besides the implementations of the \Scratch blocks, the events that \whisker tests consist of depend on temporal values as well, for example when deciding how long to send a key press to the program under test.
%
To avoid diverging program executions, all temporal arguments $\tempArg$ of \whisker events are translated into steps~$\reqSteps$.
Furthermore, to ensure all events can have impact on the \Scratch VM, we enforce a minimal step duration of one step.
Thus, given a step duration of \SI{30}{\milli\second} and a \event{KeyPress} event with a duration of \SI{29}{\milli\second}, the \event{KeyPress} event is translated into an event that presses the key for exactly $\reqSteps = \frac{29}{30} \rightarrow 1$ step.

\subsection{Virtual Sound}
\label{section:virtualSound}

The hat block \begin{scratch} \blockinit{when \selectmenu{ loudness} $>$ \ovalnum{} } \end{scratch} as well as the sensing block \ovalsensing{loudness} both check for the presence of a given sound level.
When these blocks are executed, the \Scratch VM tries to determine the current sound level in the user's environment by accessing the device's microphone.
In case a microphone has been detected, the sound level is determined by calculating the \emph{root mean square} (RMS) of the sound wave measured by the microphone.
Then, the RMS value is scaled into the range $[0, 100]$, with $0$ indicating that no noise has been detected and $100$ representing the highest possible sound level.

In order to test \Scratch projects without microphone (e.g., on a compute server), we added virtualised sound to the \Scratch VM using a new variable \emph{virtual sound}.
The virtual sound variable is directly accessible via the \whisker framework and thus can be utilised to simulate sound levels in the range of $[0, 100]$, mirroring the RMS value range. The sound event in \whisker can send sound with a given volume for a predefined number of steps to the program under test.
As soon as the defined number of steps have elapsed, \whisker sets the virtual sound level to a value of $-1$ to indicate the end of the simulated sound sequence.
To guarantee that a given program recognises the simulated sound, we further modified the \Scratch VM to always check for the presence of virtual sound first before trying to access a microphone.
However, to still allow the \Scratch VM to fall back to its default behaviour using a physical microphone to detect sound, we use a value of $-1$ to indicate that no virtual sound is currently intended to be sent to the \Scratch VM.

\newcommand{\eventtype}{\ensuremath{\mathit{ev}}\xspace}%

\newcommand{\ev}{\ensuremath{e}\xspace}%
\newcommand{\events}{\ensuremath{E}\xspace}%
\newcommand{\enq}{\ensuremath{\circ}\xspace}%
\newcommand{\programState}{\concrete}%

\section{Test Generation for \Scratch}\label{sec:event-selection}

\begin{figure}[tb]
  \centering
  \subfloat[Stage with sprites]{\includegraphics[width=0.32\textwidth]{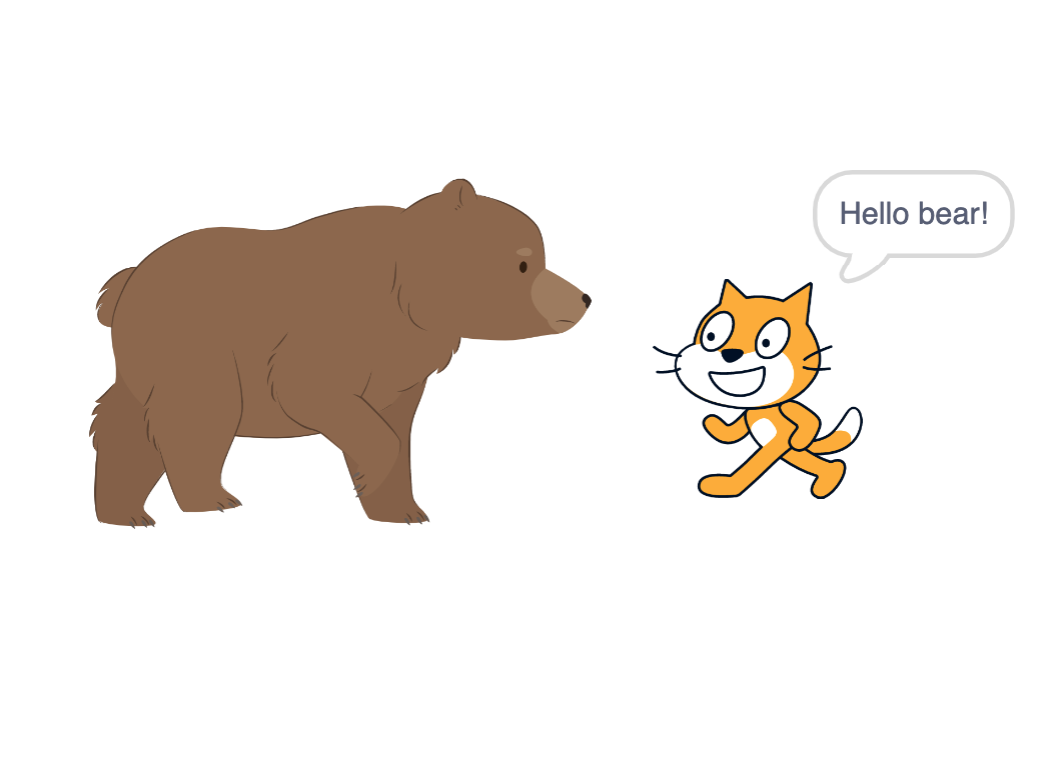}}
  \hfill
  \subfloat[Script of the cat]{\includegraphics[width=0.3\textwidth]{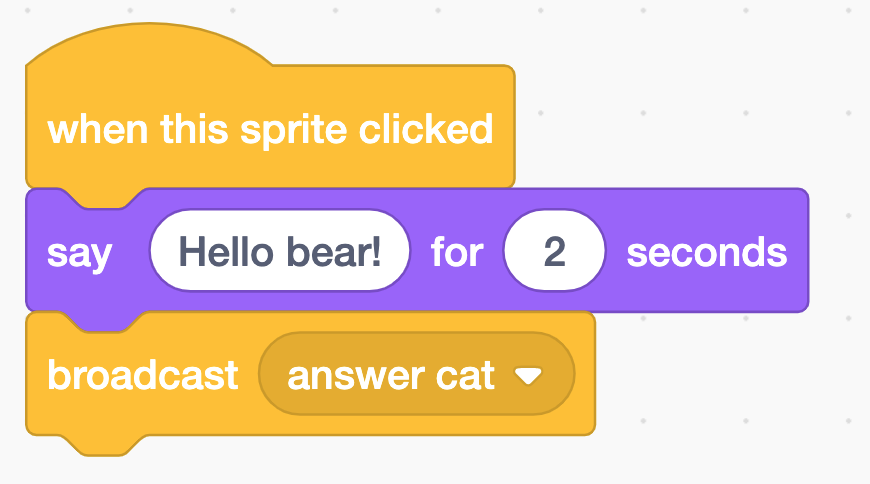}}
  \hfill
  \subfloat[Script of the bear]{\includegraphics[width=0.3\textwidth]{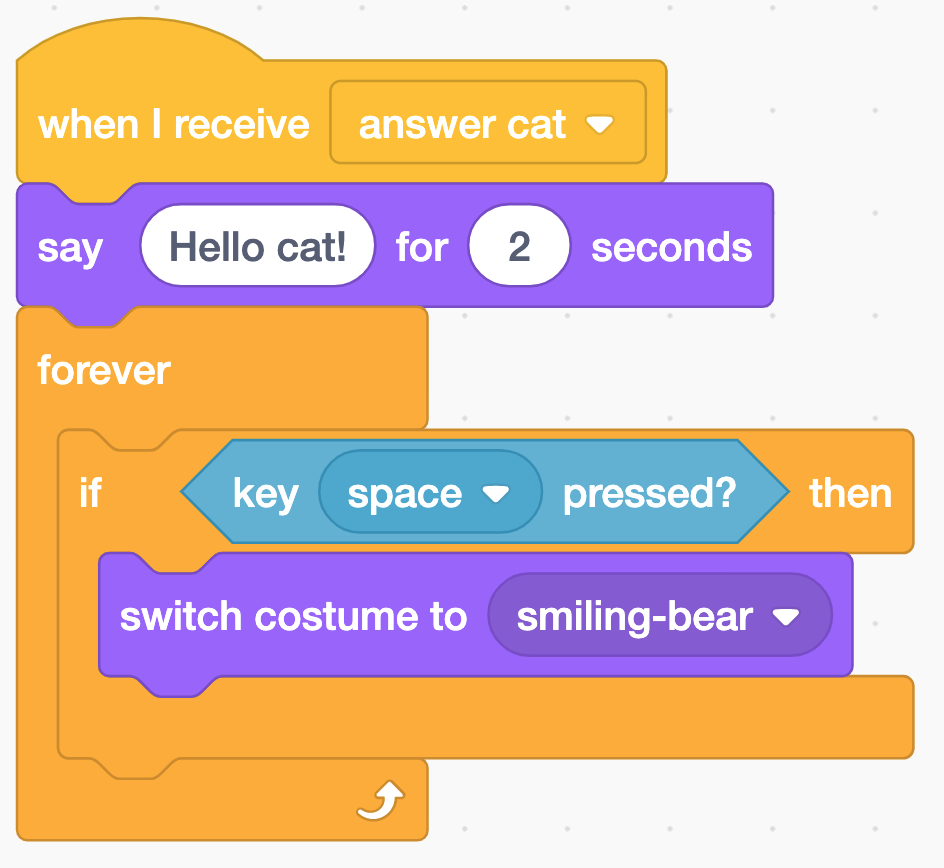}\label{fig:bearScript}}
  \caption{Example \Scratch program: The cat says ``Hello bear!'' when clicked and broadcasts the message ``answer cat''. The bear receives this message and then says ``Hello cat!''. Afterwards, when the space key is pressed the bear switches its costume to ``\texttt{smiling bear}''.}
  \label{fig:example2}
\end{figure}

A \Scratch program processes streams of user inputs (e.g., keyboard and mouse
events) to update program states. As an example, consider the \Scratch program
shown in \cref{fig:example2}, which contains two sprites, a \emph{cat} and a
\emph{bear}. The program starts when a player clicks on the cat sprite. The cat
first greets the bear by saying ``Hello bear!'' for 2~seconds. Afterwards, the
script of the bear receives the message that they need to answer the cat, and then
greets the cat back by saying ``Hello cat!'' for 2~seconds. Then, if the
user presses the space key, the bear will change to a ``\texttt{smiling bear}'' costume.
This program includes two scripts and two types of input events: clicking on
the cat sprite and pressing the space key. Consequently, testing this program
would involve repeatedly sending either of these two events to the program.

\begin{table}[tb]
\caption{Events supported by \whisker.}
\label{tab:events}
\begin{tabularx}{\textwidth}{lp{2cm}X} \toprule
Event & Parameter list & Description \\ \midrule

\event{Greenflag} & --- & Starts \Scratch program execution.  \\

\event{KeyPress} & Key and duration & Presses a certain key for a specified duration.  \\

\event{ClickSprite} & Sprite & Clicks on a sprite or one of its clones. \\

\event{ClickStage} & --- & Clicks on some location that is not occupied by sprites. \\

\event{TypeText} & Text & Enters a string from the pool of (1) all strings used in direct comparisons with the \ovalsensing{answer} block,
(2) a randomly generated string, and (3) fixed seed strings (`0, `10', `Hello').
 \\

\event{TypeNumber} & Number & Inserts the supplied number into a text field. \\

\event{MouseDown} & Status & Toggles the mouse button status (pressed/not pressed). \\

\event{MouseMove} & \( (x, y) \) coordinates & Moves the mouse to the target coordinates. \\

\event{MouseMoveTo} & Sprite & Moves the mouse to the position of the target sprite. \\

\event{DragSprite} & Sprite/colour/edge and angle &  Drags the sensing sprite to a sensed sprite/colour or randomly selected edge on the canvas. Additionally, if angle $\leq 360$ the determined dragging location is moved along the angle direction by the size of the sensing sprite. \\

\event{Sound} & Volume and duration & Sends virtualised sound to the \Scratch VM for a specified duration \\

\event{Wait} & Duration  & Waits for a specified duration. \\ \bottomrule

\end{tabularx}
\end{table}

In order to systematically generate test inputs for \Scratch
programs, we derived a representative set~\events of input events and
encoded them into \whisker:
\begin{align*}
\events = \{ &\event{Greenflag}, \event{KeyPress}, \event{ClickSprite},
                \event{ClickStage}, \event{TypeText}, \event{TypeNumber}, \event{MouseDown}, \\
               &\event{MouseMove}, \event{MouseMoveTo}, \event{DragSprite},
                \event{Sound}, \event{Wait}
            \}
\end{align*}
Some events require parameters, such as \( (x,y) \) coordinates
for \event{MouseMove}. Thus, a user event is fully defined by a tuple
$(e, v)$ consisting of an event type $e \in \events$ and a list of parameter
values $v = \langle v_1, v_2, \ldots, v_i \rangle$.
%
\Cref{tab:events} summarises the supported events and their parameters.

\subsection{Test Generation}
\SetKwFunction{RandomTestGeneration}{randomTestGeneration}
\SetKwFunction{random}{random}
\newcommand{\plus}{\mathbin{+\!\!\!+}}
\newcommand{\minus}{\setminus}

\begin{algorithm}[t]
  \Input{number~\( N \in \mathbb{N} \) of test cases to generate}
  \Output{test suite~\( T = \List{t_{1}, \ldots, t_{N}} \) (a sequence of test cases)}

  \Fn{\RandomTestGeneration{N}}{
    $ T \gets \List{} $\;

    \For{\( i \gets 1 \) \KwTo \( N \)}{
      $e \gets \random(\events \minus \List{\event{Greenflag}})$\;
      \( \List{p_{1}, \ldots, p_{n}} \gets \) parameter list of~\( e \)\;
      $v \gets \List{}$\;
      \For{\( j \gets 1 \) \KwTo \( n \)}{
        \( V_{j} \gets \) value domain of parameter~\( p_{j} \)\;
        $v \gets v \plus \List{\random(V_{j})} $\;
     }

     $T \gets T \plus \List{(e, v)} $\;
    }

    \Return{$T$}\;
  }

  \caption{Random test generator}
  \label{alg:random_test}
\end{algorithm}

A test case  $t = \langle e_1, e_2, e_3, \ldots e_i \rangle$ consists of a sequence of events and their parameters. In the simplest form of test generation, a test case may be produced by combining several randomly selected events from the sequence of available events~\events until the desired test case length has been reached.
The \RandomTestGeneration \cref{alg:random_test} repeatedly chooses an event from \events and determines parameters for the chosen event using the \random function, which randomly selects a single element from a set of elements~$S$.
Note we are excluding the \event{Greenflag} event from \events since by
design \whisker automatically sends this event at the beginning of every test
execution.
While straight-forward to implement, this strategy might be
ineffective for test generation as events can be chosen from~\events for which no
corresponding handler exists in the project under test.
For example, in \cref{fig:example2}, sending events of category \event{MouseDown} or \event{Sound} is pointless since the program cannot respond to these events.
Thus, more fine-grained event extraction methods which filter irrelevant events from \events are needed.

\subsection{Event Extraction}
The inclusion of irrelevant events can be avoided by considering only events
for which event handlers exist in the source code. Event handlers can take
on two different forms in \Scratch. First, there are event handlers in terms of
hat-blocks starting a script~\( s \) (such as \begin{scratch} \blockinit{when I
receive \selectmenu{ answer cat} } \end{scratch} and \begin{scratch}
\blockinit{When this sprite clicked} \end{scratch} in Fig.~\ref{fig:example2}).
Second, it is also possible to query the state of mouse and keyboard through
sensing blocks. For example, in the program shown in \cref{fig:example2},
the sensing block \boolsensing{key \ovalsensing*{ space} pressed?} observes if the space key is pressed.
Thus, a corresponding \event{KeyPress} event should be included in the sequence of available events \events.
\Cref{tab:block2event} summarises the event handling blocks for all the supported user events.
\setscratch{scale=0.7} 
\begin{table}[tb]
	\caption{\label{tab:block2event}Mapping of \Scratch blocks to the corresponding \whisker event handlers}
\begin{tabularx}{\textwidth}{lX}
	\toprule
	Block & Event \\ \midrule

	\begin{scratch}
	\blockinit{When \selectmenu{space} key pressed}
	\end{scratch}
	\boolsensing{key  \ovalsensing*{space} pressed}
	& \event{KeyPress} \\

	\begin{scratch}
	\blockinit{When this sprite clicked}
	\end{scratch}
	 & \event{ClickSprite}\\

	\begin{scratch}
	\blockinit{When stage clicked}
	\end{scratch}
	& \event{ClickStage}\\

	\begin{scratch}
	\blocksensing{ask \ovalnum{What's your name?} and wait}
	\end{scratch}
	& \event{TypeText}\\

  \begin{scratch}
  \blocksensing{ask \ovalnum{What's your age?} and wait}
  \end{scratch}
  followed by numeric  operations on \ovalsensing{answer}
  & \event{TypeNumber} \\

	\boolsensing{mouse down}
	& \event{MouseDown}\\

	\ovalmove{mouse x}
	\ovalmove{mouse y}
	\ovalmove{distance to \ovalsensing*{mouse-pointer}}
	& \event{MouseMove}\\

	\begin{scratch}
	\blockpen{pen down}
	\end{scratch}
	\begin{scratch}
	\blockmove{point towards \ovalsensing*{mouse-pointer}}
	\end{scratch}
	\boolsensing{touching  \ovalsensing*{mouse-pointer}}
	& \event{MouseMove}\\

	\boolsensing{touching  \ovalsensing*{mouse-pointer}}
	& \event{MouseMoveTo}\\

	\boolsensing{touching  \ovalsensing*{Sprite 1}}
	\boolsensing{touching color \ovalsensing{\pencolor{red!75!black}}}
	\boolsensing{touching edge}

	& \event{DragSprite}\\

	\begin{scratch}
	\blockinit{when \selectmenu{loudness} \textgreater\ \ovalnum{10}}
	\end{scratch}
	\ovalsensing{loudness}
	& \event{Sound}\\

	\bottomrule

\end{tabularx}
\end{table}
\setscratch{scale=0.5} 

\SetKwFunction{StaticEventExtraction}{staticEventExtraction}
\newcommand{\staticevents}{\ensuremath{\events_{\text{static}}}\xspace}

\newcommand{\dynamicevents}{\ensuremath{\events_{\text{dynamic}}}\xspace}
\SetKwFunction{DynamicEventExtraction}{dynamicEventExtraction}

Extracting only events for which corresponding event handlers exist in the \Scratch program may still include irrelevant events:
On the one hand, sensing blocks cannot process events if their scripts are inactive.
For example, in \cref{fig:example2} there is no point in sending  \event{KeyPress space} events when the forever loop in script \cref{fig:bearScript} is not executing.
On the other hand, if an event which triggers a hat block of a currently active script is sent to the program, the execution of that script is halted and starts anew.
Therefore, some sensing blocks potentially never become activated because their scripts are constantly restarted.

Consequently, \DynamicEventExtraction
 considers the source code of the \Scratch program as well as the current program state \programState, as shown in \cref{alg:dynamic_extraction}.
The filtered events 
are formed by extracting events for which corresponding sensing blocks occur within active scripts and collecting events from hat blocks whose scripts are currently inactive.
For example, in Fig.~\ref{fig:example2}, at the beginning of the program execution, the event set only includes \event{ClickSprite(cat)}.
After 2 seconds, when the script in bear starts to execute, \event{ClickSprite(cat)} is removed from the event set, and \event{KeyPress} is added to the event set.
This allows the input events to only include highly relevant events at each moment of program execution.
Furthermore, because only active sensing blocks are considered, loose sensing blocks in unconnected scripts that have no hat block responsible for starting the execution of the script are automatically removed 
as well.
Lastly, whenever an event $ e \in \{ \event{TypeText}, \event{TypeNumber} \} $ is extracted, we follow the lead of the \Scratch environment, setting the focus of the user interface to the input field, and further restrict the sequence of available events to $\List{ e, \event{Wait} }$.

\begin{algorithm}[tb]
  \Input{the current program state \programState from which events will be extracted}
  \Output{set of events ~\dynamicevents}

  \Fn{\DynamicEventExtraction{\( \programState \)}}{
    \( \dynamicevents \gets \List{} \)\;

    \BlankLine

    \( A \gets \) currently active scripts in~\( \programState \)\;
    \ForEach{script $a \in A$} {
      $h \gets$ hat block of \( a \)\;
      \ForEach{block \( b \in a \)}{
        \If{\( b \neq h \)}{
          \( e \gets \) extract event from \( b \) including inferable parameters\;
          $\dynamicevents \gets \dynamicevents \plus \List{ e } $\;
        }
      }
    }

    \BlankLine

    \( I \gets \) inactive scripts in \( \programState \)\;
    \ForEach{script $i \in I$} {
      $h \gets$ hat block of \( i \)\;
      \( e \gets \) extract event from $h$ including inferable parameters\;
      $\dynamicevents \gets \dynamicevents \plus \List{ e } $\;
    }

  \BlankLine

  \uIf{$\event{TypeText} \in \dynamicevents$} {
   $\dynamicevents \gets \List{ \event{TypeText}, \event{Wait} } $\;
  }\ElseIf{$\event{TypeNumber} \in \dynamicevents$} {
    $\dynamicevents \gets \List{ \event{TypeNumber}, \event{Wait} } $\;
  }

  \BlankLine

  \Return{\( \dynamicevents \)}\;
  }

    \caption{\label{alg:dynamic_extraction}Dynamic Event Extraction}
\end{algorithm}

%
%
%
%
%
%
%


\begin{table}[tb]
	\caption{\label{tab:inferredParameter} Inferable and non-inferable parameters mapped to their respective events.}
\begin{tabularx}{\columnwidth}{XXX}
	\toprule
	Event & Inferable parameter & Non-inferable parameter\\
	\midrule

 	\event{KeyPress} & Key & Duration \\

 	\event{ClickSprite} & Sprite  & ---\\

 	\event{ClickStage} & --- & --- \\

 	\event{TypeText} & Text & ---\\

 	\event {TypeNumber} & --- & Number \\

 	\event{MouseDown} & Press-status & ---\\

 	\event{MouseMove} & --- & \( (x, y) \) coordinates \\

 	\event{MouseMoveTo} & Sprite & --- \\

 	\event{DragSprite} & Sprite/colour/edge & Angle \\

 	\event{Sound} & Volume, fixed duration & ---\\

 	\event{Wait} & --- & Duration \\

	\bottomrule

\end{tabularx}
\end{table}

As the \DynamicEventExtraction observes the current program state, it is also capable of inferring specific event parameters automatically.
For instance, if a \event{DragSprite} event is chosen due to a \boolsensing{touching  \ovalsensing*{target sprite}} block, the target sprite's position can precisely be determined from the program state.
An overview of all events classified into inferable and non-inferable parameters is shown in \cref{tab:inferredParameter}.
Since the \DynamicEventExtraction extracts information about the current state of the program, the execution of the program has to be interleaved with the selection of events, as shown in \cref{alg:dynamic-random-test}.
Furthermore, because we assume the \event{Greenflag} event to be similar to the initiation of a regular program execution, the event is always sent at the beginning of test execution and never after the test execution has started. 

\SetKwFunction{RandomTestGenerationDynamic}{randomTestGeneration\textsubscript{dynamic}}

\begin{algorithm}[tb]
  \Input{program~\( p \) for which to generate tests}
  \Input{number~\( N \in \mathbb{N} \) of tests to generate}
  \Output{test suite~\( T \)}

  \Fn{\RandomTestGenerationDynamic{\( p \), \( N \)}}{

    \( \programState \gets \) start~\( p \) by sending \event{Greenflag} and obtain concrete program state\;
    $ T \gets \langle \rangle $\;
    \For{$i \gets 1$ \KwTo $N$} {
      $\dynamicevents \gets \DynamicEventExtraction(\programState)$\;
      $e \gets \random(\dynamicevents)$\;
      \( params \gets \) parameter list of \( e \)\;
      \( \List{params_{1}, \ldots, params_{n}} \gets \) inferable parameters in~\( params \)\;
      $v \gets \List{}$\;
      \For{$j \gets 1 $ \KwTo $n$}{
         \( V_{j} \gets \) value domain for~\( params_{j} \)\;
         \( v \gets v \plus \List{\random(V_{j})} \)\;
      }
      $T \gets T \plus \List{(e, v)}$\;
      $\programState \gets$ send \( (e,v) \) to \( p \);
    }

    \Return{$T$}\;
  }

  \caption{Random test generator using dynamic event extraction}
  \label{alg:dynamic-random-test}
\end{algorithm}

\subsection{Assertion Generation}\label{sec:assertions}

\begin{table}[tb]
	\caption{\label{tab:assertions} Types of assertions supported by the \whisker test generator.}
\begin{tabularx}{\columnwidth}{llX}
	\toprule
	Assertion & Target & Expected value\\
	\midrule

 	\event{Backdrop} & Stage & Costume (=backdrop) number \\

 	\event{Clone count} & Sprite  & Number of clones \\

 	\event{Costume} & Sprite, Clone & Costume number \\

 	\event{Direction} & Sprite, Clone & Heading $\pm 1$ degree\\

 	\event{Graphics effect} & Sprite, Clone, Effect type & Value \\

 	\event{Layer} & Sprite, Clone & Layer number\\

 	\event{List} & Sprite, Stage & Length \\

 	\event{Position} & Sprite, Clone & $(x, y)$ coordinates $\pm 5$ pixels\\

 	\event{Say} & Sprite, Clone & Existence, type, and contents of speech bubble \\

 	\event{Size} & Sprite, Clone & Size attribute\\

 	\event{Touching} & $2 \times$ Sprite or Clone & Whether or not the sprites touch \\

 	\event{Touching edge} & Sprite, Clone & Whether or not the sprite touches an edge\\

 	\event{Variable} & Sprite, Stage & Variable value\\

 	\event{Visibility} & Sprite, Clone & Visibility attribute\\

 	\event{Volume} & Sprite, Clone & Volume attribute\\

	\bottomrule

\end{tabularx}
\end{table}

The test generation algorithm produces tests that consist of sequences of
events. In order to be able to detect faults, a \whisker test case also
requires an \emph{observer} (\cref{sec:whisker}) that checks the observed
behaviour against the expected behaviour. Since we envision that a common
application scenario for \whisker is that tests are generated on a model
solution and then executed on student solutions, we operationalise the \whisker
observer in terms of \emph{regression test assertions} that capture the state
of the model solution.
An assertion is a Boolean function that takes the program state $P$ as input,
and checks one of the properties against an expected value. If the value
deviates, the assertion fails the test case. The assertions implemented in
\whisker are listed in \cref{tab:assertions}: Each assertion is implemented in
terms of code to check the value at runtime, and can synthesise JavaScript code
that implements the observer in the generated \whisker test.

The assertion generation algorithm is based on the approach proposed by
\citet{xie2006augmenting}, which essentially adds an assertion for every
attribute after every step of a test, with expected values derived from the
current version of the program. Since the number of possible assertions for a
\whisker test is proportional to the number of sprites, clones, and events in a
test case, a direct application of the approach by \citet{xie2006augmenting}
would lead to a huge number of assertions, many of which would be irrelevant
for the program under test. While a common approach to filter relevant
assertions is using mutation analysis~\citep{fraser2011mutation}, the long test
execution times of \whisker tests together with the many executions required by
the mutation analysis render this approach impracticable for \whisker tests. We
therefore reduce assertions as follows: We execute each test event by event;
after each event, we determine the values for all possible assertions in the
current state, and compare them against the values of the same assertions in
the previous state. Only if the value of an assertion changes from the previous
to the current state, this assertion is added to the current state.

\section{Test Generation Algorithms}
\label{sec:algorithm}

%

\newcommand{\codon}{\ensuremath{\mathsf{c}}\xspace}
\newcommand{\eventCodon}{\ensuremath{\mathsf{\codon_e}}\xspace}%
\newcommand{\parameterCodon}{\ensuremath{\mathsf{\codon_p}}\xspace}%
\newcommand{\numberParameterCodons}{\ensuremath{\mathsf{n_p}}\xspace}%
\newcommand{\numberCodonGroups}{\ensuremath{\mathsf{n_g}}\xspace}%
\newcommand{\crossoverIndex}{\ensuremath{\mathsf{\psi}}\xspace}%
\newcommand{\relativeCrossoverIndex}{\ensuremath{\mathsf{\psi_r}}\xspace}%

\subsection{Encoding \Scratch Tests Using Grammatical Evolution}
\label{sec:evolution}

A prerequisite for applying search algorithms is a representation amenable to the modifications required by different search
operators. For our application context we aim to evolve test cases, which
consist of sequences of events. One challenge that applies to such sequences is
that there can be dependencies between different events within the sequence.
For example, assume two successive click events on the same sprite, where the
execution of the first click event causes the sprite to be hidden---since the
sprite is hidden, no second click can be performed on the sprite. Consequently,
events cannot be performed in arbitrary order, and search operators that
manipulate events may lead to invalid sequences. Rather than directly encoding
test cases as sequences of events we therefore use an encoding inspired by
Grammatical Evolution~\citep{o2001grammatical}, where the mapping from genotype to
phenotype is performed using a problem-specific grammar $G = \langle T, N, P,
n_s\rangle$. Here, $T$ is a set of terminals, which are the items that will appear in
the resulting phenotype; $N$ are non-terminals, which are intermediate elements
associated with the production rules $P: N \rightarrow (N \cup T)^{*}$; the
element $n_s \in N$ is the start symbol, which is used at the beginning of the
mapping process.


The genotype is typically represented as a list of bits or integers; we use an integer representation (\emph{codons}).
Since a codon can represent the next event that will be executed or determine a non-inferable parameter of an event, we differentiate between \emph{event codons} \eventCodon and \emph{parameter codons} \parameterCodon.
The mapping of a list of codons to the phenotype creates a derivation of the grammar as follows: Beginning with the first production of starting symbol $n_s$ of the grammar, for each non-terminal $x$ on the right-hand side of the production we choose the $r$th production rule for $x$. 
Given an event codon \eventCodon and $n$ productions for non-terminal $x$, the number $r$ of the production rule to choose is determined as follows:
\begin{equation}\label{eq:codonMapping}
  r  = \eventCodon \bmod n
\end{equation}
In our application scenario, $n$ represents the number of available events (\cref{sec:event-selection}).
The resulting number $r$ is then used as index for selecting an event from the available events~\events.
If a look-up of \cref{tab:inferredParameter} indicates that the selected event contains $j$ non-inferable parameters, the subsequent $j$ parameter codons \parameterCodon of the genotype are then queried to determine the required parameters based on a unique parameter mapping for each event type \eventtype.

Note that a single change of one event codon might result in the selection of a different event that consumes more or less non-inferable parameters than the previous event.
Such a change of consumed parameter codons would potentially lead to a diverging interpretation of the remaining codons because former event codons might be treated as parameter codons and vice versa.
Thus, mutating a single event codon could result in an entirely different test execution, leading to considerable jumps within the fitness landscape.
To avoid this problem, we assign each event codon a fixed number of parameter codons \numberParameterCodons, regardless of the number of non-inferable parameter codons a given event consumes.
The fixed number of assigned parameter codons \numberParameterCodons is defined by the maximum number of consumed codons across all processable events of a given project \staticevents.
A genotype is then incrementally translated into a phenotype by transforming each event codon to the corresponding \Scratch event using \cref{eq:codonMapping} and consuming the event-specific number of required event parameter $j \leq \numberParameterCodons$.

\cref{alg:codon-decoding} describes the simultaneous decoding and execution of a codon sequence and demonstrates
how concrete parameter values are chosen.
The \event{DragSprite} event constitutes a special case since the dragging location defined by the event extractor tends to show unintended side effects. Consider, for example, a game in which the player has to reach a specific position at the end of a maze without touching a wall. If the \event{DragSprite} event moves the player sprite to the goal position in order to trigger code related to winning the game, the player sprite might also overlap with a wall, leading to a Game Over state instead.
To avoid these side effects, the \event{DragSprite} event consumes an additional parameter codon to slightly move the determined dragging location in the direction of the parameter codon value, which is interpreted as an angle in the range of $[0, 360]$.
Moreover, for \event{KeyPress} and \event{Wait} events, the user can specify an upper bound for the respective wait or keypress duration.

\SetKwFunction{DecodeAndExecute}{decodeAndExecute}
\SetKwFunction{pick}{pick}

\begin{algorithm}
  \SetKwFunction{extractEvents}{extractEvents}

  \Input{program~$p$}
  \Input{list \( \List{c_{1}, \ldots, c_{l}} \) of codons}
  \Input{event extractor~\extractEvents}
  \Input{user configuration $u$}
  \Output{test case~\( t \)}
  \Fn{\DecodeAndExecute{$p$, \( \List{c_{1}, \ldots, c_{l}} \), $\extractEvents$}}{
    $t \gets \langle \rangle $; $i \gets 0 $\;
    $\textsf{UpperBoundKeyPressDuration} \gets u$\;
    $\textsf{UpperBoundWaitDuration} \gets u$\;
    $n_p \gets \textsf{maxConsumedCodons}(\staticevents)$\;
     \( \programState \gets \) start~\( p \) by sending \event{Greenflag} and obtain concrete program state\;
    \While{$i <  l$} {
      $E \gets \extractEvents(\programState)$\;
      $n \gets |E|$\;
      $e \gets $ pick element $c_i \bmod n$ from $E$\;
      	\uIf(\tcp*[f]{Determine key press duration}){\(eventType(e) = KeyPress\)}{
        $\textsf{duration} \gets c_{i+1} \bmod \textsf{UpperBoundKeyPressDuration}$\;
        $\textsf{setKeyPressDuration(duration)}$\;
      }
      \uElseIf(\tcp*[f]{Determine x and y position}){\(eventType(e) = MouseMove\)}{
	$\textsf{x} \gets (c _{i+1} \bmod \textsf{StageWidth}) - (\textsf{StageWidth} / 2)$\;
	$\textsf{y} \gets (c _{i+2} \bmod \textsf{StageHeight}) - (\textsf{StageHeight} / 2)$\;
	$\textsf{setMouseMovePosition(x, y)}$\;
      }
       \uElseIf{\(eventType(e) = DragSprite\)}{
        $\textsf{angle} \gets c_{i+1}$\;
        		\If(\tcp*[f]{Move x \& y in the direction of angle}){\(angle < 360\)}{
		$\textsf{Sprite} \gets \textsf{getSpriteToDrag()}$\;
		$(x, y) \gets \textsf{getDragLocation()}$\;
        		$\textsf{x} \gets \textsf{x} + \textsf{horSize(Sprite)} * \textsf{cos}(angle)$\;
		$\textsf{y} \gets \textsf{y} + \textsf{vertSize(Sprite)} * \textsf{sin}(angle)$\;
		$\textsf{setDragLocation(x, y)}$\;
      		}
	}
      \ElseIf(\tcp*[f]{Determine wait duration}){\(eventType(e) = Wait\)}{
        $\textsf{duration} \gets c_{i+1} \bmod \textsf{UpperBoundWaitDuration}$\;
        $\textsf{setWaitDuration(duration)}$\;
      }
      $t \gets t \plus \List{e}$\tcp*{Add event including its parameters to test case}
      $\programState \gets$ send \( e \) to \( p \)\tcp*{Send event including its parameters to \Scratch VM}
      $i \gets i+n_p+1$
    }
    \Return{$t$}\;
  }
    \caption{Decoding and execution of a codon sequence}
    \label{alg:codon-decoding}
\end{algorithm}


Each time an event has been inferred from the current event codon \eventCodon, the respective event is sent to the \Scratch VM, which updates its state based on the received event.
Then, the decoding of the codon sequence moves on to the next unused event codon of the genotype, skipping all unused parameter codons \parameterCodon of the previous event codon.
Overall, we define an implicit grammar where the starting production for a test case of length $l = \textsf{len}(codons) / (\numberParameterCodons + 1)$ is given by the following formula:
\[
  testcase ::= \; input_1 \; input_2 \; \ldots \; input_l
\]
%


Consider the following example chromosome that was generated for the program depicted in \cref{fig:example2}, with $\staticevents = \langle \event{Wait}, \event{ClickSprite}(Cat), \event{KeyPress}(Space) \rangle$ and $\numberParameterCodons = 1$:
\[
      \begin{aligned}
         T  = \; & \langle [4 \; 3] \; [5 \; 8] \; [2 \; 9]\rangle
      \end{aligned}
\]
For a better visualisation of the codon groups, we placed each event codon together with its \numberParameterCodons parameter codons inside rectangular brackets.
After sending the \event{Greenflag} event to the \Scratch VM, the resulting initial program state provides a choice of two events: a \event{Wait} event as well as a \event{ClickSprite} event on the cat sprite. Using the first event codon value $4$ and the two available events $\langle \event{Wait}, \event{ClickSprite}(Cat) \rangle$, we compute $4 \bmod 2 = 0$ and thus select the \event{Wait} event from the set of available events. Since the \event{Wait} event requires a parameter that denotes the duration, the next codon $3$ is interpreted as the event's parameter, i.e., as the number of steps to wait for. Moving on, the next event codon is $5$. In our case, waiting does not have an impact on the list of available events, which means we have $\langle \event{Wait}, \event{ClickSprite}(Cat) \rangle$ as our sequence of available events again. Due to $5 \bmod 2 = 1$, we choose to click on the cat sprite. Since the \event{ClickSprite} event does not require additional parameter codons, we skip the reserved parameter 8. After executing the \event{ClickSprite} event, the event handler script broadcasts a message that triggers the receiver script in the bear sprite, which in turn waits for a press of the space key in an infinite loop. Thus, when interpreting the next event codon ($2$), there are three possible events to choose from: $\langle \event{Wait}, \event{ClickSprite}(Cat), \event{KeyPress}(Space)\rangle$. Because $2 \bmod 3 = 2$, the chosen event is to press the space key. Because the \event{KeyPress} event requires a single additional codon to determine the press duration, the following reserved parameter codon is consumed to define a press duration of $9$. In order to explore the search space and derive new chromosomes, we apply mutation and crossover.

\subsubsection{Mutation}\label{sec:mutation}
During mutation, event codons are grouped with their \numberParameterCodons assigned parameter codons, which results in $\numberCodonGroups = \textsf{len}(codons) / (\numberParameterCodons + 1)$ codon groups.
Codon groups are then traversed and mutated with a probability of $1 / \numberCodonGroups$.
If a codon group is mutated, a single mutation operator out of the following three operators is chosen randomly.
Each description of a mutation operator succeeds an exemplary mutant that results from applying the respective mutation operation at the codon group $[5,8]$ of the genotype $T= \langle [4 \; 3] \; [5 \; 8] \; [2 \; 9]\rangle$.
\begin{itemize}
\item Add a novel codon group in front of the selected codon group, consisting of $\numberParameterCodons + 1$ randomly generated codon values:  $T  = \langle [4 \; 3] \; [1 \; 7] \; [5 \; 8] \; [2 \; 9]\rangle$
\item Iteratively modify every codon value of the selected codon group by sampling new codon values from a gaussian distribution that has its mean value set to the respective codon value: $T  = \langle [4 \; 3]  \; [4 \; 10] \; [2 \; 9]\rangle$
\item Delete the selected codon group: $T  = \langle [4 \; 3]  \; [2 \; 9]\rangle$
\end{itemize}
Since each codon group is mutated with a probability of $1 / \numberCodonGroups$, we apply, on average, one mutation operation to every parent.

\subsubsection{Crossover}\label{sec:crossover}
Crossover takes as input two parents and produces two children by combining the codons of both parents at specific codon positions $\crossoverIndex_1$ and $\crossoverIndex_2$.
Similar to the mutation operator, the crossover operator starts by grouping the codons of both parents into codon groups of sizes $\numberCodonGroups_1$ and $\numberCodonGroups_2$.
To derive the crossover positions $\crossoverIndex_1$ and $\crossoverIndex_2$ for both parents, we first randomly select a relative crossover point \relativeCrossoverIndex in the range of $[0, 1]$, with 0 representing the first codon group and 1 representing the last codon group of any given parent, and map \relativeCrossoverIndex to the corresponding codon group for each parent individually.
A new offspring is generated by combining the first parent's codon groups from 0 to $\crossoverIndex_1 - 1$ with the codon groups residing at the index positions $ \crossoverIndex_2$ to $\numberCodonGroups_2$ of the second parent.
Finally, a second child is produced by swapping the first and second parent.
For example, consider the two parents
\[
      \begin{aligned}
T_1  = \; & \langle [0 \; 1] \; [2 \; 3] \; [4 \; 5] \; [6 \; 7] \; [8\; 9] \rangle \\
T_2  = \; & \langle [10 \; 11] \; [12 \; 13] \; [14 \; 15] \rangle
      \end{aligned}
\]
and a randomly chosen relative crossover point $\relativeCrossoverIndex = 0.5$, which is mapped to the crossover positions $\crossoverIndex_1 = 2$ and $\crossoverIndex_2 = 1$.
Then, using the derived crossover positions, the crossover operator produces the two children:
\[
      \begin{aligned}
T_{12} = \; & \langle [0 \; 1] \; [2 \; 3] \; [12 \; 13] \; [14 \; 15] \rangle \\
T_{21} = \; & \langle [10 \; 11] \; [4 \; 5] \; [6 \; 7] \; [8\; 9] \rangle
      \end{aligned}
\]

\subsection{Fitness Function}
\label{sec:fitness}

Fitness functions offer a means of distinguishing ``good'' and ``bad''
chromosomes. We aim for statement coverage, such that the fitness function
estimates how close a given execution was to reaching a statement. This
estimate is traditionally calculated by considering distances in the control
flow (approach level~\citep{WBS01}) and distances of the executions of
individual conditional statements (branch distance~\citep{Kor90}):
\begin{itemize}
\item Approach Level: A target statement can be nested arbitrarily deep in
conditional (e.g., if-then-else) parts of the program. The control flow only
reaches the target if it takes specific branches at these decision nodes.
Intuitively, the approach level measures the minimum number of decision nodes
by which control flow missed the target.
\item Branch Distance: If the control flow takes the wrong branch at any of the
dependent decision nodes, then the branch distance estimates how close the
underlying conditional statement was from taking the opposite branch.
\end{itemize}
The fitness calculation in \whisker is based on these concepts, but requires
 adaptations.

\subsubsection{Interprocedural Control Flow and Control Dependence Graphs}

\label{sec:cfgcdg}

The approach level metric was designed for procedural code containing
potentially deeply nested code constructs. In contrast, \Scratch programs tend
to consist of many small scripts, which communicate through events and
messages. To counter this discrepancy, we create an interprocedural control
flow graph and control dependency graph for \Scratch programs, and use this for
calculating approach levels.
A given target \Scratch program consists of a number of scripts; for each script we derive the control flow graph (CFG), defined as $CFG = (\ctrllocations \cup \{entry, exit\}, \controlflows)$, i.e., a directed graph consisting of control flow locations $\ctrllocations$ as well as dedicated $entry$ and $exit$ nodes,  and edges based on the control flow $\controlflows$ between these nodes. We combine these intraprocedural CFGs to an interprocedural super-CFG as follows:
\begin{itemize}
	\item For each event handler  \begin{scratch}\blockinit{when ...}\end{scratch}, we add an artificial node with edges to the event handler (\emph{hat block}) as well as the $exit$ node. 
	We further add an edge from $entry$ to this artificial node for event handlers of user inputs. 
	\item For each 
	\inlineFigure{0.115}{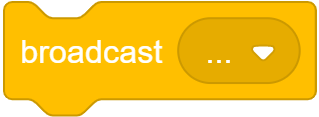}
	or
	\inlineFigure{0.115}{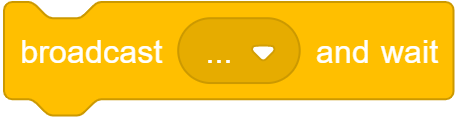}
	statement, we add an edge from the
broadcast to all scripts that start with a matching receive event handler block
\begin{scratch}\blockinit{when I receive \selectmenu{...}}\end{scratch}.
	\item For each \begin{scratch}\blockcontrol{create clone}\end{scratch} statement, we add an edge to all scripts that start with a matching \begin{scratch}\blockinit{when I start as a clone}\end{scratch} event handler block for the corresponding sprite.
	\item For each \begin{scratch}\blocklook{next backdrop}\end{scratch} and
\begin{scratch}\blocklook{switch backdrop to \ovallook*{...}}\end{scratch}
statement, we add an edge from that block to all scripts that start with a
matching \begin{scratch}\blockinit{when backdrop switches to
\selectmenu{...}}\end{scratch} block for the corresponding
backdrop (or all such event handlers if the name of the backdrop is not known).
	\item For each \emph{procedure call} statement \begin{scratch}\blockmoreblocks{call to custom block \ovalnum{X}}\end{scratch}, we add an edge from the call to the start block of the procedure (custom block), and a return edge from its end to the successor node in the calling script. If there are multiple calls to a custom block, all calls lead to the same start block, and there are multiple return edges from the end of the procedure.
	\item For \begin{scratch}\blockcontrol{wait until
 \boolsensing{cond}}\end{scratch} we add an edge to the exit node, since the rest
of the execution is dependent on the condition being satisfied.
\end{itemize}

\begin{figure}[tb]
\centering
    \subfloat[Control Flow Graph\label{fig:cfg}]{\includegraphics[scale=0.22]{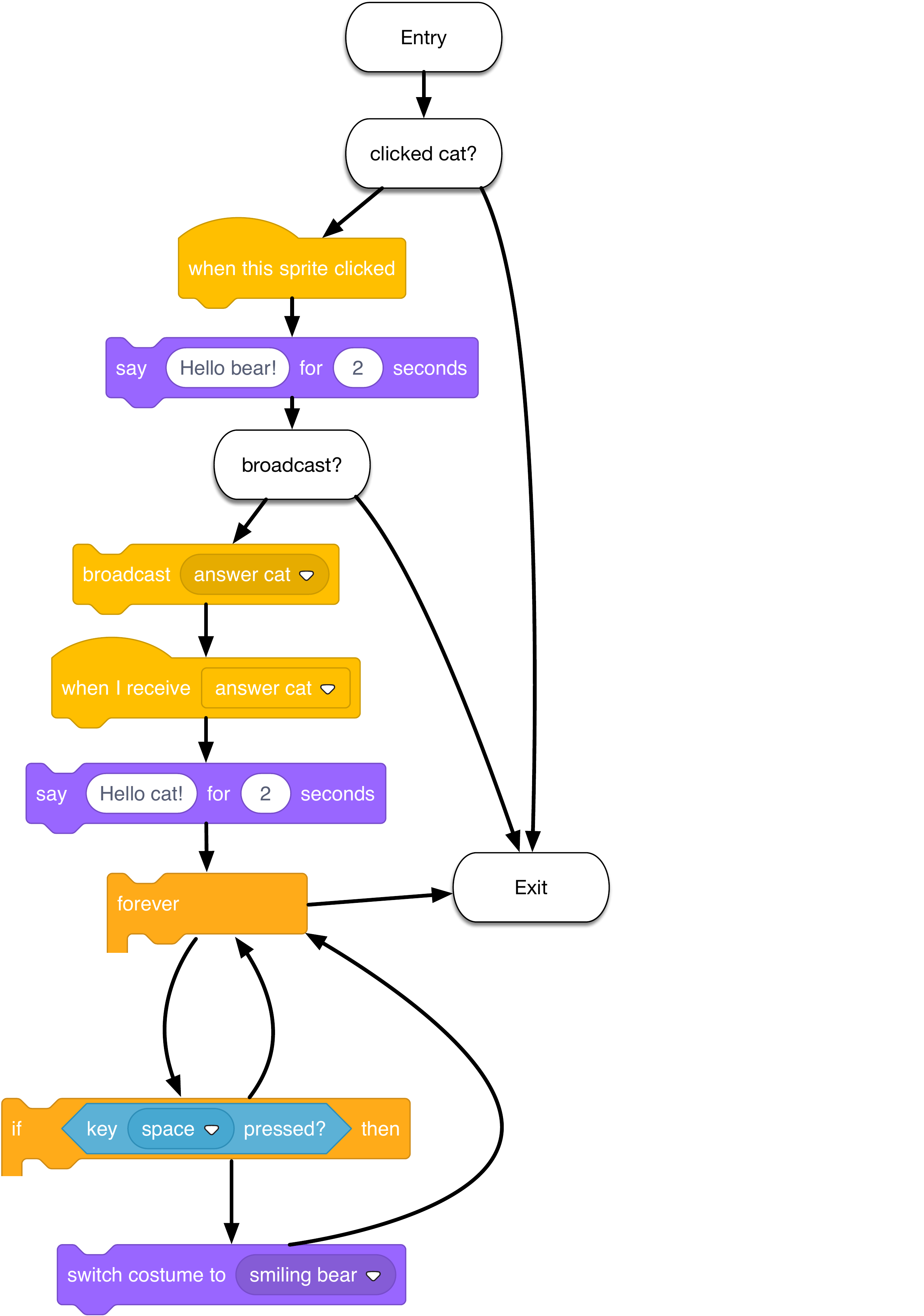}}
	\hfill
    \subfloat[Control Dependence Graph \label{fig:cdg}]{\includegraphics[scale=0.22]{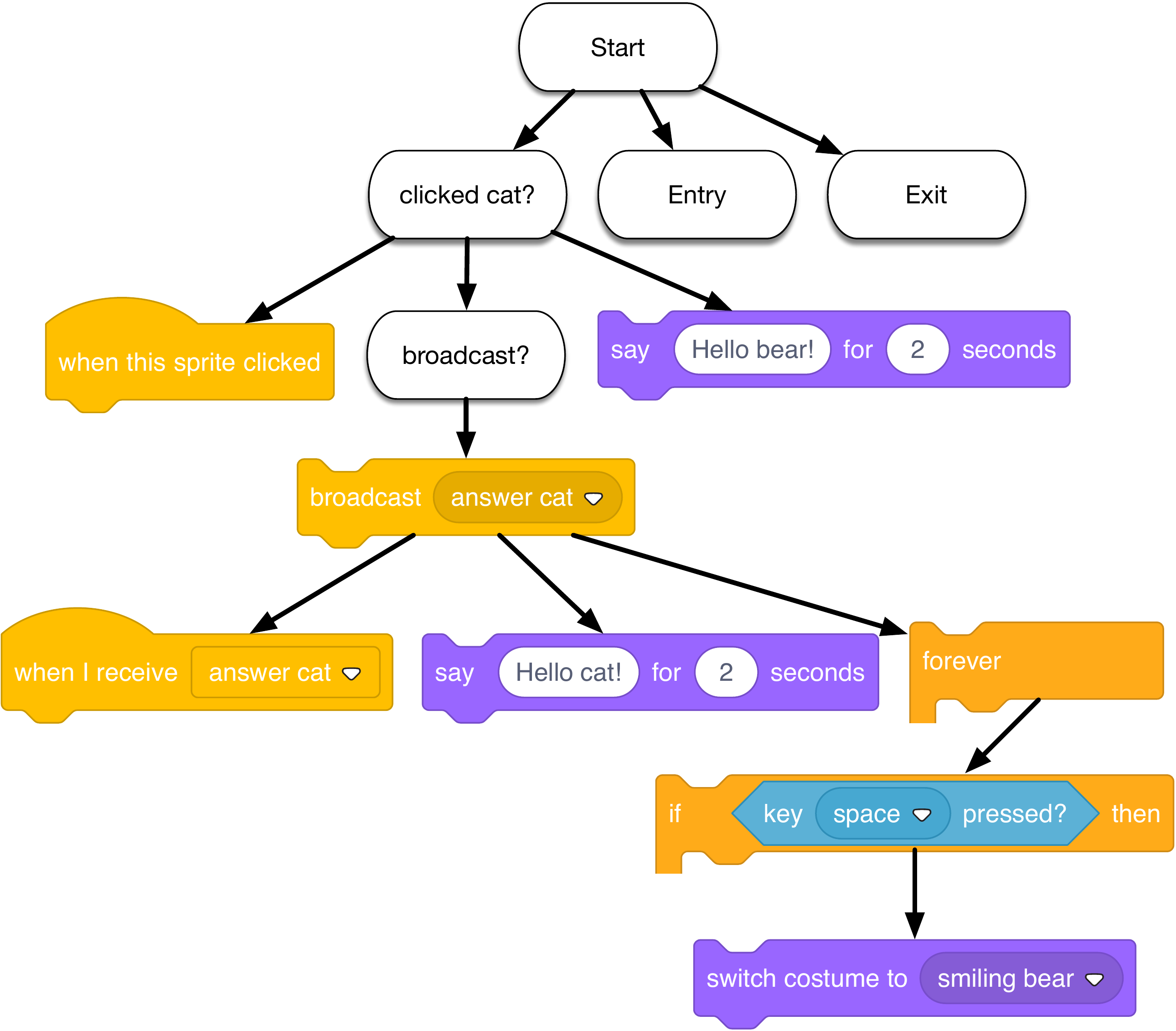}}
\caption{Interprocedural control flow graph and control dependence graph created for the example program from \cref{fig:example2}.}
\end{figure}

Figure~\ref{fig:cfg} shows the interprocedural CFG for the program in \cref{fig:example2}. This CFG contains two artificial event nodes (\textit{clicked cat?}, \textit{broadcast?}), each of which effectively is a branching statement depending on whether the event occurs. These branches turn the occurrence of events into control dependencies of the statements in the event handler code.

\subsubsection{Approach Level}

\begin{figure}[tb]
	\centering
    \subfloat[Code]{\includegraphics[scale=0.2]{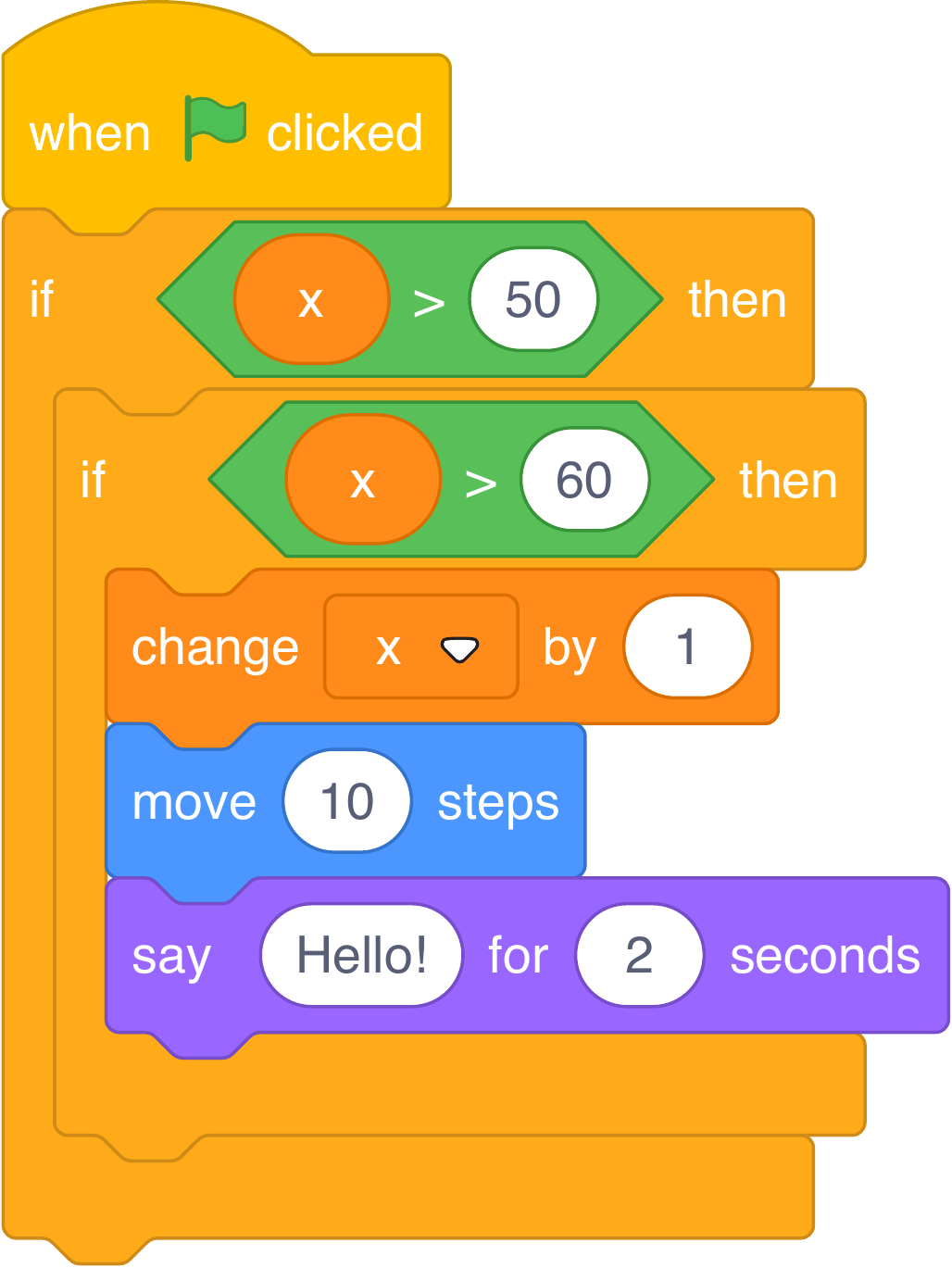}}
    \hfill
    \subfloat[Control Flow Graph]{\includegraphics[scale=0.2]{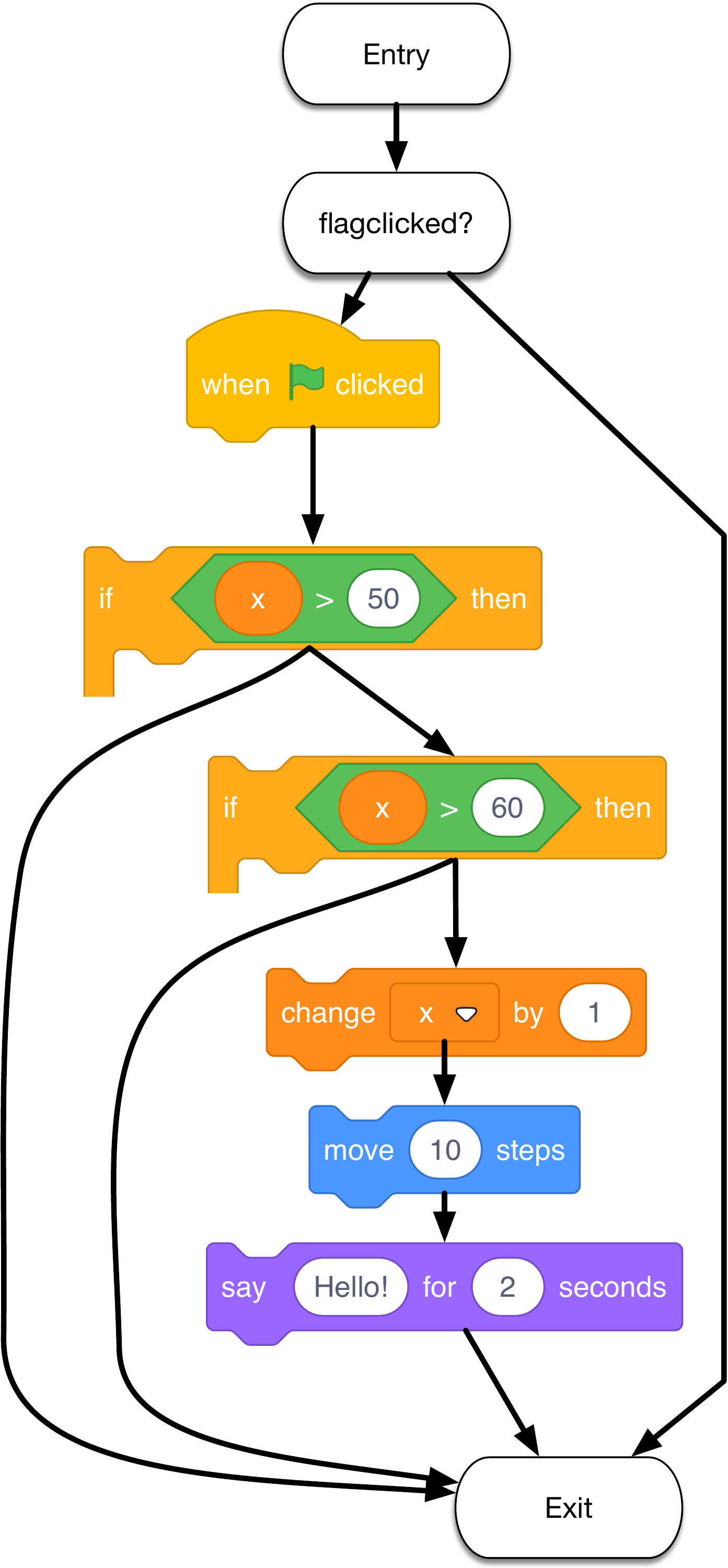}}
    \hfill
    \subfloat[Control Dependence Graph]{\includegraphics[scale=0.2]{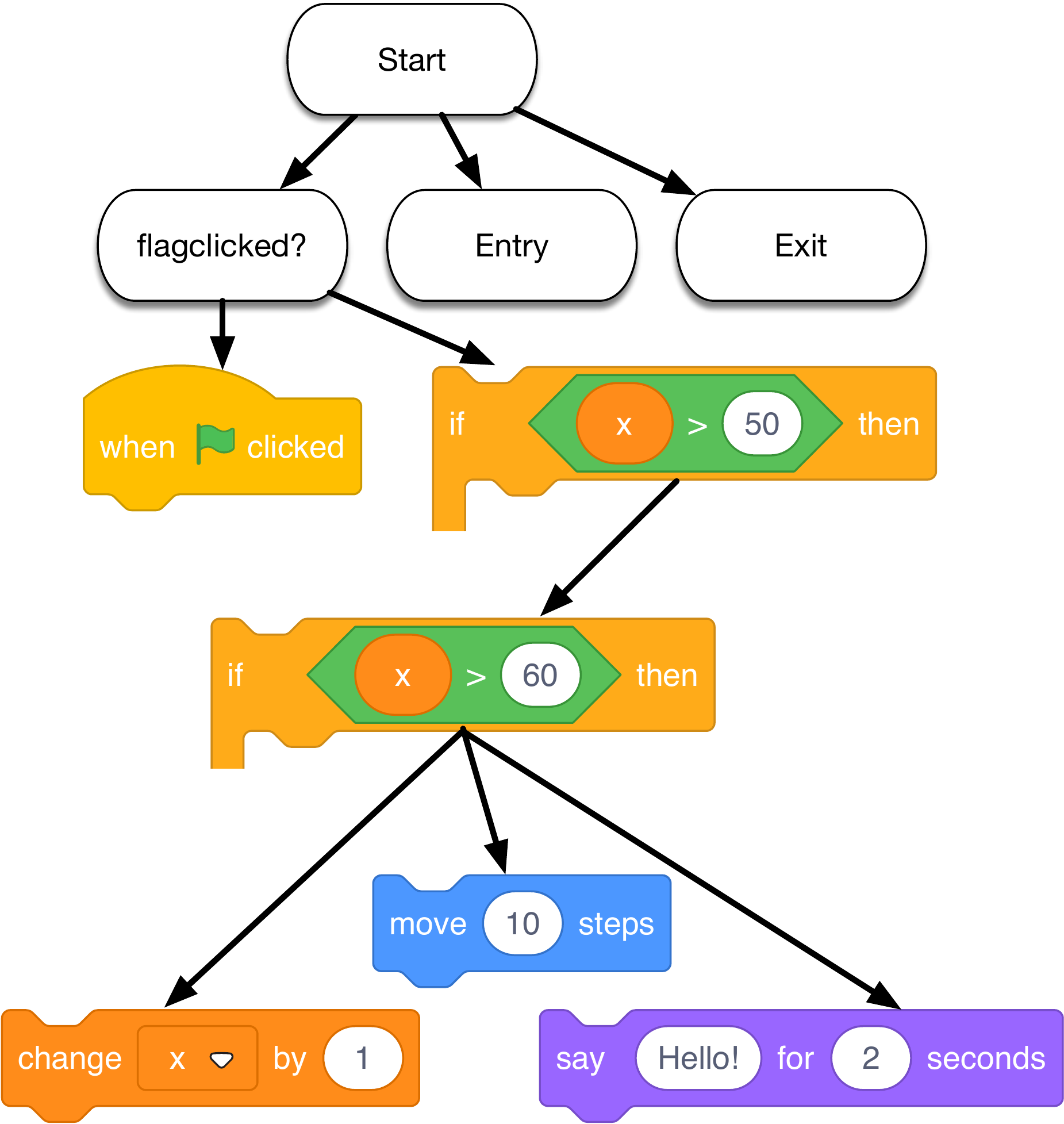}}
	\caption{\label{fig:fitness_example} Example program to illustrate aspects of the fitness function.}
\end{figure}

The approach level is calculated using the control dependence graph, which can be derived directly from the interprocedural CFG. \Cref{fig:cdg} shows the CDG for the example program shown in \cref{fig:example2}; inter-dependencies, for example caused by message broadcasts, are captured in this graph.
\Cref{fig:fitness_example} shows a simple example program, where the \emph{say} block is only executed if the two control dependencies checking $x$ against $50$ and then against $60$ are satisfied. If $x$ is less than $50$, then the approach level after executing this script will be 1; if $x$ is larger than $50$ but less than or equal to $60$, then the approach level is $0$.

To measure the approach level, each statement fitness function pre-computes the
distance in the CDG for each node to the target node. We instrument program
executions by extending the \Scratch VM such that information about executed
branching statements is added to execution traces. Execution traces consist of
block traces, which collect specific block-related data such as the block
type and argument values of conditional statements. Given an execution trace,
we iterate over the covered nodes and determine the minimum distance observed
along the trace using the pre-computed approach levels.

\subsubsection{Control Flow Distance}

Traditional programs execute their procedures from start to end. In contrast,
\Scratch programs tend to execute for long durations because of the animations
they tend to contain, while \whisker tests at the same time impose strict time
limits (defined implicitly by the number and duration of the \event{Wait} events in a
test). When a \whisker-test reaches its end it is terminated, which
may interrupt the execution at any point in the control flow. 
Such an interrupted execution might be following the correct path in the
control flow, such that the branch distance of those executed control dependencies
is 0 (i.e., the correct branch was taken). The search would now
receive no guidance towards reaching the next control dependency or the target
statement. In order to counter this problem, we introduce the \emph{control
flow distance} metric, which informs the fitness function how close an
execution was to reaching a target node within a sequence of statements; the
target node might either be the target of the fitness function itself, or the
next control dependency on the path to it.

\SetKwFunction{ControlFlowDistance}{controlFlowDistance}
\SetKwFunction{ApproachLevel}{approachLevel}
\SetKwFunction{BranchDistance}{branchDistance}

\begin{algorithm}[tb]
\Input{A CFG~\( G \) with control locations~\( L \)}
\Input{Set of covered locations~\( C \subset L \)}
\Input{Target node \( t \in L \) for which to compute the control flow distance}
\Output{The minimal distance \( d \in \mathbb{N} \) between \( n \) and a covered node in \( C \)}

\Fn{\ControlFlowDistance{\( G \), \( C \), \( t \)}}{
  \( d \gets 0 \)\tcp*{initialise distance}
  \( Q \gets \List{ (t, d) } \)\tcp*{initialise work queue}
  \( V \gets \{ t \} \)\tcp*{initialise set of visited nodes}
  
  \While{\( Q \neq \List{} \)}{
    \( \List{(n, d), \ldots} \gets Q \)\tcp*{get first element of \( Q \)}
    \( Q \gets Q \setminus \List{ (n,d) } \)\;
    \If{\( n \in C \)}{
      \Return{d}\;
    }
    \( V \gets V \cup \{ n \} \)\;
    \( P \gets \) immediate predecessors of~\( n \) in~\( G \)\;
    \ForEach{predecessor \( p \in P \)}{
      \If{\( p \notin V \)}{
        \( Q \gets Q \plus \List{ (p, d+1) } \)\;
      }
    }
  }
  
  \Return{d}\;
}

\caption{Control flow distance computation via breadth-first traversal}
\label{alg:cfgdistance}
\end{algorithm}

%
The computation of the control flow distance is outlined in
\cref{alg:cfgdistance}: Given a CFG~\( G \) 
with control locations~\( L \), and a set of already covered locations~\( C
\subset L \), we perform a \emph{backwards} breadth-first search
starting from the desired target node~\( t \in L \), and compute the minimal
distance between \( t \) and any covered statement.
The function is called using the coverage information contained in a trace. The target node passed as parameter is the target node itself if the approach level is 0. Otherwise, we determine the control flow distance towards each successor control dependency, and select the minimum value.
In \cref{fig:fitness_example}, if the variable \emph{x} is larger than 60, the execution may still be interrupted before the \emph{say} block is executed. Therefore, once the second if-condition has been evaluated, the control flow distance to the \emph{say}-block is 2, and after the \emph{change} statement it is 1.

\subsubsection{Branch Distance Instrumentation}
\label{sec:branchDistance}

The branch distance estimates how close a conditional statement was to
evaluating to a specific outcome (true or false). We extended the \Scratch VM
such that for each conditional statement the branch distances are calculated
and traced. For each conditional statement, the execution trace contains
information about the minimum branch distances (for evaluation to true and to
false) observed during an execution. To calculate the branch distance we first
select the closest control dependence, as determined when calculating the
approach level, and then select the minimum branch distance of the outgoing
edge that would take the execution closer to the target node.
Suppose the target is to reach the \emph{say} block in
\cref{fig:fitness_example}, then if $x$ is, for example, $42$, then the branch
distance is computed based on the first if-condition as $|50-42| + 1 = 9$. If the
first if-condition evaluates to true, for example with $x = 55$, then the
second if-condition serves to calculate the branch distance $|60-55| + 1 =6$.

The instrumentation applies the regular equations known from the
literature~\citep{Kor90}; for example, given an equality comparison
\booloperator{\ovalvariable{x} = \ovalnum{42}},
the distance for this condition to evaluate to true is $0$ if $x$ equals $42$
and otherwise $|x - 42|$; the distance for this condition to evaluate to false
is $0$ if $x$ is already different from $42$, otherwise it is $1$. The
instrumentation of the \Scratch VM implements this for all standard relational
and logical operators.

Due to the game-like nature of many \Scratch programs, a common task is to
check for interactions between sprites, e.g. whether they are touching. To this
end, \Scratch provides dedicated \emph{sensing blocks}. These can be used as
conditions for if-then-else or loop blocks, but are also often found in
combination with \begin{scratch}\blockcontrol{wait until
\boolempty[3em]}\end{scratch} blocks, which are encoded as branching statements
in the CFG. Therefore, branch distance needs to be calculated and traced for
all sensing blocks, but the equations presented for standard operators cannot
be applied, and we require a novel definition of branch distance for sensing
blocks.

A dichotomous notion (e.g., using $0$ or $1$ depending on whether a sensing block
reports true or false) leads to challenging plateaus in the fitness landscape, which reduces the
effectiveness of the search. This is a well-known problem to the test generation
community. It has been shown that altering the fitness landscape and restoring lost gradients 
can lead to better guidance and consequently improvements in the search~\citep{certaintyBooleans}.

The key to this is the observation that many sensing blocks query the location
of objects on the stage. For example, \boolsensing{touching color \ovalnum{} ?}
or \boolsensing{color \ovalnum{} touching color \ovalnum{} ?} blocks check if the
current sprite or one of its colours is touching the other target colour. We
transform these conditions by checking if the \emph{Euclidean distance} between
the subjects is 0 on the canvas, and use \emph{it} as the branch distance. This fits
the traditional notion: if the condition is true, both the Euclidean distance and the
branch distance to the true-branch are $0$, and the distance to the false-branch is
$1$. On the other hand, if the condition is false, we define the distance to the
false-branch as $0$, and use the Euclidean distance for the true-branch. This way,
sprites or colours that are closer together will have smaller distance values than
the ones that are further apart.


Similarly, if the condition checks if a sprite is touching the edge of the
stage (\boolsensing{touching \ovalsensing*{edge} ?}), we can gather the position
information and calculate the distance to all four edges, and use the minimal
distance as the branching distance. If the condition checks if a sprite is 
\boolsensing{touching \ovalsensing*{mouse pointer} ?} or 
\boolsensing{touching \ovalsensing*{sprite} ?}, we use the distance between the
sprite and the mouse pointer or the target sprite, respectively, as the branch
distance.

The repeat-times 
block also represents a special case since it does not evaluate a condition
expressed in code. We therefore instrumented these loops such that the
branch distance to exiting the loop is represented by the remaining number of
loop iterations, and the false distance is only non-0 when the loop is exited.
To ensure that these loops are considered as control dependencies, we add an
edge in the CFG from the loop header to the exit node.

The occurrence of events (e.g., greenflag, key press, sprite click) is
encoded in artificial branching nodes in the CFG (cf.
\cref{sec:cfgcdg}). The occurrence or absence of events is encoded with branch
distances of $1$ or $0$, depending on whether or not the event occurred.

\subsubsection{Time-Dependent Statements}

\Scratch contains several time-dependent statements, such as explicit waits
\begin{scratch}\blockcontrol{wait \ovalnum{x} seconds}\end{scratch}, think/say blocks \begin{scratch}\blocklook{think/say \ovalnum{} for \ovalnum{x} seconds}\end{scratch}
glide-animations \begin{scratch}\blockmove{glide \ovalnum{x} secs to} \end{scratch}, audio playback with a
certain duration \begin{scratch}\blocksound{play sound \selectmenu{x} until done}\end{scratch}, or text to speech translation \begin{scratch}\blocksound{speak \ovalnum{x}}\end{scratch}. 
The control flow distance only provides limited guidance in terms of the number
of statements remaining to be covered in a sequence. To better capture time in
the fitness function, we model time-dependencies explicitly in the CFG, and
include information on remaining times in the branch distance. 

To achieve this, we add artificial edges for each time-dependent statement 
to the exit node in the CFG, turning them into control dependencies of all
successor statements, thus including them in the calculation of the approach
level. We extend the instrumentation of the \Scratch VM such that traces
include branch distances for time-dependent statements: If a time-dependent
statement was fully executed, then the true distance is $0$ and the false
distance is $1$; if the execution was interrupted before the statement
completed, then the true distance is defined as the remaining time (and the
false distance is $0$).

\subsubsection{Overall Fitness Function}

\SetKwFunction{Fitness}{fitness}

The overall fitness function for a specific target node is a combination of Approach Level, Branch Distance and Control Flow Distance:
The approach level value is an integer, and to avoid creating deceiving fitness landscapes it needs to dominate the other measurements. We therefore normalise branch distance and control flow distance in the range $[0,1]$ using the normalisation function $\alpha(x) = x/(1+x)$~\citep{arcuri_it_2013}.
If the branch distance is greater than 0, then we set the control flow distance to the maximum value of $1$, since the execution first needs to change the evaluation of the last control dependency before progressing in the CFG matters. 
To ensure the dominance of the approach level, we thus need to multiply it by a factor of $2$.
Finally, we determine the fitness \( f = \Fitness(t) \in \mathbb{R} \) of a test $t$ for a given target location as described in \cref{alg:fitness}. 
Since our overall objective is to achieve full coverage,
one such fitness function is created for each block in the program.

\begin{algorithm}[tb]
\Input{A test case \( t \)}
\Input{Control flow graph \( G \) (implicitly)}
\Input{Target location \( s \) (implicitly)}
\Output{The fitness for \( t \)}
\Fn{\Fitness{\( t \)}}{
  \( C \gets \) set of locations in \( G \) covered by \( t \)\;
  \( a \gets 2 \times \ApproachLevel(G, C, s) \)\;
  \( b \gets \alpha(\BranchDistance(G, C, s)) \)\;
  \uIf{\( b > 0 \)}{
    \( c \gets 1 \)\;
  }
  \Else{
    \( c \gets \alpha(\ControlFlowDistance(G, C, s)) \)\;
  }
  \Return{\( a + b + c \)}\;
}

\caption{Fitness computation for test cases}
\label{alg:fitness}
\end{algorithm}

%




%

\subsection{Search Algorithms}
\label{sec:algorithms}
\label{subsec:random-search}

\SetKwFunction{randomSearch}{randomSearch}
\SetKwFunction{generateRandomCodons}{generateRandomCodons}
\begin{algorithm}[tb]
  \Input{\Scratch program $p$} 
  \Output{Test suite $T$}

  \Fn{\randomSearch{\( p \)}}{
    $G \gets$ coverage goals in \( p \)\tcp*{Set of all goals}
    $C \gets \emptyset$\tcp*{Set of already covered goals}
    $T \gets \emptyset$\tcp*{Test suite}
  
    \BlankLine
  
    \While{stopping condition not reached} {
      $\List{c_{1}, \ldots, c_{k}} \gets \generateRandomCodons(p)$\;
      \( t \gets \DecodeAndExecute(p, \List{c_{1}, \ldots, c_{k}}, \extractEvents) \)\;
      
      $C' \gets $ goals covered by \( t \)\;
      \If(\tcp*[f]{Check if \( t \) covers new goals}){$C' \minus C \neq \emptyset$}{
        $T \gets T \cup \{ t \}$\;
        $C \gets C \cup C'$
      }
    }
    
    \BlankLine
    
    \Return{$T$}\;
  }

  \BlankLine
  
  \Fn{\generateRandomCodons{p}}{
    \( \overline{C} \gets \List{} \)\;
    \( i \gets 0 \)\;
    \( k \gets \random(\{ 1, 2, \ldots \}) \)\;
    \While{\( i < k \)}{
      \( c \gets \random(\{ 0, 1, \ldots \}) \)\;
      \( \overline{C} \gets \overline{C} \plus \List{c} \)\;
      \( i \gets i + 1 \)\;
    }
    \Return{\( \overline{C} \)}\;
  }

  \caption{Random search}
  \label{alg:random_suite}
\end{algorithm}

Given the encoding and fitness function, it is possible to apply any
meta-heuristic search algorithm to the problem of test generation for \Scratch.
Random search~(\cref{alg:random_suite}) is the simplest conceivable global
search algorithm. As a global search algorithm, it considers the \emph{entire}
search space (in contrast to neighbouring chromosomes in a local search
algorithm), trying to cover as many statements at a time as possible. It
operates by repeatedly sampling a test~\( t \) at random, and adds it to the
test suite~\( T \) if it covers a new target. This process continues until a
given search budget is exhausted, after which \( T \) is returned. Due to its
simplicity, random search is often used as a baseline for comparison. As many
\Scratch programs tend to be small, it is also possible that random search will
often be sufficient in order to generate adequate test suites.

\label{subsec:MOSA}

\SetKwFunction{MOSA}{MOSA}

\SetKwFunction{randomPopulation}{randomPopulation}
\SetKwFunction{updateArchive}{updateArchive}
\SetKwFunction{preferenceSorting}{preferenceSorting}
\SetKwFunction{sortByCrowdingDistance}{sortByCrowdingDistance}
\SetKwFunction{localSearch}{localSearch}

\SetKwFunction{DecodeAndExecuteAll}{decodeAndExecuteAll}

\begin{algorithm}[tb]
  \Input{program~\( p \) for which to generate tests}
  \Input{population size~\( N \in \mathbb{N} \)}
  \Output{test suite~\( A \) (an ``archive'' of test cases)}
	\Fn{\MOSA{\( p \), \( N \)}}{
   		$ P \gets \randomPopulation(p, N) $\;
      \( T \gets \DecodeAndExecuteAll(p, P) \)\;
       
    	$ A \gets \updateArchive(\emptyset, T) $\;
    	\While{stopping condition not satisfied} {
    		$ O \gets \textsf{generateOffspring}(P) $\;
        \( T \gets \DecodeAndExecuteAll(p, O) \)\;
	    	$ A \gets \updateArchive(A, T) $\;
	    	$ \List{F_{0}, F_{1}, F_{2}, \ldots} \gets \preferenceSorting(P \cup O) $\;
        $ P \gets \emptyset $\;
	    	$ i \gets 0 $\;
	    	\While{$ |P| + |F_i| \leq N $} {
				$ P \gets P \cup F_i $\;
				$ i \gets i + 1 $\;
			}
	    	$ \List{c_{1}, c_{2}, \ldots} \gets \sortByCrowdingDistance(F_i) $\;
        \( P \gets P \cup \{ c_{j} \mid j = 1, 2, \ldots, (N - |P|) \} \)\;
        \( (P, A) \gets \localSearch(P, A) \)\;
		}
  		\Return{$ A $}\;
  	}
    
    \BlankLine
    
    \Fn{\randomPopulation{\( p \), \( N \)}}{
      \( P \gets \emptyset \)\;
      \( i \gets 0 \)\;
      \While{\( i < N \)}{
        \( c \gets \generateRandomCodons(p) \)\;
        \( P \gets P \cup \{ c \} \)\;
      }
      \Return{\( P \)}\;
    }
    
    \BlankLine
    
    \Fn{\DecodeAndExecuteAll{\( p \), \( C \)}}{
      \( T \gets \emptyset \)\;
      \ForEach{codonSequence \( c \in C \)}{
        \( t \gets \DecodeAndExecute(p, c, \extractEvents) \)\;
        \( T \gets T \cup \{ t \} \)\;
      }
      \Return{\( T \)}\;
    }
    
    \caption{Many-Objective Sorting Algorithm~\citep{MOSA}}
    \label{alg:MOSA}
\end{algorithm}

\SetKwFunction{MIO}{MIO}

\SetKwFunction{updateArchive}{updateArchive}
\SetKwFunction{random}{random}
\SetKwFunction{mutate}{mutate}
\SetKwFunction{getArchivePopulation}{getArchivePopulation}
\SetKwFunction{constructTestSuite}{constructTestSuite}
\SetKwFunction{chooseTarget}{chooseTarget}
\SetKwFunction{calculateDynamicParameter}{calculateDynamicParameter}
\SetKwArray{archive}{\ensuremath{\overline{A}}}

\begin{algorithm}[tb]
  \Input{program~\( p \)}
  \Input{set~\( K \) of targets in~\( p \)}
  \Input{start of the focused phase~\( F \in [0,1]\)}
  \Input{initial population size per target~\( n_0 \in \mathbb{N}\)}
  \Input{final population size per target~\( n_f \in \mathbb{N}\)}
  \Input{initial random selection probability~\( r_0 \in [0,1]\)}
  \Input{final random selection probability~\( r_f \in [0,1]\)}
  \Input{initial maximum mutation count~\( m_0 \in \mathbb{N}\)}
  \Input{final maximum mutation count~\( m_f \in \mathbb{N}\)}
  \Output{test suite~\( T \)}

	\Fn{\MIO{\( p \), \( K \), \( F \), \( n_0 \), \( n_f \), \( r_0 \), \( r_f \), \( m_0 \), \( m_f \)}}{
   		$ \archive \gets \emptyset^{|K|} $\;
   		$ B \gets $ depleted search budget\;
   		$ n \gets n_0$\;
   		$ r \gets r_0$\;
   		$ m \gets m_0$\;
   		\While{stopping condition not satisfied}{
   			\If {$B < F$}{
   				$ n \gets \calculateDynamicParameter(n_0, n_f, B, F)$\;
   				$ r \gets \calculateDynamicParameter(r_0, r_f, B, F)$\;
   				$ m \gets \calculateDynamicParameter(m_0, m_f, B, F)$\;
            }
   			\uIf{\( \archive = \emptyset^{|K|} \) or with probability~\( r \)} {
                $ \List{c_{1}, \ldots, c_{k}} \gets \generateRandomCodons(p) $\;
                \( t \gets \DecodeAndExecute(p,\List{c_{1}, \ldots, c_{k}}, \extractEvents) \)\;
                $ \archive \gets \updateArchive(\archive, t, n) $\;
                $ \archive \gets \localSearch(\{ t \}, \archive) $\;
            }
        \Else{
          $ k \gets \chooseTarget(K) $\;
          $ t \gets \random(A_k)$\;
          $ i \gets 0 $\;
          \While{$i < m$} {
            $ t' \gets \mutate(t) $\;
            $ \archive \gets \updateArchive(\archive, t', n) $\;
            \If {$t'$ has a better fitness value than $t$ for target $k$} 
            {
              $ t \gets t' $\;
            }
            $ i \gets i + 1 $\;
          }
          $ \archive \gets \localSearch(\{ t \}, \archive) $\;
        }
        $ B \gets $ updated search budget\;
	  }
	  $ T \gets \constructTestSuite(\archive, K)$\;
  	  \Return{$T$}\;
  	}
  	
  	\BlankLine
    
    \Fn{\calculateDynamicParameter{\( x_0 \), \( x_f \), \( B \), \( F \)}}{
    			\Return{$ x_0 + (x_f - x_0) \cdot \frac{B}{F} $}\;
    }
    
    \caption{Many Independent Objective Algorithm \citep{MIO}}
    \label{alg:MIO}
\end{algorithm}

The aim of automated test generation is to maximise the achieved code coverage, which is a task that lends itself to a multi-objective problem representation, where every single statement is an individual optimisation goal in its own right. Hereby, it is not uncommon to encounter 
conflicting goals (e.g., statements in 
\texttt{if}-branches vs.\ \texttt{else}-branches), and depending on the size of~\( p \), the 
number of goals might range from tens to hundreds, or possibly even more.
This poses scalability challenges to ``traditional'' many-objective algorithms, such as 
the well-known NSGA-II~\citep{NSGA-II}. Due to the so-called dominance resistance phenomenon,
the proportion of non-dominated solutions increases \emph{exponentially} with the number of goals to 
optimise, thus degrading the search to a random one in the process.

We therefore employ the Many-Objective Sorting Algorithm (\MOSA)~\citep{MOSA} 
(outlined in \cref{alg:MOSA}). \MOSA is a modified variant of NSGA-II that caters
to mentioned peculiarities of test generation. Most notably, it introduces a
so called \emph{preference criterion} that allows us to assign a preference among 
non-dominated solutions based on how ``close'' they come to covering a \emph{new}, 
\emph{previously uncovered} target. This way, the number of targets to consider at a 
time is reduced and the search budget is directed towards the targets that are still left 
to be covered.
As a notable deviation from the original 
algorithm~\citep{MOSA}, we have extended the search algorithm to a memetic algorithm~\citep{fraser2015memetic} by applying a local search to the new parent 
population at the end of every generation (and also updating the archive if necessary). A local search algorithm explores the local neighbourhood of a candidate solution in a more focused way than a global exploration would. In particular, this post-processing step is necessary to address the challenge of finding suitable test execution durations specific
to \Scratch. 

\label{subsec:MIO}

The number of targets to cover in a \Scratch program can be very large, and even 
though \MOSA tries to address this problem, it might still struggle to achieve \SI{100}{\percent}
coverage. For example, certain statements may be infeasible, and it is thus not
worthwhile trying to cover them. For this reason, the Many Independent Objective algorithm
(\MIO)~\citep{MIO}, outlined in \cref{alg:MIO}, tries to strike a balance 
between exploration and exploitation by focusing on those goals that are most promising, 
given the resources available. It has been specifically designed for test generation and 
is based on the (1+1)~EA evolutionary algorithm, but also maintains an archive of candidate solutions for each coverage objective. Similar to \MOSA, we extended MIO with local search by applying the local search defined in \cref{sec:localSearch} with a certain probability after each generation.



\newcommand{\lastImprovedCodon}{\ensuremath{\mathsf{c_I}}\xspace}%

\label{sec:localSearch}

We have turned MOSA and MIO into memetic algorithms by integrating local search operators~\citep{fraser2015memetic}, which take an existing test case as input and explore its \emph{neighbours} by applying operator-specific changes to its genotype.
If a local search operator manages to generate an improved test case, which is measured based on the pursued goal of the applied operator, the original test case is replaced with the modified one.

\subsubsection{Extension Local Search}
\SetKwFunction{ExtensionLocalSearch}{extensionLocalSearch}
\SetKwFunction{generateParameterCodons}{generateParameterCodons}

\begin{algorithm}[tb]
  \Input{list of codons~\( \List{c_{1}, \ldots, c_{l}} \)}
  \Input{event extractor~\extractEvents}
  \Input{program p}
  \Input{user configuration $u$}
  \Output{modified test case~\( t' \)}
  \Fn{\ExtensionLocalSearch{$\List{c_{1}, \ldots, c_{l}}$, $p$, $\extractEvents$, $u$}}{
  $ (\textsf{MaxCodonLength}, \textsf{NewEventProbability}) \gets u$\;
  $ \numberParameterCodons \gets \textsf{getNumParameterCodons}(t') $\;
  $ \textsf{done} \gets \textsf{False} $\;
  $ j \gets 0 $\;
  $ t' \gets \DecodeAndExecute(p, \List{c_{1}, \ldots, c_{l}}, \extractEvents)$\;
  $ \programState \gets \textsf{fetch program state of}  \ p \ \textsf{after executing} \ t $\;
  $ E_{prev} \gets \extractEvents(\programState) $\;
  \While{$\textsf{isActive}(\programState)$ and  $\textsf{length}(c) \leq \textsf{MaxCodonLength}$ and $\neg\textsf{done}$} {
  	$ E \gets \extractEvents(\programState) $\;
	
	\tcp{Apply \event{TypeTextEvent}/\event{TypeNumberEvent}.}
	\uIf{$\textsf{\event{TypeTextEvent}} \in E$}{
	$ e \gets \textsf{getRandomTypeTextOrNumberEvent}(E) $\;
	}
	
	\tcp{Apply a newly found event.}
	\uElseIf{$E_{prev} \neq E$ and $\random(\{0, \ldots 1\}) \leq \textsf{NewEventProbability}$}{
	$ e \gets \textsf{getRandomNovelEvent}(E) $\;
	}
	
	\tcp{Apply \event{WaitEvent}.}
	\Else{
	$ e \gets \textsf{\event{WaitEvent}} $\;
	}
		
	$ c \gets c \plus \textsf{indexOf}(e, E) \plus \generateParameterCodons(\numberParameterCodons) $\;
	$ t' \gets t' \plus \List{e} $\;
	$\programState \gets \textsf{send} \  e \  \textsf{to} \  p $\;
  \If{$\textsf{fitnessHasNotImproved}(t')$}{
    $ \textsf{done} \gets \textsf{True} $\;
  }
	
	$ E_{prev} \gets E $\;
	
  }
  \Return{$t'$}\;
  }
  
  \BlankLine
  \Fn{\generateParameterCodons{$ \numberParameterCodons $}}{
  	$  i \gets 0  $\;
	$ c_p \gets \List{} $\;
      	\While{$ i <  \numberParameterCodons $}{
        		$ c_p \gets c_p \plus \random(\{ 1, 2, \ldots \}) $\;
      }
      \Return{$c_p$}\;
    }

    \caption{Extension Local Search}
    \label{alg:extension-ls}
\end{algorithm}

\Scratch projects often contain blocks that pause the execution of a program for a certain amount of time (\cref{sec:Scratch-Programs}).
Due to these, some program statements can only be reached after waiting for an extended period of time.
Extension local search aims to overcome these execution halting blocks by repeatedly adding \event{WaitEvents} to a given test case, eventually making previously hard to reach blocks more accessible.
In order to append and execute \event{WaitEvents} at the end of a test case, the operator first has to obtain the \Scratch VM's state after executing the original test case.
For this purpose, the extension local search algorithm, shown in \cref{alg:extension-ls}, starts with the re-execution of the original test case.

In the while-loop that follows, the operator repeatedly checks for the presence of \event{TypeTextEvents}/\event{TypeNumberEvents} and novel events that were not present in the list of events $E$ during the previous iteration of the loop. 
If a \event{TypeTextEvent} or \event{TypeNumberEvent} is found, it is preferably selected over \event{WaitEvents} since the program execution halts until the user has given an answer to a posed question, which is indicated to the user through an UI-focus switch that highlights a text field.
Hence, until an answer has been given in the form of a \event{TypeTextEvent}/\event{TypeNumberEvent}, adding additional \event{WaitEvents} usually does not contribute in exploring novel program states.
Newly found events, on the other hand, are preferred because these events become available due to overcoming certain execution halting blocks, and therefore promise to lead to novel program states.
However, newly discovered events are only selected with a specific probability, as otherwise, program states potentially hiding behind even longer wait durations would most likely remain out of reach.
A prominent example of such a novel event scenario is a \emph{ClickSpriteEvent}, representing a click on a button that gets only clickable after an introductory animation has finished.

Furthermore, to save time the operator checks after every iteration if the overall fitness of the test case has improved and stops if no improvement could be observed.
In addition to the lack of observable fitness improvements, the algorithm also stops if the maximum codon length defined by the user has been reached or if the program has stopped due to the execution of a \begin{scratch}\blockstop{stop \selectmenu{all}}\end{scratch} block or having reached the end of all program scripts $s$.

Since the extension local search operator aims to discover novel program states by extending the genotype of a test case, the operator can only be applied if a genotype has not reached its full length yet.
Finally, the original test case is replaced with the extended one if the operator discovered new statements not previously covered by the existing test case.

\subsubsection{Reduction Local Search}
Reduction local search aims to reduce the codon length by removing genes that did not contribute to improving the fitness. 
For that purpose, every time a test case is executed, we save the codon position \lastImprovedCodon that points to the last codon group after which no further fitness improvements have been observed.
Using the index \lastImprovedCodon, the reduction local search operator generates a new test case by cloning the codon groups located at the positions $[0, \lastImprovedCodon]$, excluding all codons occurring after \lastImprovedCodon.
A significant difference to the extension local search operator is that reduction local search does not re-execute the original test case since \lastImprovedCodon is already saved during the execution of the original test case.

The operator is only applied to genotypes for which $\lastImprovedCodon < \numberCodonGroups$ is satisfied, with \numberCodonGroups representing the total number of codon groups contained in the genotype.
Furthermore, since \lastImprovedCodon points to the codon group after which no further fitness improvements have been observed, it is ensured that no covered statements are lost in the process of reducing the size of the genotype.
Hence, every reduced test case is guaranteed to be an improvement over the original test case in terms of codon size and will therefore replace it in the search algorithm's population.

The benefit of applying reduction local search to a given test case is twofold: 
First, removing codon groups can save valuable search time due to not re-executing events that do not contribute in discovering new program states.
Second, since the presented mutation operator mutates each codon group with a probability of $1 / \numberCodonGroups$, reduction local search forces the mutation process to focus on relevant codon groups by increasing their mutation probability.

\subsection{Test Minimization}

Although MIO and MOSA both use minimization as a secondary criterion, the final
test suite may contain test cases that are not minimal. We therefore apply a
post-processing step that removes redundant events from test cases. The
minimization algorithm produces test cases that are 1-minimal: A test case $T =
\langle e_1, e_2, \ldots e_n \rangle$ of length $n$, where $e_i$ can be
interpreted as either a codon group or as an event, is 1-minimal with respect
to a coverage goal represented by fitness function $f$, if for all $i$ the test
case $T' = \langle e_1, e_{i-1}, e_{i+1}, \ldots, e_n \rangle$ has $f(T') > f(T)$.
That is, removing any of the events leads to the test case no longer satisfying
the coverage goal. We use the minimization algorithm implemented in
\textsc{EvoSuite}~\citep{fraser2012whole}: For each test case $T$ we iterate
over all $e_i$ starting from the last event, produce a test case $T'$ without
that event, and measure its fitness; if the fitness is not worse, then $e_i$ is
discarded and $T = T'$. In theory, a more efficient algorithm such as delta
debugging could be used to increase the performance of the
minimization~\citep{leitner2007efficient}.

\section{Experiments}
\label{sec:evaluation}

\newcommand{\atwelve}{\ensuremath{\hat{A}_{12}}\xspace}
\newcommand{\pval}{\ensuremath{p\text{-value}}\xspace}

%

To provide a better understanding of the problem of test generation for
\Scratch, we aim to answer the following questions:
\begin{itemize}
\item[\textbf{RQ1:}] How much can test execution be accelerated reliably?
\item[\textbf{RQ2:}] Can \Scratch projects be trivially covered?
  %
  %
\item[\textbf{RQ3:}] What is the best test generation algorithm for \Scratch programs?

\item[\textbf{RQ4:}] How effective are generated tests at detecting faults?

\end{itemize}

\subsection{Experimental Setup}
\label{subsec:evaluation-setup}

\label{sec:subjects}


\subsubsection{Dataset 1: Projects with Manually Created Tests}

In order to answer RQ1 we require test cases with test assertions, such that
they can detect potential differences in program executions when different
program acceleration factors are used. The dataset consists of the Fruit
Catching game used in the original \whisker study~\citep{TestingScratch} as
well as 14 projects used by \citet{greifenstein2021effects} in a study with
teachers in education; these include interactive projects with user input (such
as the games Pong and Snake) as well as animations and stories.
The projects contain a total of 94 manually crafted \whisker tests (11.3 on
average per project) with a total of 185 assertions. The projects are typical
in size and complexity of the projects created by learners, with 29--98 (mean: 49) connected, reachable statements.

\subsubsection{Dataset 2: 1000 Random Projects (\textsc{Random1000})}
\label{subsec:Random1000}

 \Scratch is among the most popular platforms for programming beginners, and
backed by a large online ecosystem and community.
We created a dataset of publicly shared \Scratch projects by
mining projects as follows: Each \Scratch project has a unique ID which is
sequentially increasing as new projects are created. By probing project IDs we
determined that starting roughly from ID 400.000.000 we can reliably retrieve
projects in the format of \Scratch 3, whereas below we frequently encountered
version 2 projects. While \whisker can also handle projects saved in version 2
of \Scratch, our static analysis tool
\textsc{LitterBox}~\citep{fraser2021litterbox} used in related research on the
same dataset~\citep{searchBasedRefactoringScratch} requires version 3 projects.
 We then uniformly sampled project IDs in the range of 400.000.000 and
700.000.000 (i.e., a number larger than the latest projects at the time of this
writing), and downloaded batches of 1000 starting from the random ID using the
REST-API provided by the \Scratch
webserver. Projects can only be downloaded if they are publicly
shared. This process resulted in a dataset of 2.2 million projects, of which
1.500.937 are not remixes of other projects, i.e., variations of already
uploaded projects. From these, we randomly sampled a subset of 1000, which
is a compromise between a large desirable set of evaluation subjects and the
resulting computational costs of running experiments in multiple different
configurations with many repetitions to counter randomness.

\begin{figure}[t]
	\centering
	\subfloat[\label{fig:stats_random1000}\rand]{\includegraphics[width=0.49\textwidth]{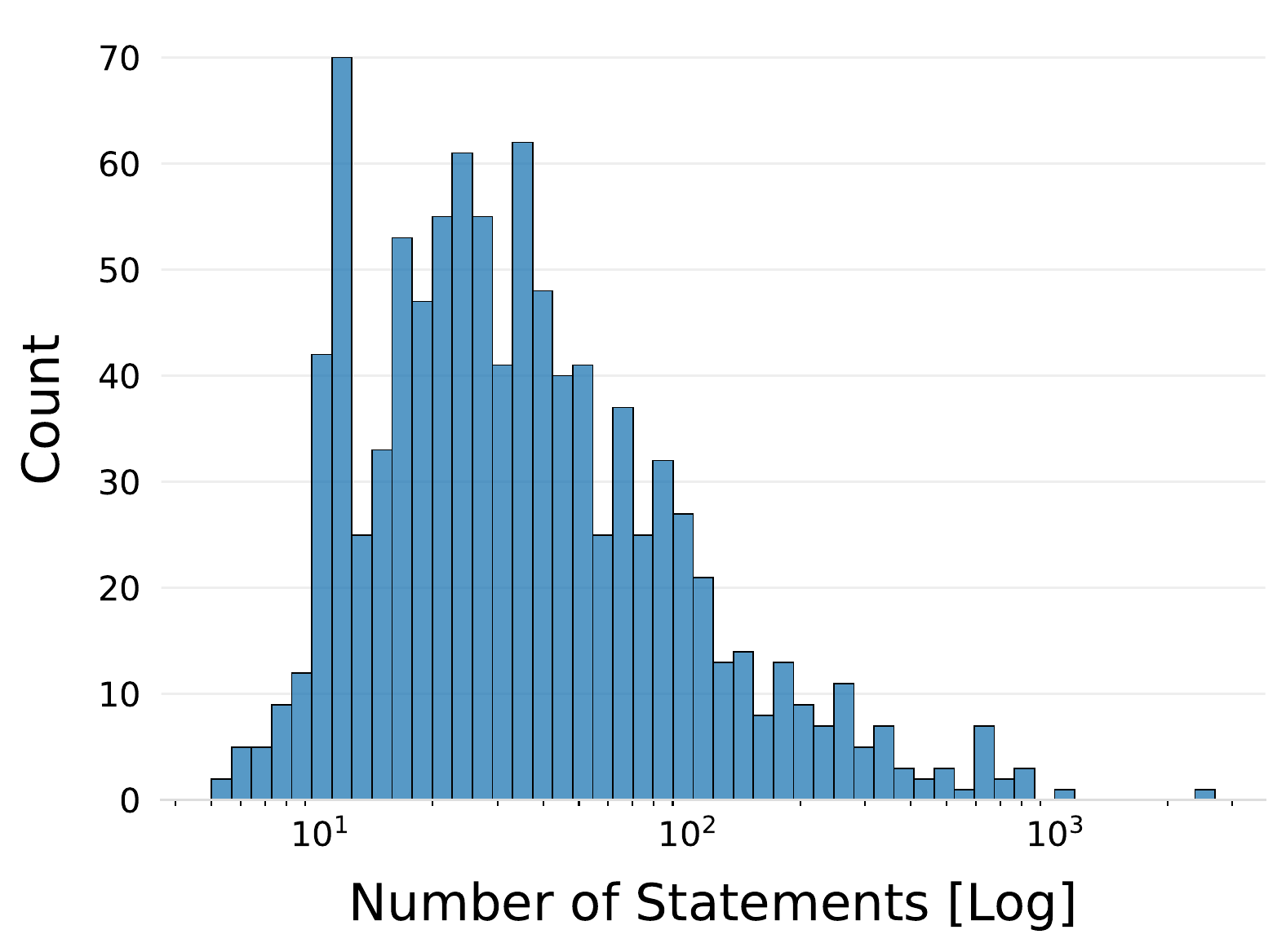}}\hfill
	\subfloat[\label{fig:stats_top1000}\topp]{\includegraphics[width=0.49\textwidth]{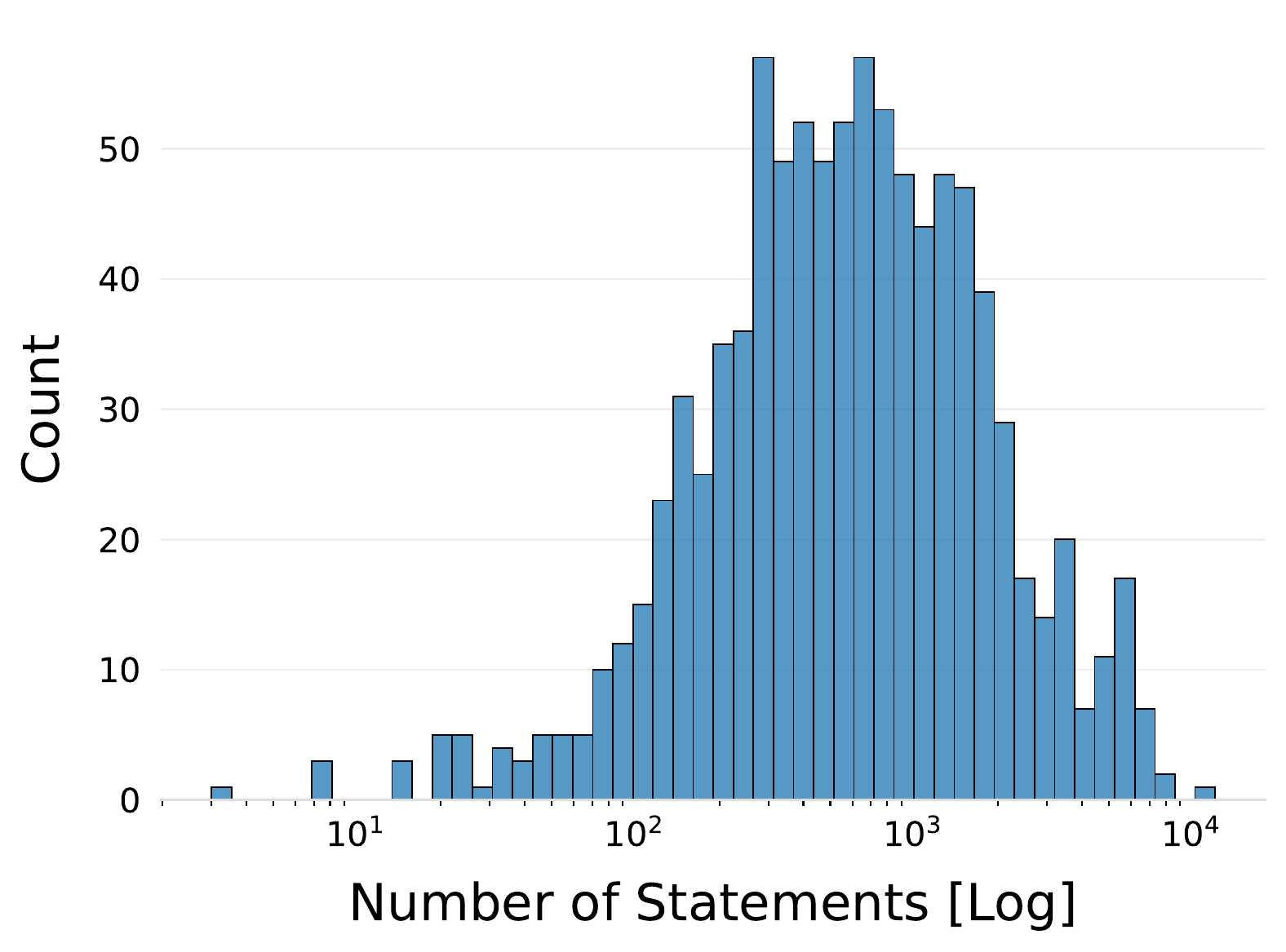}}

	\caption{Distribution of number of statements per project.}
\end{figure}

\Cref{fig:stats_random1000} shows the log-scale distribution of sizes of the
1000 projects based on the count of statement-blocks. Note that we only count executable statement blocks; i.e., we exclude any loose blocks, or blocks contained in dead code (e.g., event handlers for events not generated in the project).
The majority of projects are small and have less than 100 statements, but there
are also larger projects with up to 1082 (connected, reachable) statements.


\subsubsection{Dataset 3: Top 1000 Most Loved Projects (\textsc{Top1000})}

Many of the perils of mining GitHub open source projects do not apply to our
sampling process of \Scratch projects: For example, whereas when mining GitHub
projects it is often problematic that many projects are personal (i.e., there
is only one committer, and collaboration with others is not intended) or do not even
contain code~\citep{kalliamvakou2016depth}, all \Scratch projects per
definition are personal and contain \Scratch code since that is the only use
case.
However, since a common application scenario for \Scratch is creative use
rather than programming it is possible that a random sample may contain many
trivial projects. Thus, while the random sample is useful for external
validity, we also created a second dataset of \emph{popular} projects. Indeed
it has been shown that a focus on the star-rating in GitHub projects can have
an impact on resulting findings~\citep{maj2021codedj}.
To derive a dataset of popular projects, we crawled the \url{https://scratchstats.com/} website, which
collects live statistics on \Scratch users and projects, on 2021--12--16 and
identified and downloaded the 1000 most ``loved'' projects on \Scratch.
\Cref{fig:stats_top1000} shows that these projects are substantially
larger, with sizes ranging from 5 to 11\,530 connected and reachable statements,
with a mean of 1036 statements. On average, these projects have received
10\,315.6 loves and 432\,589 views by other \Scratch users. The
dataset includes 21 projects which are a remix of other projects in the
dataset.

\subsubsection{Implementation and Tuning}
\label{subsec:Parameter}

The ideas presented in this paper are implemented as an extension to the test
generation tool \whisker~\citep{TestingScratch}. In particular, we added the
\Scratch VM modifications presented in \cref{sec:execution}, the event selection
mechanisms outlined in \cref{sec:event-selection} and the algorithms described in
\cref{sec:algorithm}. The source code used in this study is publicly available%
\footnote{%
  \url{https://github.com/se2p/whisker} based on the commit starting with \gitId, and \newline
  \url{https://github.com/se2p/scratch-vm} based on the commit starting with \gitIdVm
}.

Every described test generation algorithm is accompanied by a set of configurable parameters that guide the search for a test suite.
To optimise these parameters, we establish an optimisation dataset, which is disjoint from the \textsc{Top1000} and \textsc{Random1000} datasets, by randomly sampling 250 projects following the same procedure as in \Cref{subsec:Random1000}.
Parameters used across all algorithms, and MOSA-specific ones, are determined by executing MOSA on all 250 projects using different values for a single parameter while fixating all other configurable parameters.
After the search has finished, we compare the achieved coverages and choose the best performing configuration.
Finally, we repeat the same procedure with MIO to optimise remaining parameters that only occur within the MIO algorithm.

 The results of the optimisation process indicates that a codon range of $[2, 20]$ works best, which means that a single test case may contain up to 20 events.
 Furthermore, a probability of 30\% for applying extension and reduction local search indicates that the search benefits from using these operators.
 For the MOSA algorithm, we chose a population size of 30 and a crossover probability of 70\%.
 Regarding MIO, an unreachable focus phase of 100\% combined with a random test generation probability starting at 90\% reveals that exploration is more beneficial than exploitation for the \Scratch problem domain.
A complete list of all used parameter configurations can be found in the \whisker repository.


\subsubsection{Environment}\label{sec:environment}

We conducted our experiments in a controlled execution environment using a Docker image
based on Debian Slim Buster and Node.js~16. The revision of \whisker used is
based on git commit \gitId. The experiments were run on a dedicated computing cluster,
with each computing node featuring one Intel Xeon E5-2690v2 CPU per node with
\SI{3.00}{\giga\hertz} and \SI{64}{\giga\byte} of RAM. Each run of \whisker was allocated one
CPU core and \SI{5}{\giga\byte} of RAM.

\subsection{Experiment Methodology}

\subsubsection{RQ1: How much can test execution be accelerated reliably?}\label{subsec:experiment-acceleration}

To confirm whether the changes introduced to the \Scratch VM allow for 
\emph{reliable} accelerated test execution without introducing any form of non-deterministic behaviour, we run the manually written tests presented in \cref{sec:subjects} with a fixed seed for the random number generator, and repeat the experiment 20 times. If the tests reveal no functional differences in the execution,
we can then validate whether the \Scratch VM can be accelerated at all by comparing the total execution time on all 15 projects using varying acceleration factors.
Finally, we want to ascertain that the accelerated execution of \Scratch programs behaves deterministically in two ways:
First, the test results of a given project for each used acceleration factor is compared with the outcomes of the non-accelerated test execution to make sure that the introduced acceleration of \Scratch programs does not alter the execution behaviour.
Second, to validate that the modified \Scratch VM does not introduce any flakiness, we also check whether all 20 experiment repetitions within the observed acceleration factors lead to the same test results.

\subsubsection{RQ2: Can \Scratch projects be trivially covered?}\label{subsec:experiment-trivial}

In order to determine whether \Scratch projects represent a test generation
challenge in the first place, we run a baseline algorithm of random testing
with dynamic event selection on the \rand and \topp
datasets. While there is no clear boundary of what constitutes a ``trivial''
project, intuitively a project does not pose a challenge to a test generator if
it is possible to consistently achieve 100\% code coverage without requiring a
large number of test executions and without using an evolutionary algorithm.
Following this intuition, we define for both datasets a threshold, which is based on the lowest number of test executions at which we encounter the first program that is not covered entirely. Using this threshold, we then report the number of trivial and non-trivial projects for each dataset individually.
\subsubsection{RQ3: What is the best test generation algorithm for \Scratch programs?}
\label{subsec:experiment-algorithms}

\Cref{sec:algorithm} describes three different algorithms for test generation;
the aim of this research
question is to determine which of these performs best. 
To answer this question, each search algorithm is applied 20 times to every project of the \rand and \topp dataset using the dynamic event selector and a search budget of 10 minutes.
Besides the achieved block coverages, we also compare the number of events contained within final test suites and the average execution time of a test case during the search by comparing averages, the \atwelve effect size and the number of projects for which one approach outperforms another one. Furthermore, we report the average block coverage achieved over time.
\subsubsection{RQ4: How effective are generated tests at detecting faults?}
\label{sec:Methodology-RQ4}

In our fourth research question, we evaluate whether the generated
tests together with regression assertions (\cref{sec:assertions}) are
able to detect faulty \Scratch programs. For that purpose, we
introduce a mutation analysis framework which implements the eight
mutation operators shown in \cref{tab:mutationOperators}, which are
based on the traditional set of sufficient mutation
operators~\citep{offutt1996experimental}. Using the mutation framework
and the tests generated during RQ3, we produce mutants for each
program and validate whether the tests with assertions are able to
detect the inserted program modifications. A mutant is detected if a
test that passes on the original program leads to a failing assertion
when executed on the mutant.

The mutation analysis requires reloading a \Scratch project once for
each of its mutants. This process revealed a problem in the memory
management of the \Scratch VM: Each restart increases the memory
consumption, and this eventually leads to a crash of the \Scratch
VM. While this memory leak is not a problem for our test and assertion
generation, where we reset the \Scratch VM state directly between test
executions, it is a problem during mutation analysis, where not only
the state but also the code need to be replaced. We therefore limit
the RQ4 experiments to the \rand projects, which are small enough such
that the memory leak does not affect the results significantly, as only 4\% of all test executions escalated into program crashes. For the \topp set, on the other hand, we observed crashes due to this issue in 58\% of all test executions. We notified the
maintainers of the \Scratch VM about this problem and seek to evaluate
the effectiveness of our automatically generated assertions on bigger
programs (e.g., the \topp set) in future studies.

Research question four is then answered by reporting the
\emph{Mutation Score}~\citep{surveyMutationAnalysis} for every applied
operator after excluding test cases which reported a false-positive
result on the respective unmodified project. In order to keep the
mutation analysis experiment within a reasonable time frame, we only
consider first-order mutants as is usually done, and set a timeout of
90 minutes for the mutation analysis of an individual project. To
avoid false-positive results due to randomised program behaviour, we
seed each test execution with the same seed that was used during the
test generation phase.

\begin{table}[tb]
	\caption{\label{tab:mutationOperators} Mutation operators.}
\begin{tabularx}{\linewidth}{p{2.5cm}Xp{3.5cm}}
	\toprule
	Operator & Description & Affected Blocks\\
	\midrule

 	Key Replacement Mutation (KRM) & Replaces the $key$ argument with a randomly chosen one. & \begin{scratch} \blockinit{When \selectmenu{key} key pressed} \end{scratch} \boolsensing{key  \ovalsensing*{key} pressed} \\

 	Single Block Deletion (SBD) & Removes a block that is neither a hat nor a branching block. & All except
 	\begin{scratch} \blockinit{\hspace{1em}...\hspace{1em}} \end{scratch}
 	\begin{scratch} \blockif{\hspace{1em}...\hspace{1em}} {\blockspace[0.4]}\end{scratch} \\

 	Script Deletion Mutation (SDM) & Deletes a script by removing a hat block.  & \begin{scratch} \blockinit{\hspace{1em}...\hspace{1em}} \end{scratch} \\

 	Arithmetic Operator Replacement (AOR) & Replaces an arithmetic operator with a randomly chose one. & \booloperator{\ovalnum{} + \ovalnum{}} \booloperator{\ovalnum{} - \ovalnum{}} \booloperator{\ovalnum{} * \ovalnum{}} \booloperator{\ovalnum{} / \ovalnum{}} \\

 	Logical Operator Replacement (LOR) & Replaces a logical operator with the opposing one. &
 	\booloperator{\ovalnum{} and \ovalnum{}} \booloperator{\ovalnum{} or \ovalnum{}}\\

 	Relational Operator Replacement (ROR) & Replaces a relational operator with a randomly chosen one. & \booloperator{\ovalnum{} $<$ \ovalnum{}} \booloperator{\ovalnum{} $>$ \ovalnum{}} \booloperator{\ovalnum{} = \ovalnum{}} \\

 	Negate Conditional Mutation (NCM) & Negates boolean blocks. & \booloperator{\hspace{3em}} \boolsensing{\hspace{3em}} \\

 	Variable Replacement Mutation (VRM) & Replaces a variable with a randomly chosen one. & \ovalvariable{my variable} \\

	\bottomrule

\end{tabularx}
\end{table}

\subsection{Threats to Validity}\label{subsec:threats}

\paragraph{Threats to internal validity.}
%
To ensure that results can be trusted, \whisker has an extensive test suite, RQ1 aims to demonstrate validity, and we manually inspected results.
Upstream changes to the \Scratch VM may require adaptation of our modifications.
However, such changes are rather unlikely, as they would also break many programs shared across the Scratch community.
Since \whisker uses randomised algorithms and results may be affected by
chance, all experiments are based on 20 repetitions and are
statistically analysed following common guidelines~\citep{arcuri2014hitchhiker}.
The performance of search algorithms depends on many parameters. In order to ensure our results are not negatively influenced by unsuitable configurations, we applied the tuning procedure described in \cref{subsec:Parameter}.

\paragraph{Threats to external validity.}
Results may not generalise beyond the specific dataset used for experiments.
However, we aimed to maximise generalisability by using two large datasets of
1000 projects each for RQ2 and RQ3, one randomly sampled from the \Scratch
website, and the other based on popularity as measured using the number of
``love'' reactions from other users. A possible source of bias is that only
projects which users chose to publicly share can be accessed this way. It is
conceivable that programs not shared publicly are more incomplete or broken.
Similarly, the 15 projects and their tests used to answer RQ1 may not suffice
to cover all possible sources of non-determinism that may occur in \Scratch
projects. The reported results in RQ4 are based on the \rand dataset and may not
generalise to more complex programs, such as the ones contained in the \topp set.

\paragraph{Threats to construct validity.}
The main metric for comparison is code coverage (coverage of statement blocks).
Code coverage is the most common metric used in practice as well as in research
in order to compare test suites as well as test generation algorithms; however,
whether and how code coverage is related to fault detection is an ongoing
debate~\citep{inozemtseva2014coverage,chen2020revisiting}.
We also evaluate whether the tests are able to detect artificial faults using mutation
analysis. Mutation scores may be skewed by equivalent mutants~\citep{programEquiv}.
Furthermore, artificially generated mutants may not be representative of real program faults~\citep{gopinath2014mutations}.
However, automated testing is not the only targeted application
scenario of \whisker; the tests are intended to enable any form of dynamic
analysis that can support the generation of hints and feedback to learners.
Code coverage is a prerequisite for any form of testing and dynamic analysis,
and so it is important to consider this as a first step.

\newcommand{\heading}[1]{\multicolumn{1}{c}{#1}}
\newcommand{\scratchVM}{\textsf{SVM}\xspace}
\newcommand{\improvedScratchVM}{\textsf{SVM\textsuperscript{+}}\xspace}

\section{Results}

\subsection{RQ1: How much can test execution be accelerated reliably?}

\begin{figure}[t]
\centering
\subfloat[\label{fig:scatterAcceleration} Execution speed differences per project ]{\includegraphics[width=0.45\textwidth]{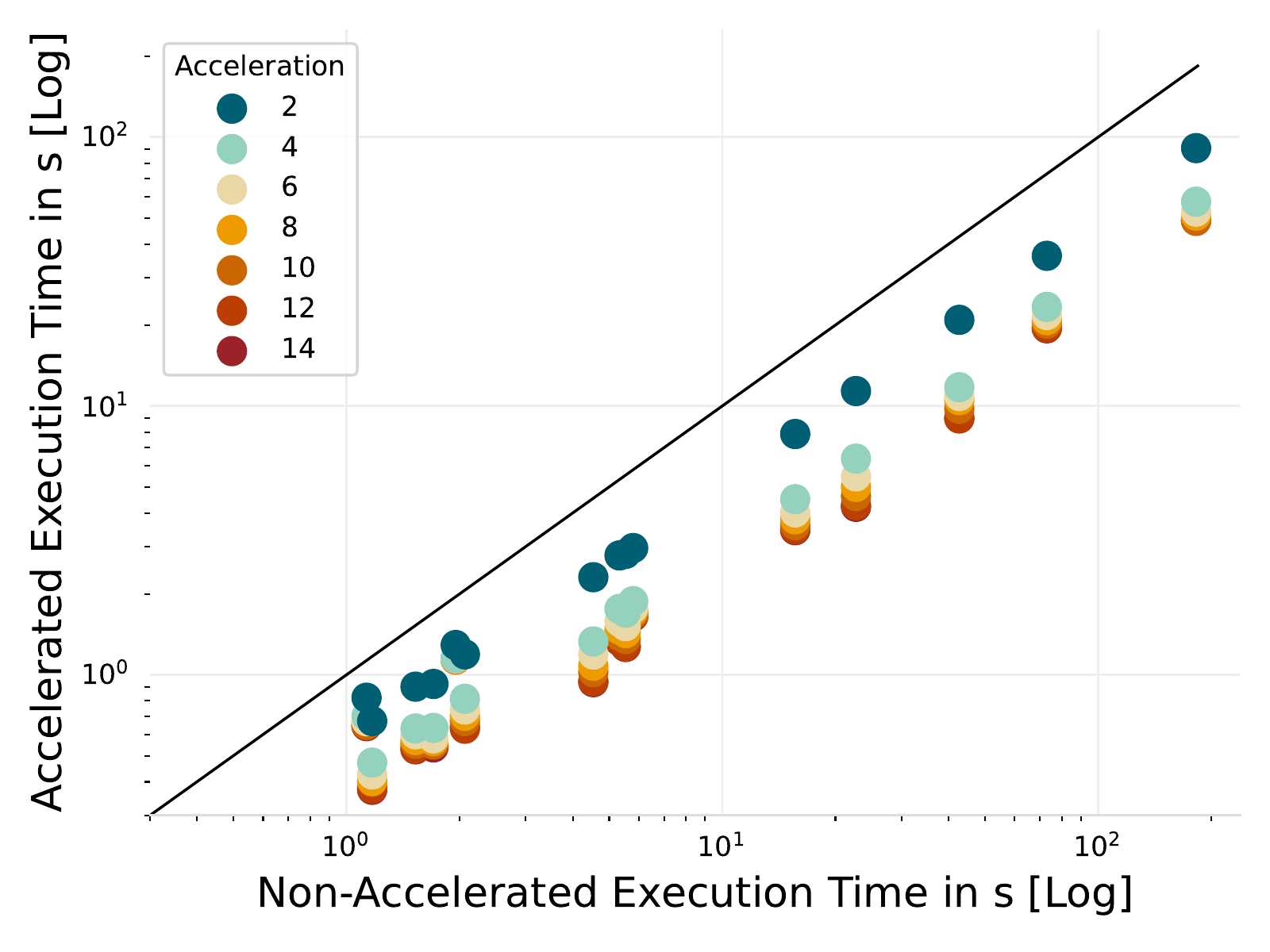}}
\subfloat[\label{fig:barAcceleration}Mean execution duration across all projects]{\includegraphics[width=0.45\textwidth]{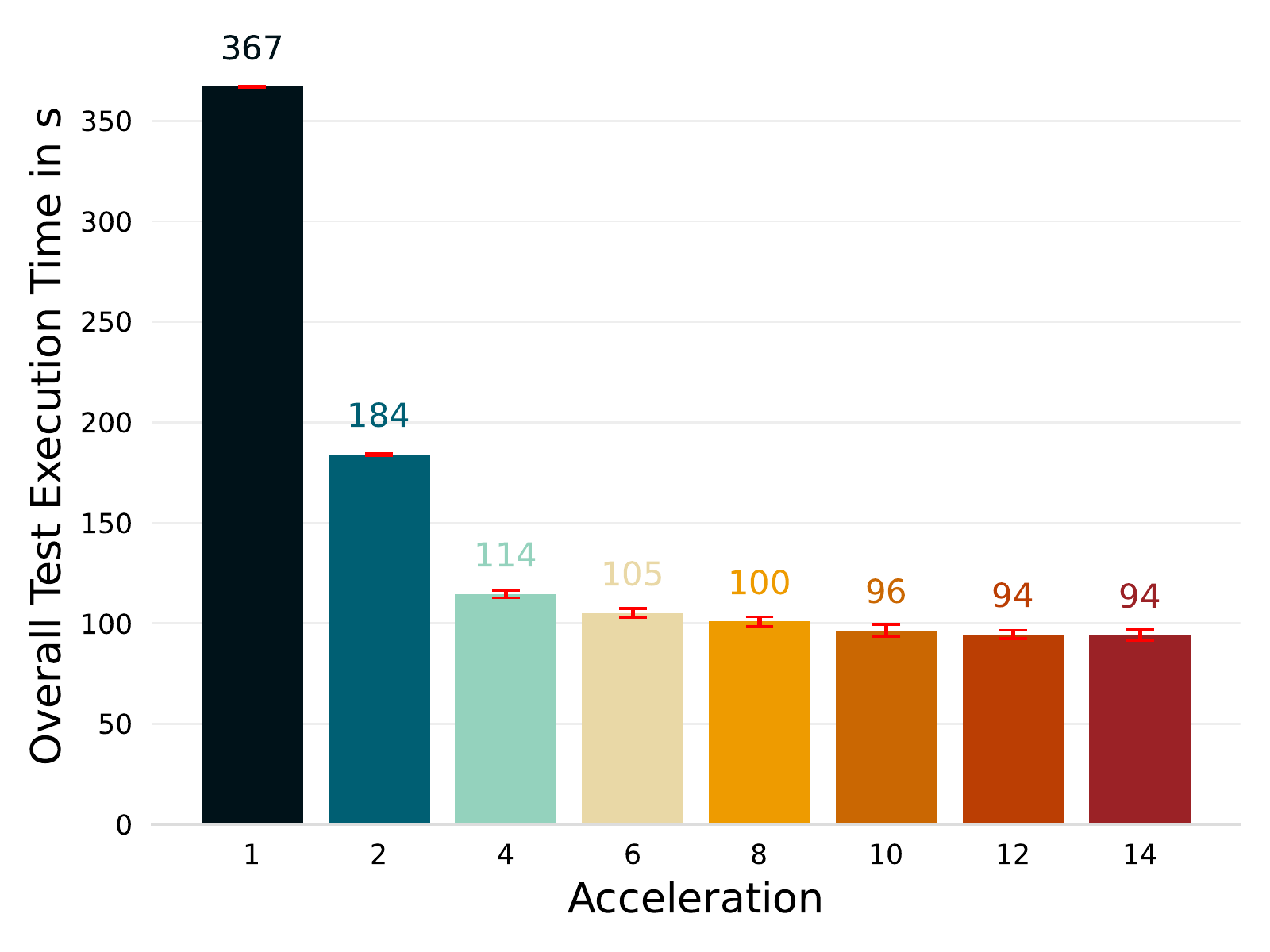}}
 \hfill
\subfloat[\label{fig:barFlakiness}Number of flaky tests without (\scratchVM) / \\ with (\improvedScratchVM) safety measures ]{\includegraphics[width=0.45\textwidth]{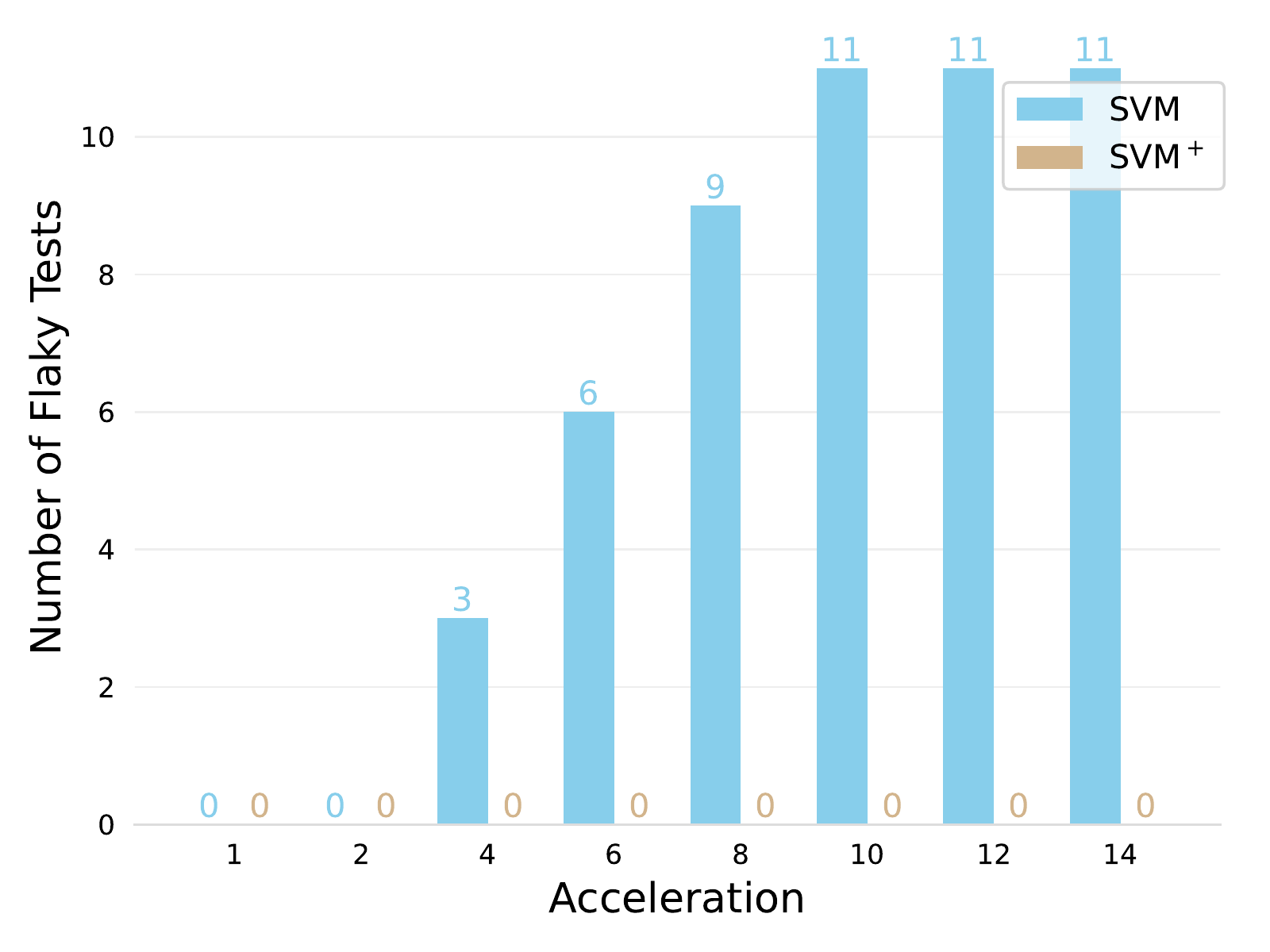}}
\subfloat[\label{fig:barPassedFailed} Number of passed and failed tests without (\scratchVM) / with (\improvedScratchVM) safety measures after excluding flaky tests ]{\includegraphics[width=0.45\textwidth]{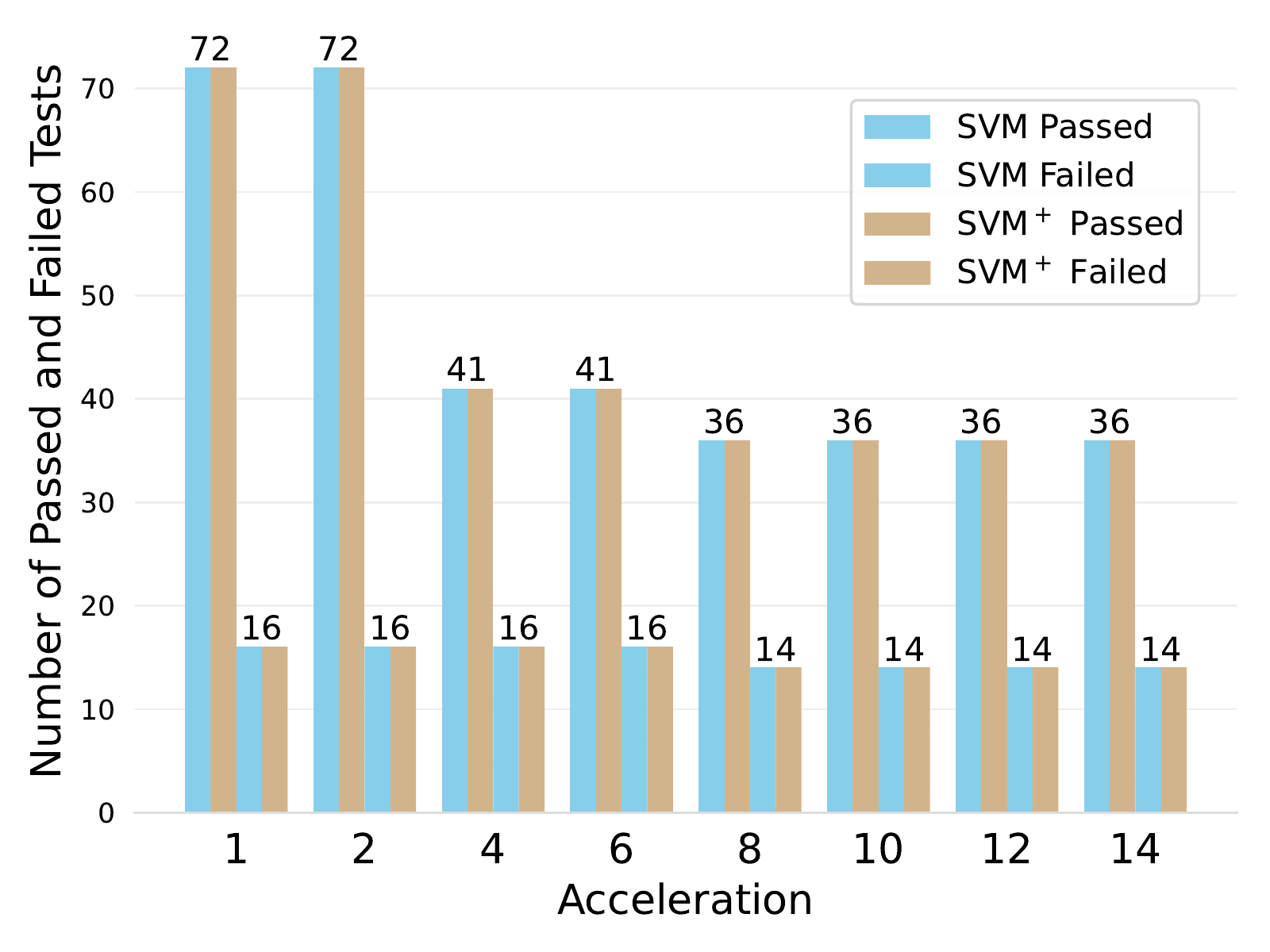}}
\caption{Comparison of execution times, including standard deviations highlighted as red error bars, and flaky tests across different acceleration factors.}
\end{figure}

To determine whether \Scratch projects can be accelerated at all, we executed a set of 15 test suites and analysed the required execution durations for each project.
As can be observed in \cref{fig:scatterAcceleration}, every recorded data point resides beneath the diagonal, indicating that all projects benefit from an increased acceleration factor.
\Cref{fig:barAcceleration} shows its direct impact, with a factor of two halving the average execution time across all projects.
However, as acceleration factors further increase, the gains in speed up are diminished significantly.
This is due to halting blocks enforcing lock durations, which start to reach the minimum duration of a single step in the VM.
The error bars displayed in red indicate that the execution speed is very consistent and does not suffer from significant variation, which shows that with the given hardware environment the results are consistent.
However, hardware with higher code execution speed would be capable of processing individual steps faster, thereby making even higher speedups feasible.

Besides validating whether \Scratch programs can be accelerated at all, RQ1 seeks to ascertain that acceleration does neither introduce flakiness nor alter the program's behaviour as a whole.
To ensure deterministic behaviour, \cref{fig:barFlakiness} compares the number of flaky tests obtained using the accelerated \Scratch VM (\scratchVM) introduced in \cref{section:acceleration} against the improved accelerated \Scratch VM (\improvedScratchVM) which contains safety measures for sources of flaky behaviour as described in \cref{sec:ensureDeterminism}.
In contrast to \scratchVM, the \improvedScratchVM does not show any signs of non-deterministic behaviour within 20 repetitions of the same test as well as between the execution of a given test in an accelerated and non-accelerated scenario.
The \scratchVM, however, shows increasingly flaky behaviour for higher acceleration factors due to reasons explained in \cref{sec:ensureDeterminism}.
Even though the \scratchVM does not show any signs of flakiness in our experimental setup, non-deterministic behaviour is very likely to occur if tests are generated and executed on different machines or if the machine's computing resources are scarce during test execution.

Finally, \cref{fig:barPassedFailed} compares the number of passed and failed tests for both VM versions after excluding flaky (c.f.\ \cref{sec:ensureDeterminism}) test results.
The results demonstrate that both variants achieve the same results, which indicates that the \improvedScratchVM behaves similarly to the \scratchVM, and errors found in one of the two versions can be reproduced in the other one.
All in all, due to the ensured determinism and limited increase in perceivable program speed up for acceleration factors greater than 10, we decided to conduct all following experiments using an acceleration factor of 10.

\summary{RQ1}{Using an acceleration factor of 10, the overall execution
duration for the test suite dataset was reduced by \SI{73.84}{\percent} without changing the test results or introducing any flakiness. Thus, program execution can be sped up reliably if appropriate safety measures against sources of non-deterministic behaviour are added to the \Scratch VM.}

\subsection{RQ2: Can \Scratch projects be trivially covered?}

\begin{figure}[t]
	\subfloat[\label{fig:relevance_coverage:random1000}\textsc{Random1000}]{\includegraphics[width=0.49\textwidth]{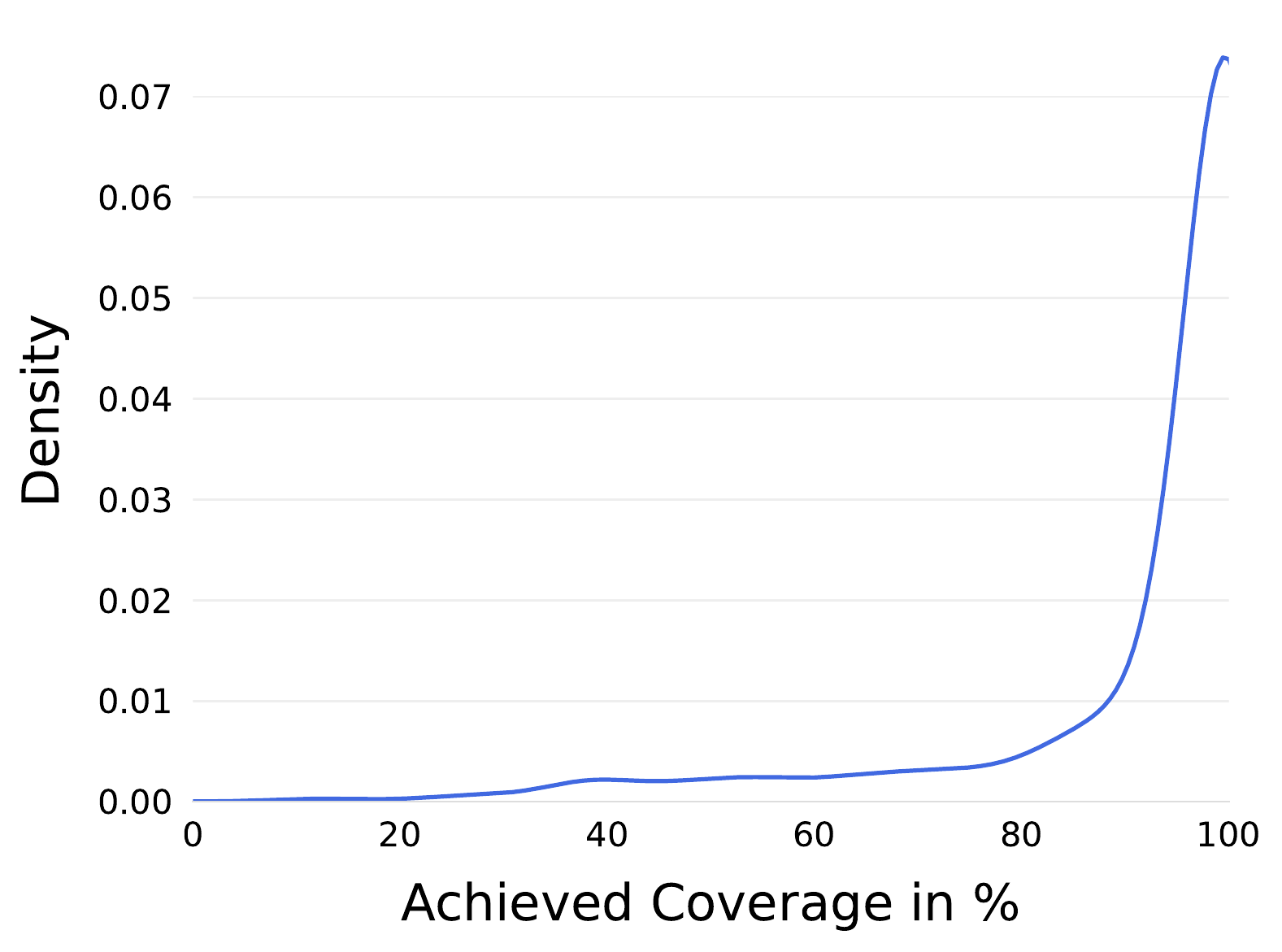}}\hfill
	\subfloat[\label{fig:relevance_coverage:top1000}\textsc{Top1000} ]{\includegraphics[width=0.49\textwidth]{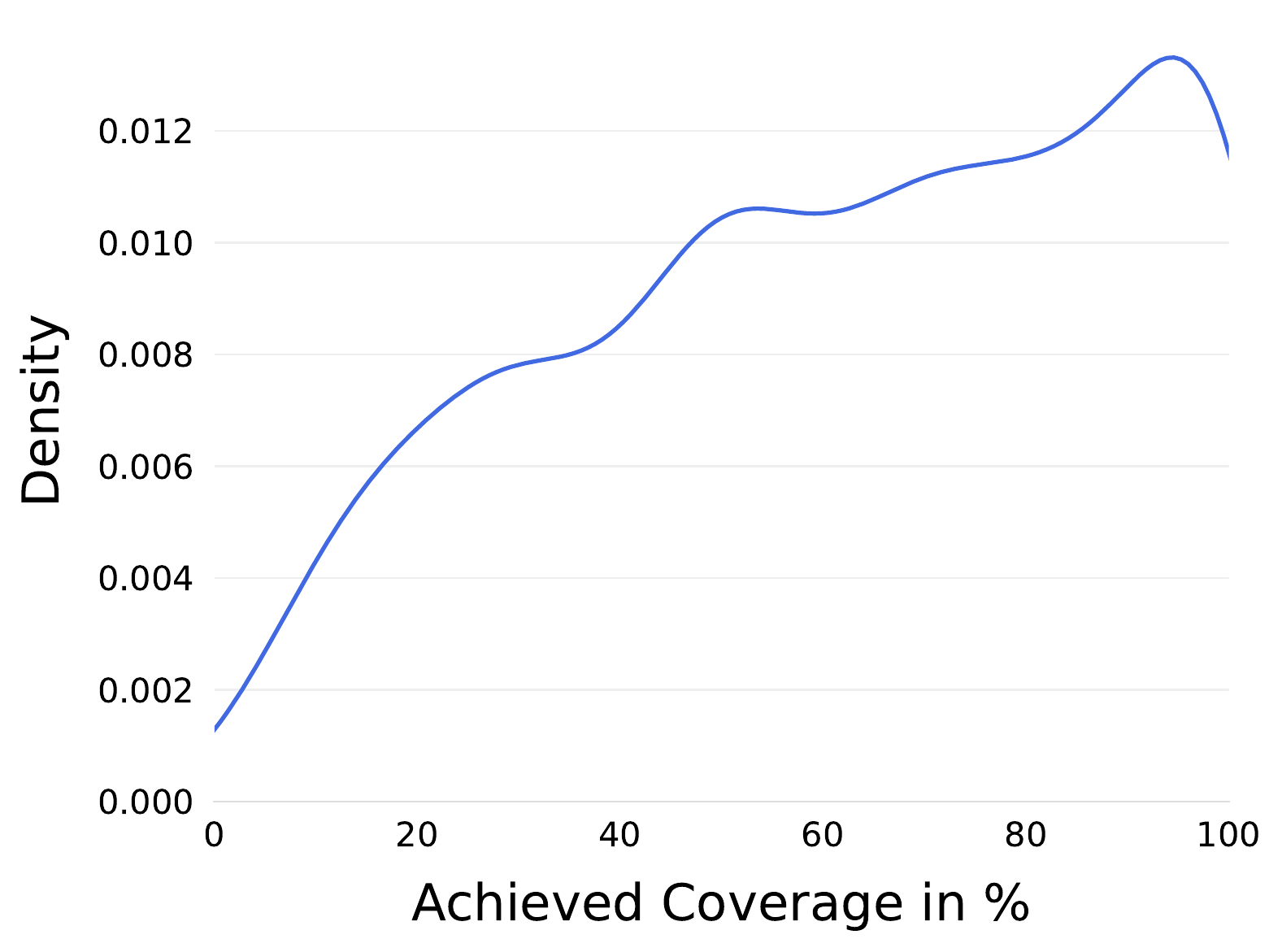}}
\caption{Distribution of coverage results for random testing.}\label{fig:relevance_coverage}
\end{figure}

\begin{figure}[t]
	\centering
	\subfloat[\label{fig:relevance:random1000}\textsc{Random1000}]{\includegraphics[width=0.49\textwidth]{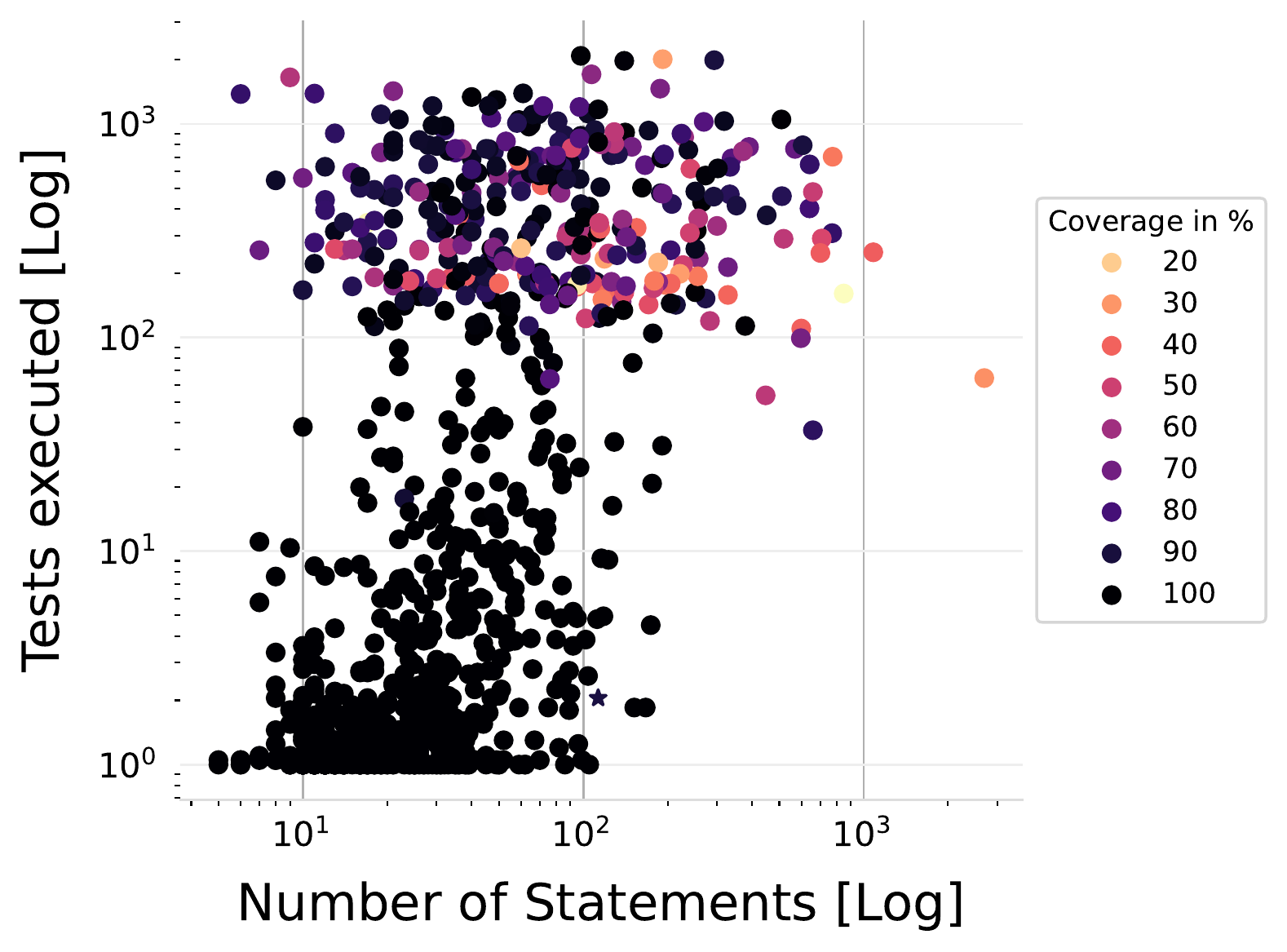}}\hfill
	\subfloat[\label{fig:relevance:top1000}\textsc{Top1000} ]{\includegraphics[width=0.49\textwidth]{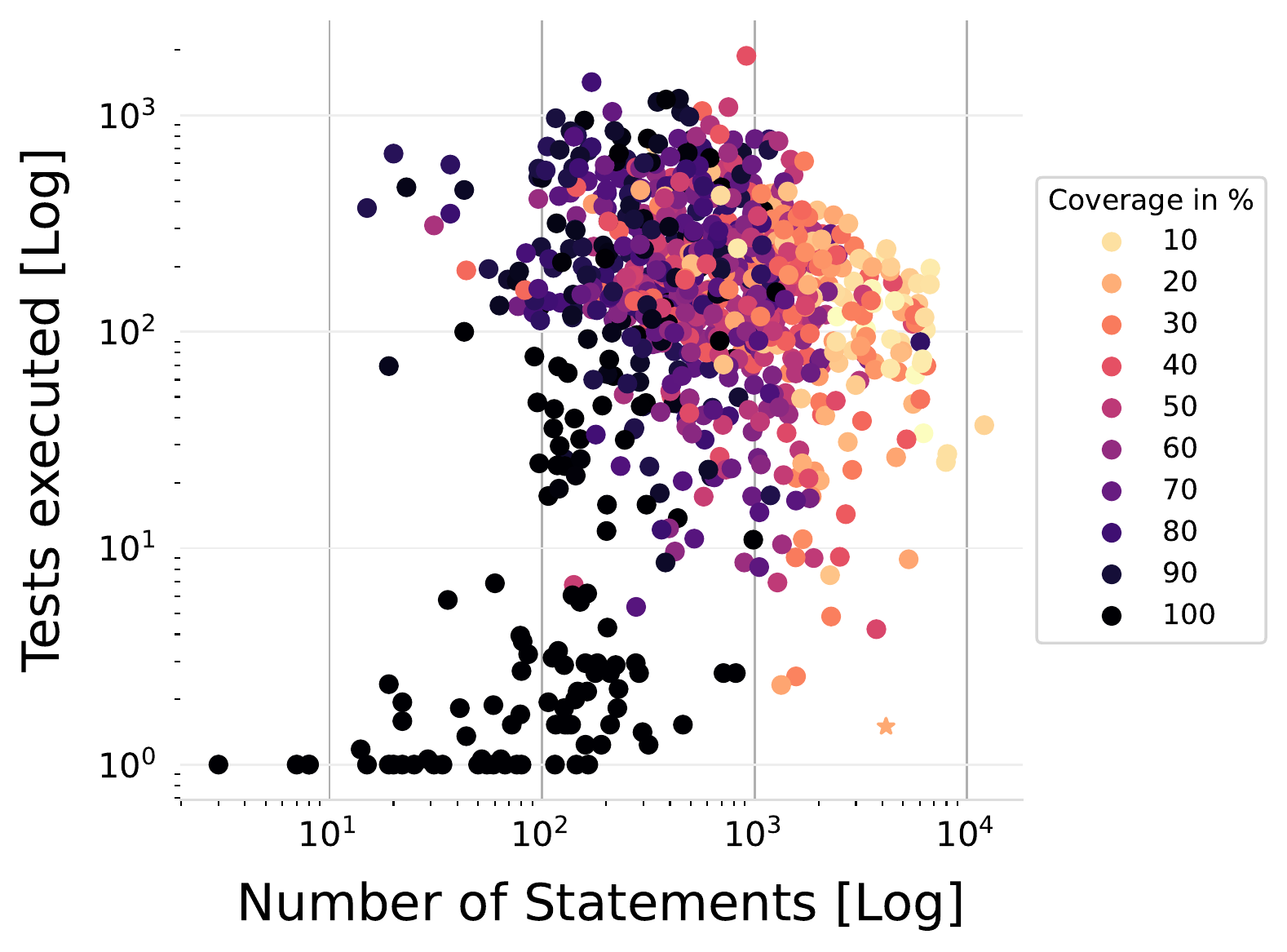}}
\caption{Size vs. tests executed vs. coverage. Executed tests threshold after which projects are treated as non-trivial is marked as a star.}\label{fig:relevance}
\end{figure}

\Scratch projects are created mainly by young learners, which raises the
question whether test generation is actually a \emph{problem}.
For the \rand set, \whisker managed to produce tests for \RandomAlgorithmsWorkingProjects/1000 projects across all used search algorithms. Most of the time, the search algorithms struggle with the remaining 17 programs due to memory and time limitations.

%
\Cref{fig:relevance_coverage:random1000} shows the distribution of coverage
results using basic random testing on \textsc{Random1000}, suggesting that the
majority of projects are fully covered. \Cref{fig:relevance:random1000}
visualises the relation of program size (measured in blocks), the number of
executed tests until the time limit or \SI{100}{\percent} coverage was
achieved, and the resulting average coverage.
The plot contains two clusters: In the bottom left there is a large cluster of projects with less
than 100 blocks, for which \SI{100}{\percent} coverage was achieved within less
than 100 executed tests. The upper half contains a second cluster where results are more varied in terms of size, tests executed, and
resulting coverage.
Based on the project with the lowest average test execution count (plotted as $\bigstar$ in \cref{fig:relevance:random1000}) for which the random test generator did not manage to reach full coverage, we define the \textsc{Random1000} dataset's threshold below which we classify projects as being trivial to be at \AlgTrivThresholdRandom \ executed tests, leading to \RandomAlgorithmsTrivialProjects \ trivial projects.

For \textsc{Top1000} the results look somewhat different
(\cref{fig:relevance_coverage:top1000}): \whisker manages to synthesise slightly fewer tests (\TopRatedAlgorithmsWorkingProjects/1000) for the more challenging programs than for the \rand dataset. While there are still many projects
covered fully, the coverage distribution shows a wider spread of coverage
values achieved, with a tendency towards an almost bimodal distribution with
one peak at about \SI{100}{\percent} and the other around \SI{50}{\percent}. \Cref{fig:relevance:top1000} confirms that the cluster of fully covered projects having less than 100 executed tests is very small for the popular projects. Furthermore, most projects have more than 100
statements, and many even more than 1000 blocks. This is considerable since the
domain-specific blocks of the language do not require many blocks to conjure
interesting behaviour. Furthermore, assembling the blocks for projects larger
than 1000 statements in 
the \Scratch code
editor is a feat in itself.
In the \topp dataset, we encounter the first project having a coverage below 100\% (plotted as $\bigstar$ in \cref{fig:relevance:top1000}) already after a test execution count of \AlgTrivThresholdTop.
Even though the difference between both thresholds is smaller than a single executed test, the \topp set contains significantly fewer trivial projects (\TopRatedAlgorithmsTrivialProjects) than the \rand set.

\begin{figure}[t]
	\subfloat[Full code of the project]{\includegraphics[scale=0.5]{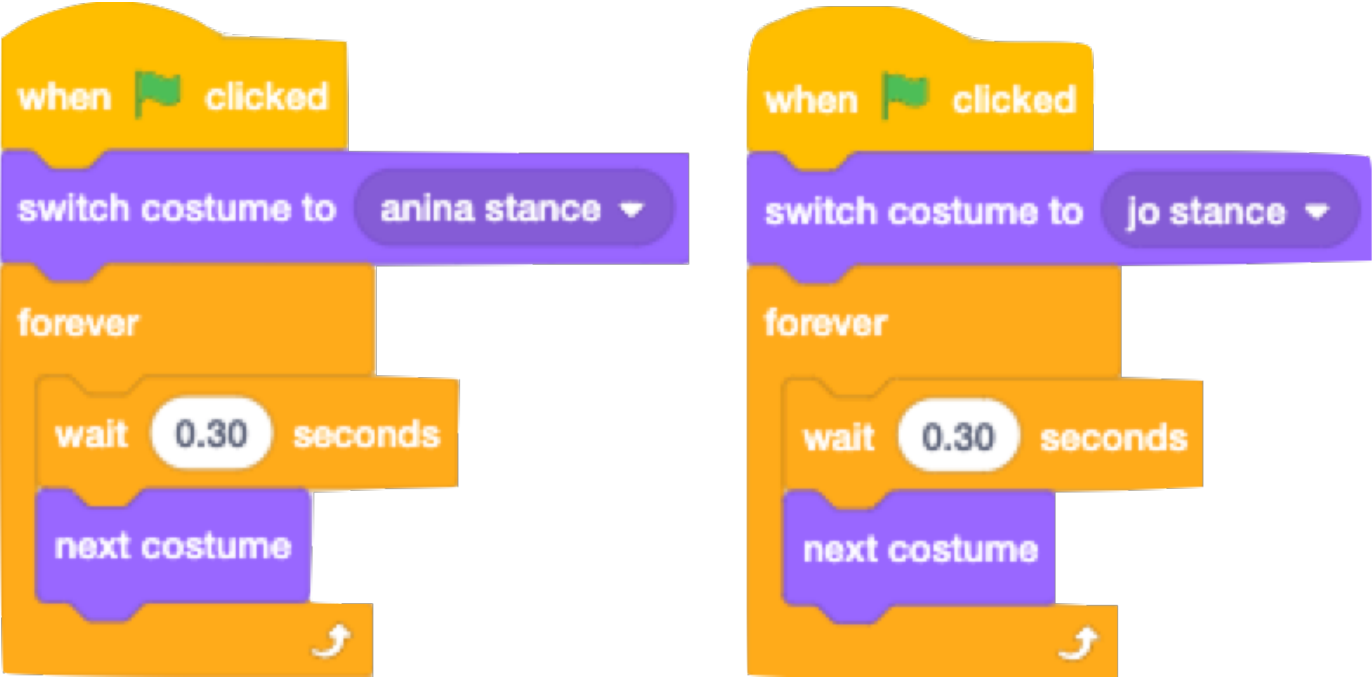}}
	\hfill
	\subfloat[Project stage]{\includegraphics[scale=0.25]{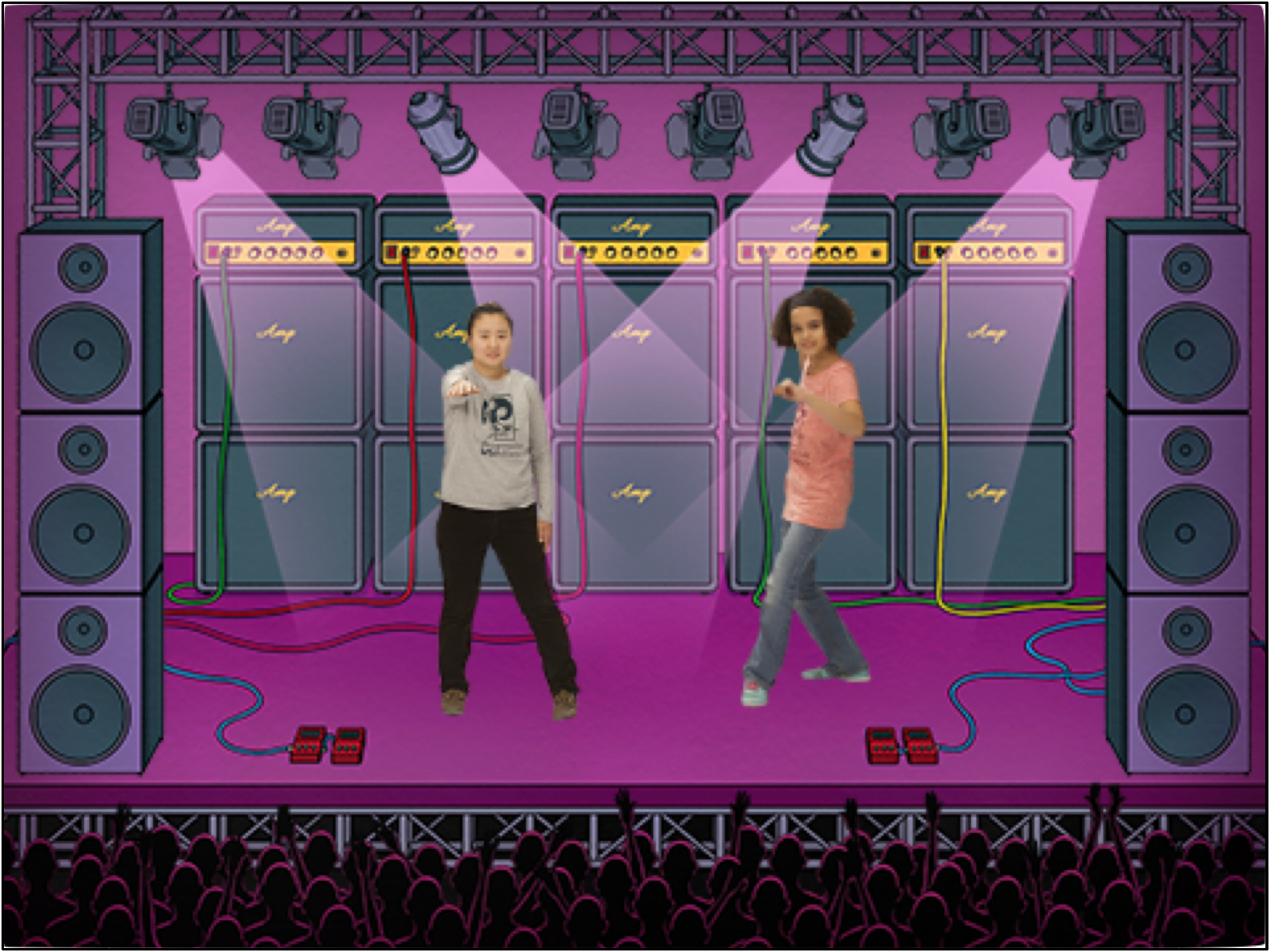}}
\caption{Trivial example project (ID 400050176): Simply clicking on the green flag will cover everything within a few execution steps.\label{fig:example_trivial_small}}
\end{figure}

\begin{figure}[t]
	\subfloat[Full code of the project]{\includegraphics[scale=0.3]{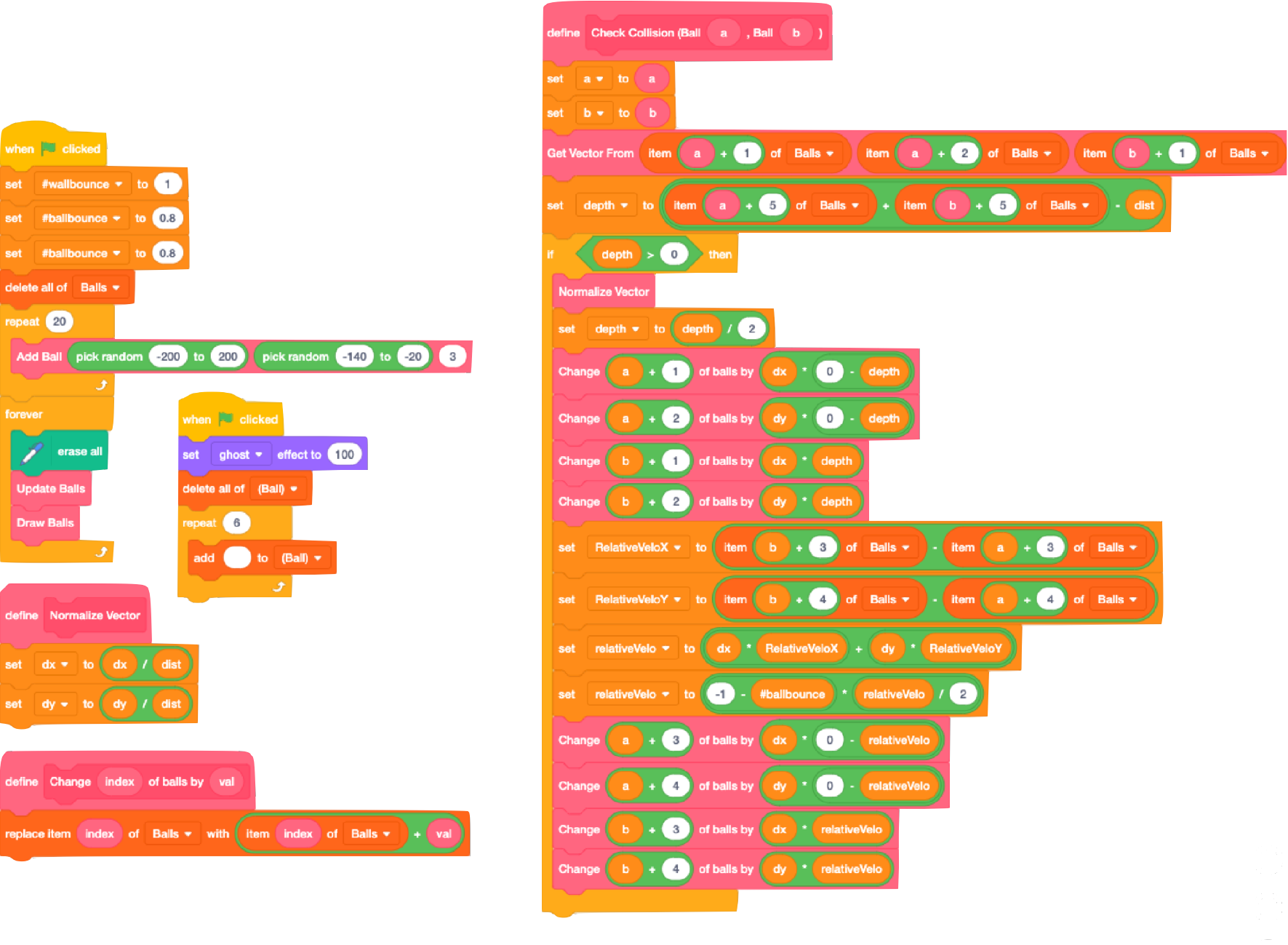}}
	\hfill
	\subfloat[Project stage]{\includegraphics[scale=0.25]{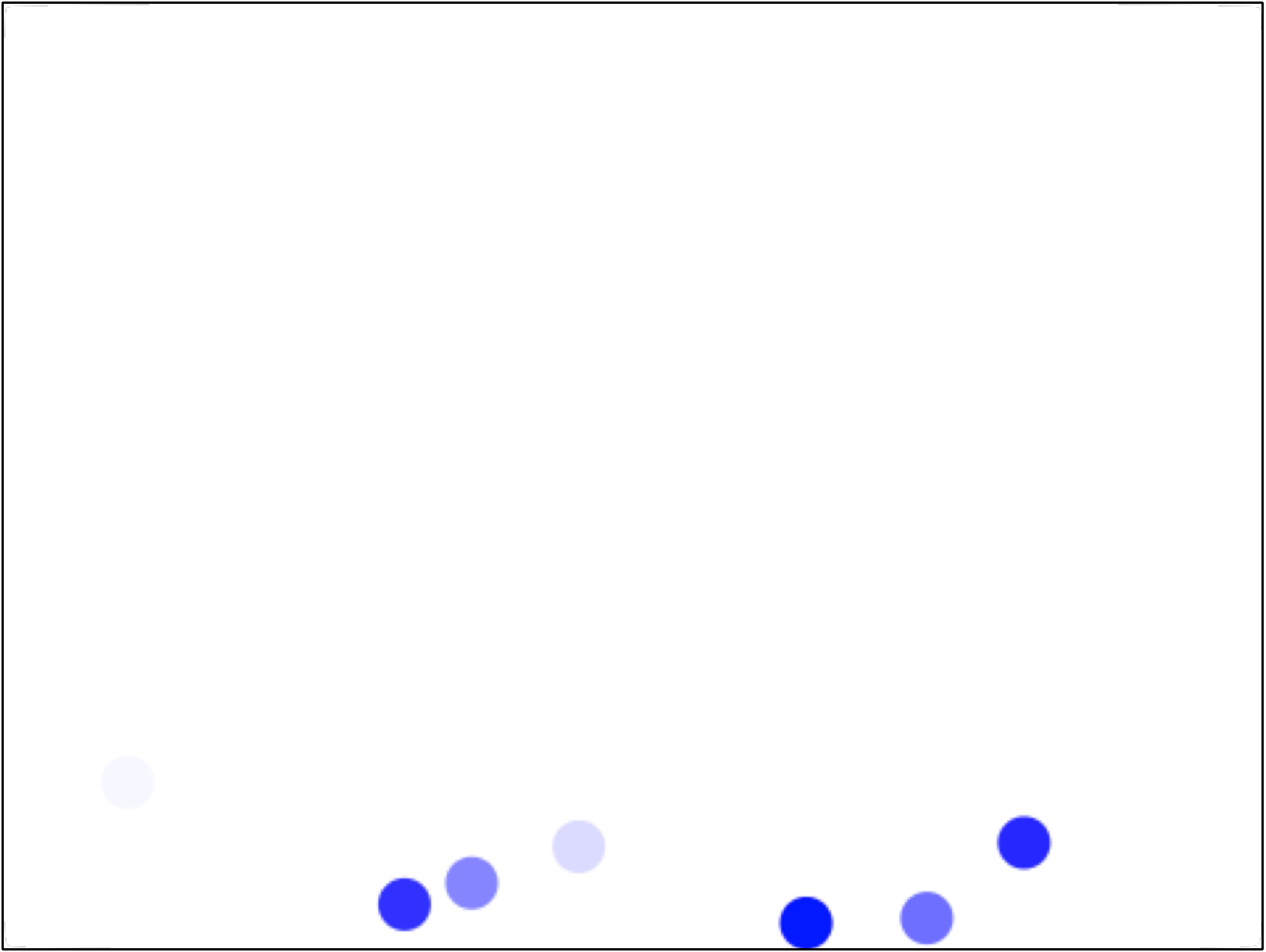}}
\caption{Trivial example project (ID 400011212): Even though there are 314 blocks representing complex vector calculations, covering them requires no interactions.\label{fig:example_trivial_large}}
\end{figure}

Some of the trivial projects are simply very small. For example,
\cref{fig:example_trivial_small} shows a project that contains two scripts with
a total of ten statements, including two loops that control the dance
behaviour of the two sprites, which is achieved by cycling through costumes.
Simply starting the program will cause all statements to be executed within a
few execution steps. On the other hand, even projects that contain more code
may be trivial if they are not interactive. For example,
\cref{fig:example_trivial_large} shows a project which performs complex vector
calculations on list datastructures, but this only serves to simulate
bouncing balls with no user interactions.

Overall, the coverage observed on \textsc{Random1000} is clearly higher than in
other domains, so it appears that many \Scratch programs are indeed trivial.
This matches the intended use case, where young learners initially take their
first steps by building small animations and story-like projects. However, a
fairly large number of projects nevertheless clearly challenges the random
tester for projects in \textsc{Random1000}, which suggests that even the
average \Scratch user may produce non-trivially covered projects. For projects
to become popular (\textsc{Top1000}) it rather seems that the level of
difficulty is on par with other domains of software.


\summary{RQ2}{Out of \RandomAlgorithmsWorkingProjects\ randomly sampled projects, \RandomAlgorithmsTrivialProjects\ achieve full coverage, requiring less than \AlgTrivThresholdRandom\ randomly generated tests.
For the \TopRatedAlgorithmsTotalProjects\ most popular projects, only~\TopRatedAlgorithmsTrivialProjects\ can be
covered fully, requiring less than \AlgTrivThresholdTop\ tests. The other \RandomAlgorithmsNonTrivialProjects\ randomly sampled projects and most (\TopRatedAlgorithmsNonTrivialProjects) of the \topp projects challenge random test generation.}

\subsection{RQ3: What is the best test generation algorithm for \Scratch programs?}
\label{sec:Results-RQ3}

\begin{figure}[t]
	\subfloat[\label{fig:algorithms_coverage:random1000}\textsc{Random1000}]{\includegraphics[width=0.49\textwidth]{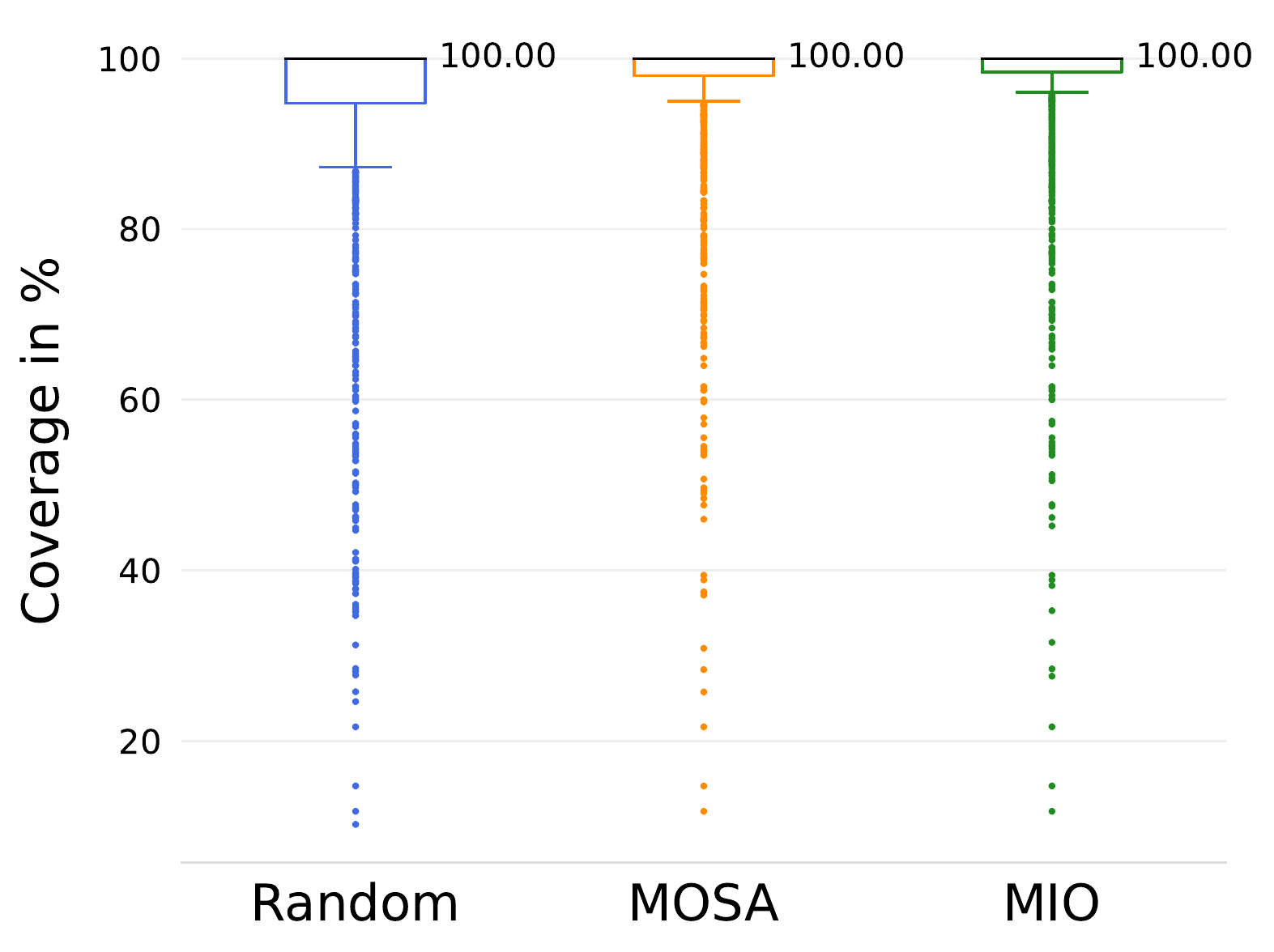}}\hfill
	\subfloat[\label{fig:algorithms_coverage:top1000}\textsc{Top1000}]{\includegraphics[width=0.49\textwidth]{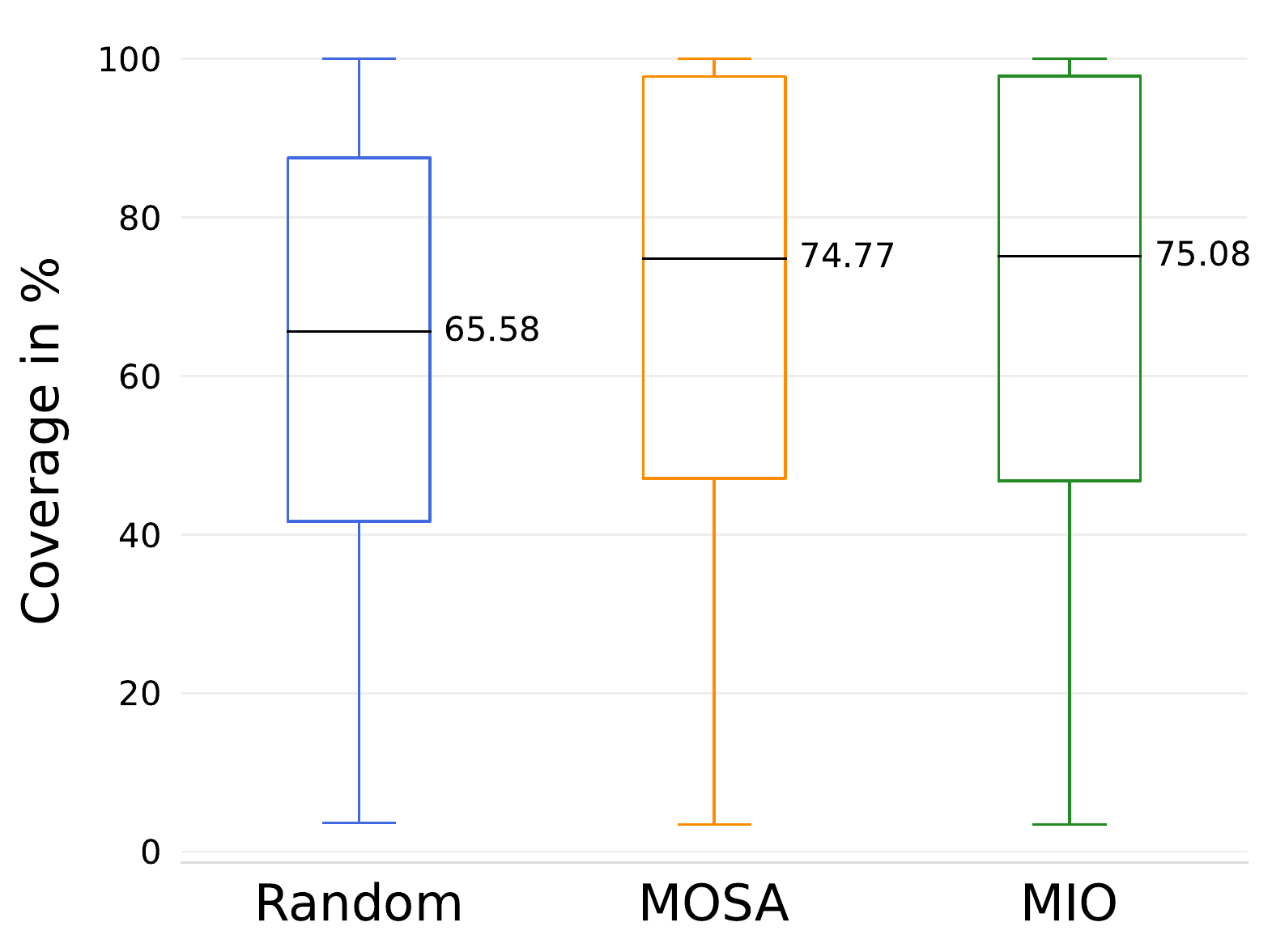}}
\caption{Overall coverage.}\label{fig:algorithms_coverage}
\end{figure}

\begin{figure}[t]
	\subfloat[\label{fig:algorithms_effects:random1000}\textsc{Random1000}]{\includegraphics[width=0.49\textwidth]{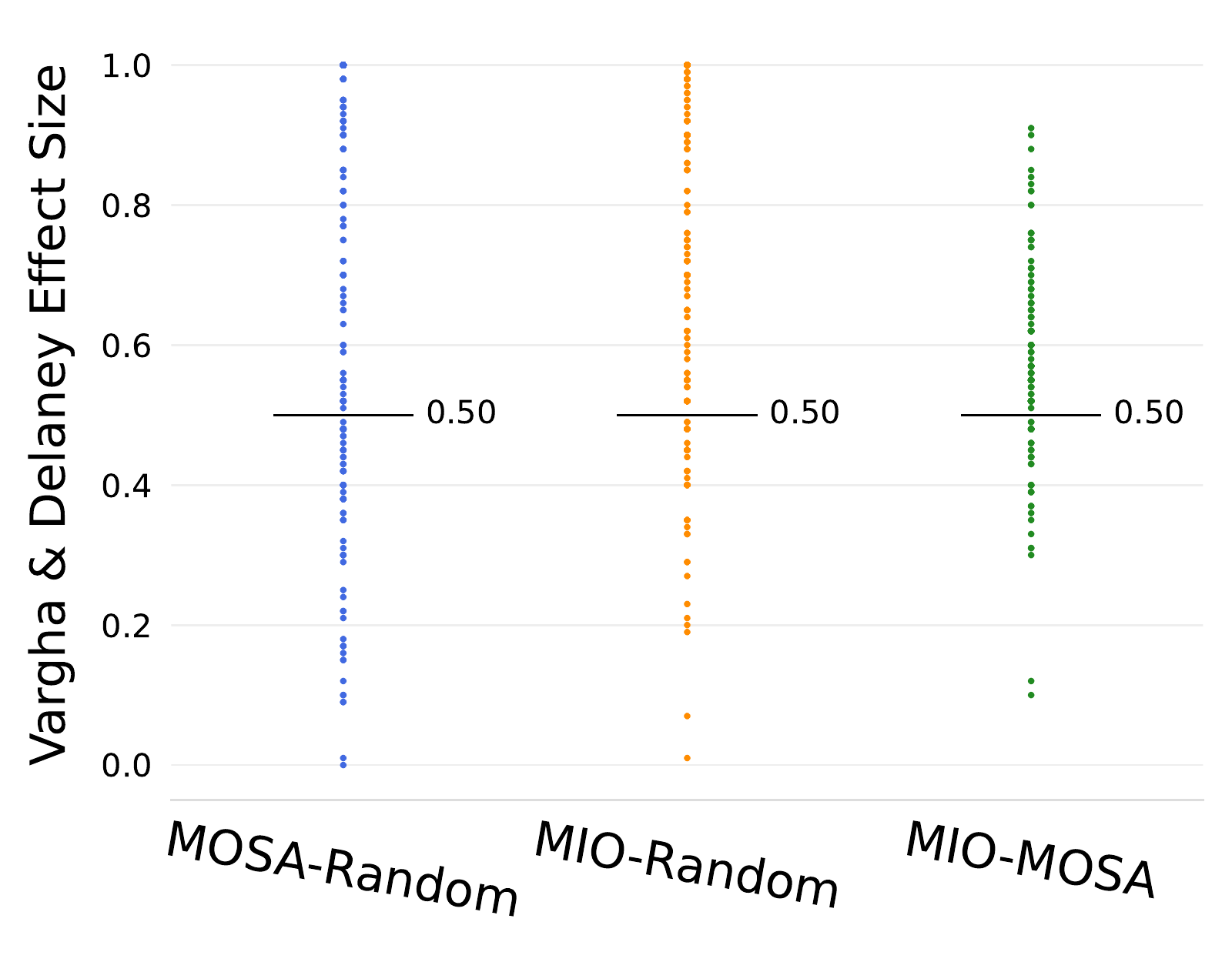}}\hfill
	\subfloat[\label{fig:algorithms_effects:top1000}\textsc{Top1000}]{\includegraphics[width=0.49\textwidth]{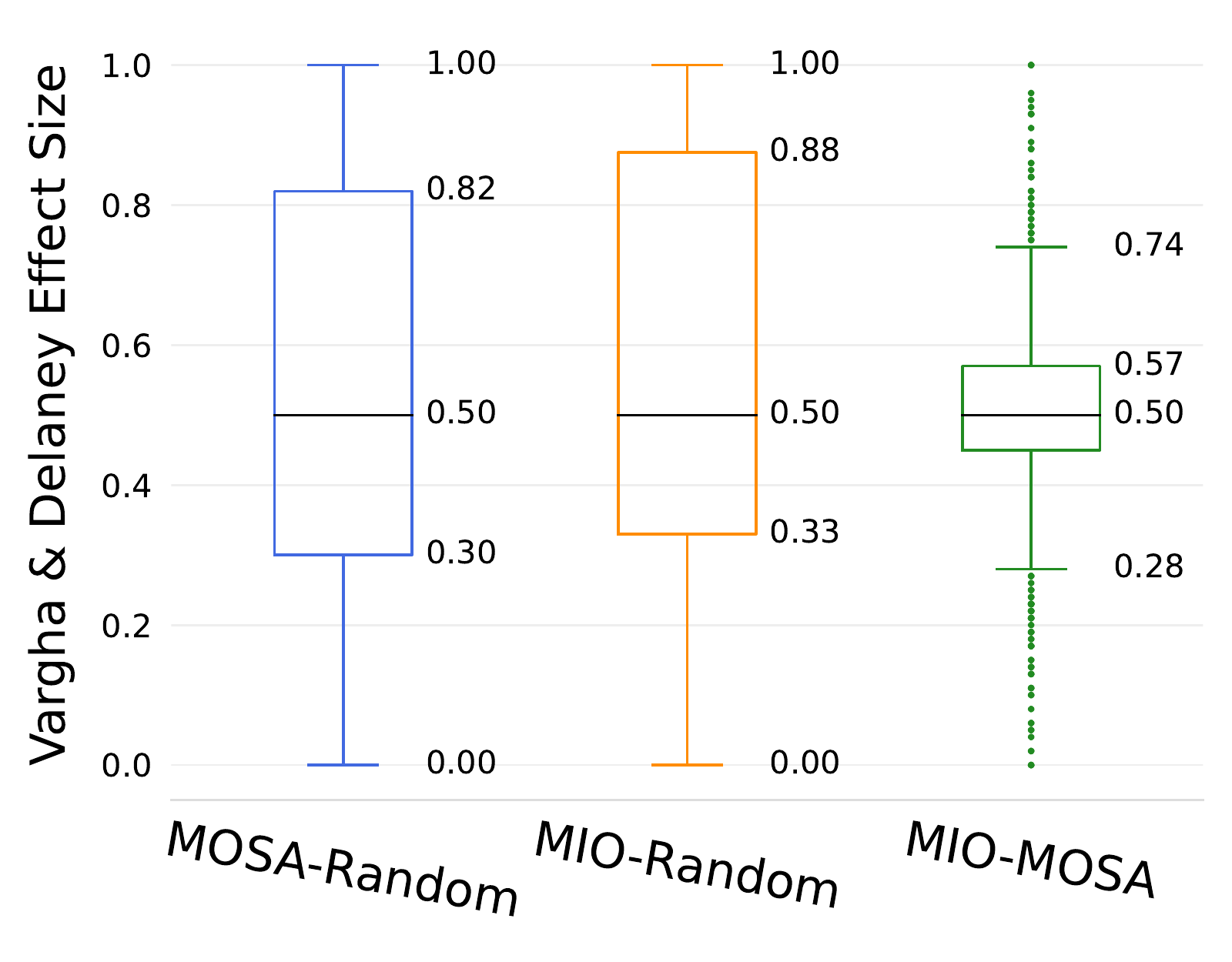}}
\caption{Effect sizes comparing algorithms.}\label{fig:algorithms_effects}
\end{figure}

The box plots in \cref{fig:algorithms_coverage} show the overall coverage
achieved by the different test generation algorithms. For \rand
(\cref{fig:algorithms_coverage:random1000}), all algorithms achieve very high
coverage with a median of \SI{100}{\percent}. This is not surprising
considering the large number of trivial projects (cf. RQ2). There is, however a
difference noticeable on average, where random test generation leads to
\SI{\RandomAlgorithmsAverageCoverageRANDOM}{\percent} coverage, MOSA achieves
\SI{\RandomAlgorithmsAverageCoverageMOSA}{\percent} coverage,
and MIO achieves \SI{\RandomAlgorithmsAverageCoverageMIO}{\percent} coverage.
For \topp (\cref{fig:algorithms_coverage:top1000}) the coverage is
substantially lower, and the differences between the algorithms are more
pronounced: random test generation leads to
\SI{\TopRatedAlgorithmsAverageCoverageRANDOM}{\percent} coverage, MOSA
achieves \SI{\TopRatedAlgorithmsAverageCoverageMOSA}{\percent} coverage,
and MIO achieves \SI{\TopRatedAlgorithmsAverageCoverageMIO}{\percent} coverage.

The average coverage values suggest that MIO is the best algorithm, and MOSA is
still better than random testing. \Cref{fig:algorithms_effects} sheds more
light on the differences by showing the distribution of Vargha-Delaney \atwelve
effect sizes. The median is 0.5 for both datasets, which the statistical
comparison summarised in \cref{tab:Algorithm-Coverage} explains: For a large
share of projects all algorithms achieve the
same level of coverage; for \rand this is often \SI{100}{\percent}
(\RandomAlgorithmsFullCoverageProjects \ projects), while for \topp only
\TopRatedAlgorithmsFullCoverageProjects \ achieve \SI{100}{\percent} coverage.
Consequently, particularly for projects similar to those in \rand very often
the chosen algorithm will make no difference.

\begin{table}[t]
\caption{Test generation comparison of the achieved block coverages. Besides the \atwelve value, we report the number of projects for which the first approach achieves significantly better, better, equal, worse and significantly worse results than the second approach. Statistically significant results are determined by using the Mann-Whitney-U test and a \pval $<$ 0.05.}
\label{tab:Algorithm-Coverage}
\begin{tabularx}{\columnwidth}{llrrrrrr}
\toprule
Testset & Comparison & \heading{\atwelve} & \heading{Sig Better} & \heading{Better} & \heading{Equal} & \heading{Worse} & \heading{Sig Worse} \\
\midrule
\rand & MOSA vs. Random & \RandomAlgorithmsVDCoverageMOSARandom & \RandomAlgorithmsCovSigBetterMOSAvsRANDOM & \RandomAlgorithmsCovBetterMOSAvsRANDOM & \RandomAlgorithmsCovEqualMOSAvsRANDOM & \RandomAlgorithmsCovWorseMOSAvsRANDOM & \RandomAlgorithmsCovSigWorseMOSAvsRANDOM \\
\rand & MIO vs. Random  & \RandomAlgorithmsVDCoverageMIORandom & \RandomAlgorithmsCovSigBetterMIOvsRANDOM & \RandomAlgorithmsCovBetterMIOvsRANDOM & \RandomAlgorithmsCovEqualMIOvsRANDOM & \RandomAlgorithmsCovWorseMIOvsRANDOM & \RandomAlgorithmsCovSigWorseMIOvsRANDOM \\
\rand & MIO vs. MOSA & \RandomAlgorithmsVDCoverageMIOMOSA & \RandomAlgorithmsCovSigBetterMIOvsMOSA & \RandomAlgorithmsCovBetterMIOvsMOSA & \RandomAlgorithmsCovEqualMIOvsMOSA & \RandomAlgorithmsCovWorseMIOvsMOSA & \RandomAlgorithmsCovSigWorseMIOvsMOSA \\
\midrule
\topp & MOSA vs. Random & \TopRatedAlgorithmsVDCoverageMOSARandom & \TopRatedAlgorithmsCovSigBetterMOSAvsRANDOM & \TopRatedAlgorithmsCovBetterMOSAvsRANDOM & \TopRatedAlgorithmsCovEqualMOSAvsRANDOM & \TopRatedAlgorithmsCovWorseMOSAvsRANDOM & \TopRatedAlgorithmsCovSigWorseMOSAvsRANDOM \\
\topp & MIO vs. Random & \TopRatedAlgorithmsVDCoverageMIORandom & \TopRatedAlgorithmsCovSigBetterMIOvsRANDOM & \TopRatedAlgorithmsCovBetterMIOvsRANDOM & \TopRatedAlgorithmsCovEqualMIOvsRANDOM & \TopRatedAlgorithmsCovWorseMIOvsRANDOM & \TopRatedAlgorithmsCovSigWorseMIOvsRANDOM \\
\topp & MIO vs. MOSA & \TopRatedAlgorithmsVDCoverageMIOMOSA & \TopRatedAlgorithmsCovSigBetterMIOvsMOSA & \TopRatedAlgorithmsCovBetterMIOvsMOSA & \TopRatedAlgorithmsCovEqualMIOvsMOSA & \TopRatedAlgorithmsCovWorseMIOvsMOSA & \TopRatedAlgorithmsCovSigWorseMIOvsMOSA \\
\bottomrule
\end{tabularx}
\end{table}

\begin{figure}[t]
	\subfloat[\label{fig:random_vs_mio:random1000}Random vs. MIO on \rand]{\includegraphics[width=0.49\textwidth]{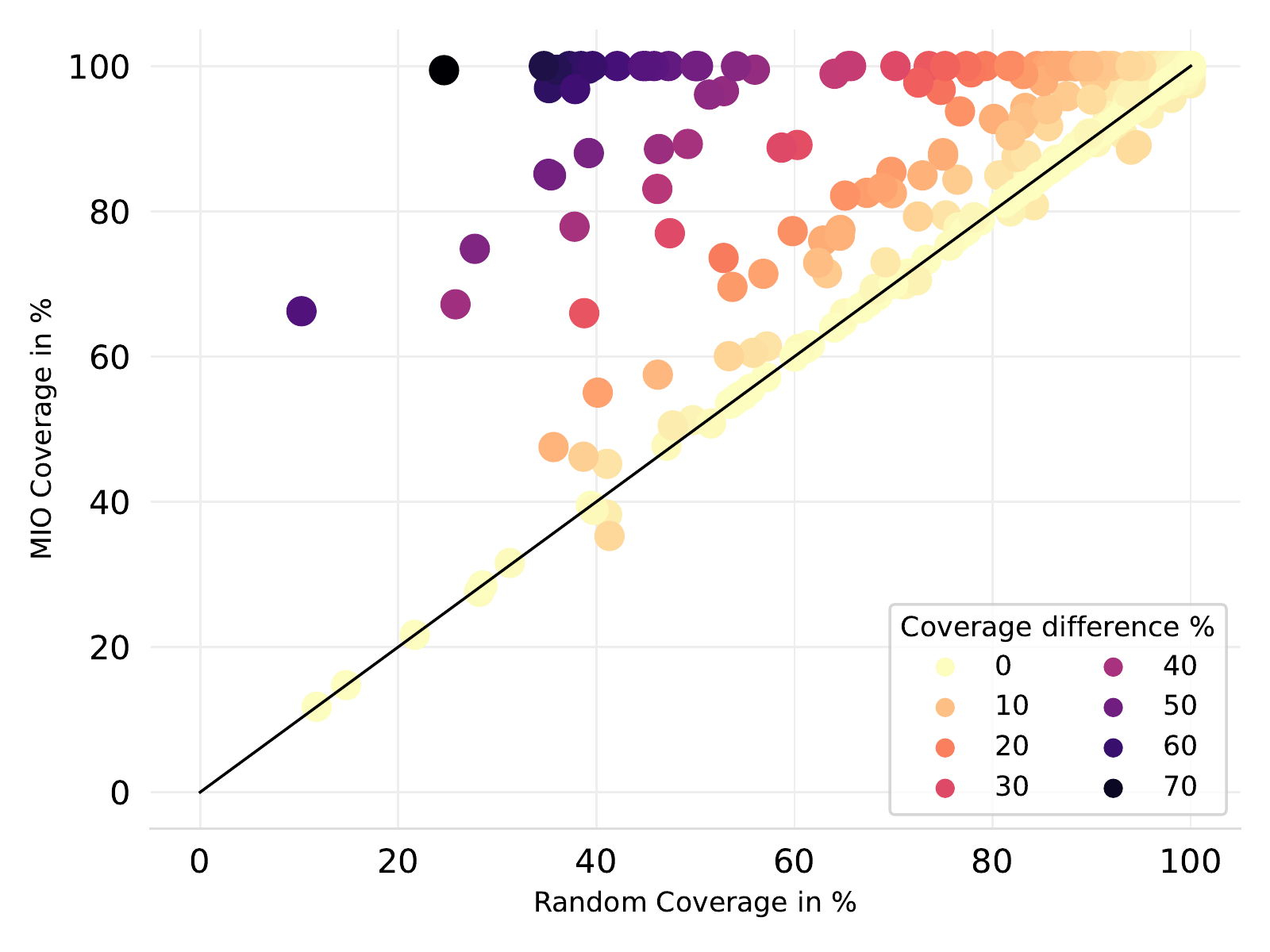}}\hfill
	\subfloat[\label{fig:random_vs_mosa:random1000}Random vs. MOSA on \rand]{\includegraphics[width=0.49\textwidth]{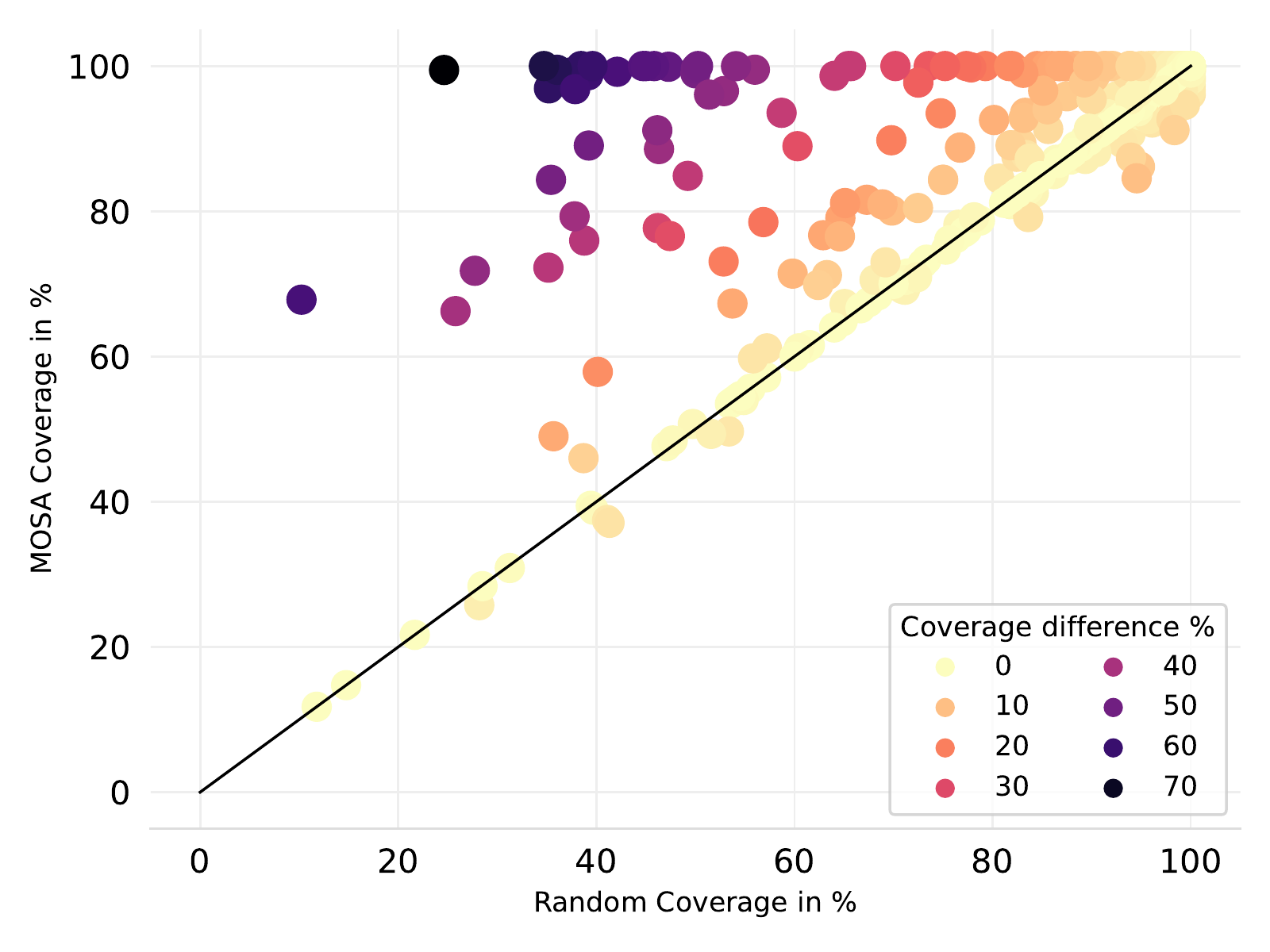}}\\
	\subfloat[\label{fig:random_vs_mio:top1000}Random vs. MIO on \topp ]{\includegraphics[width=0.49\textwidth]{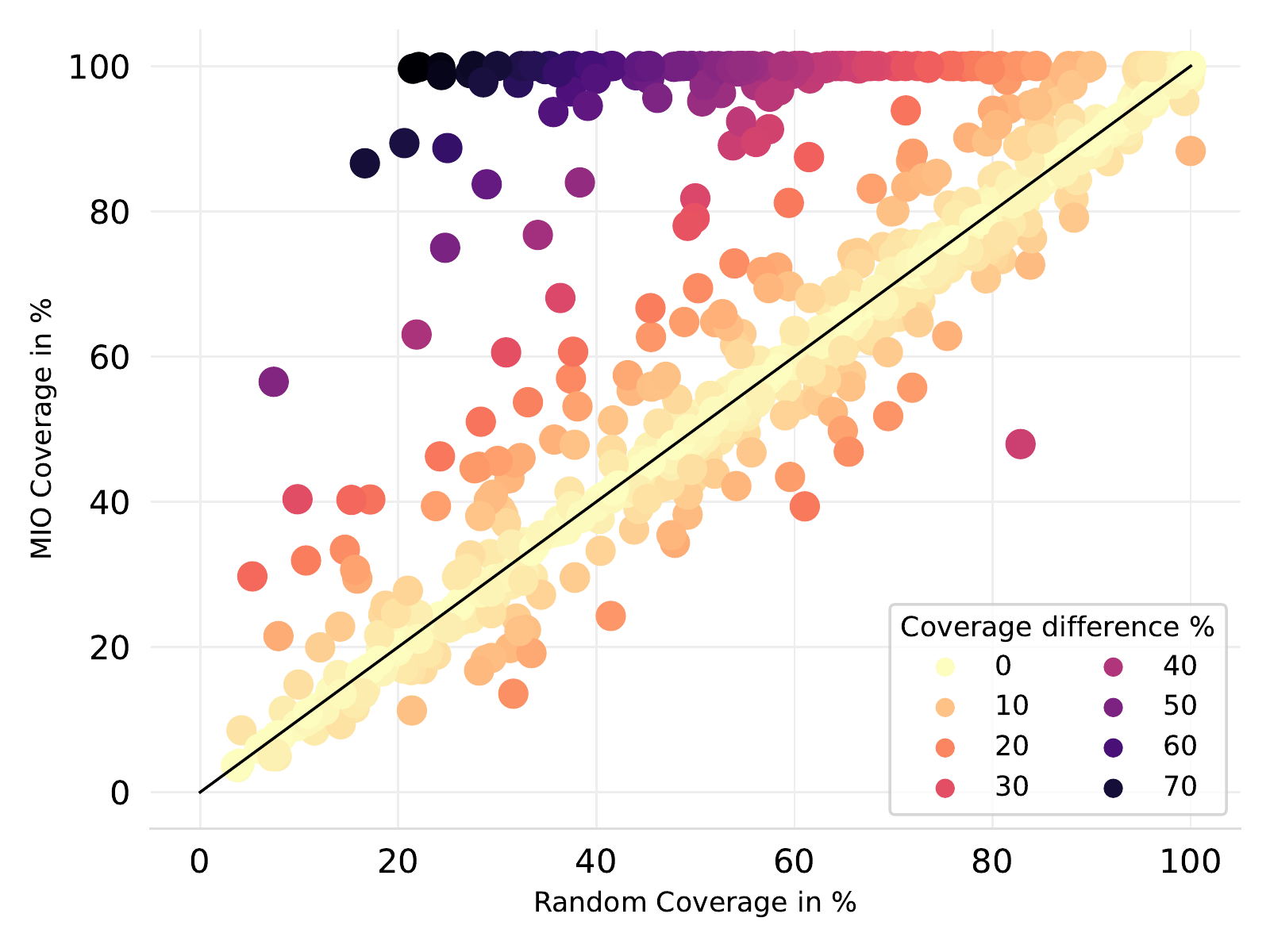}}\hfill
	\subfloat[\label{fig:random_vs_mosa:top1000}Random vs. MOSA on \topp ]{\includegraphics[width=0.49\textwidth]{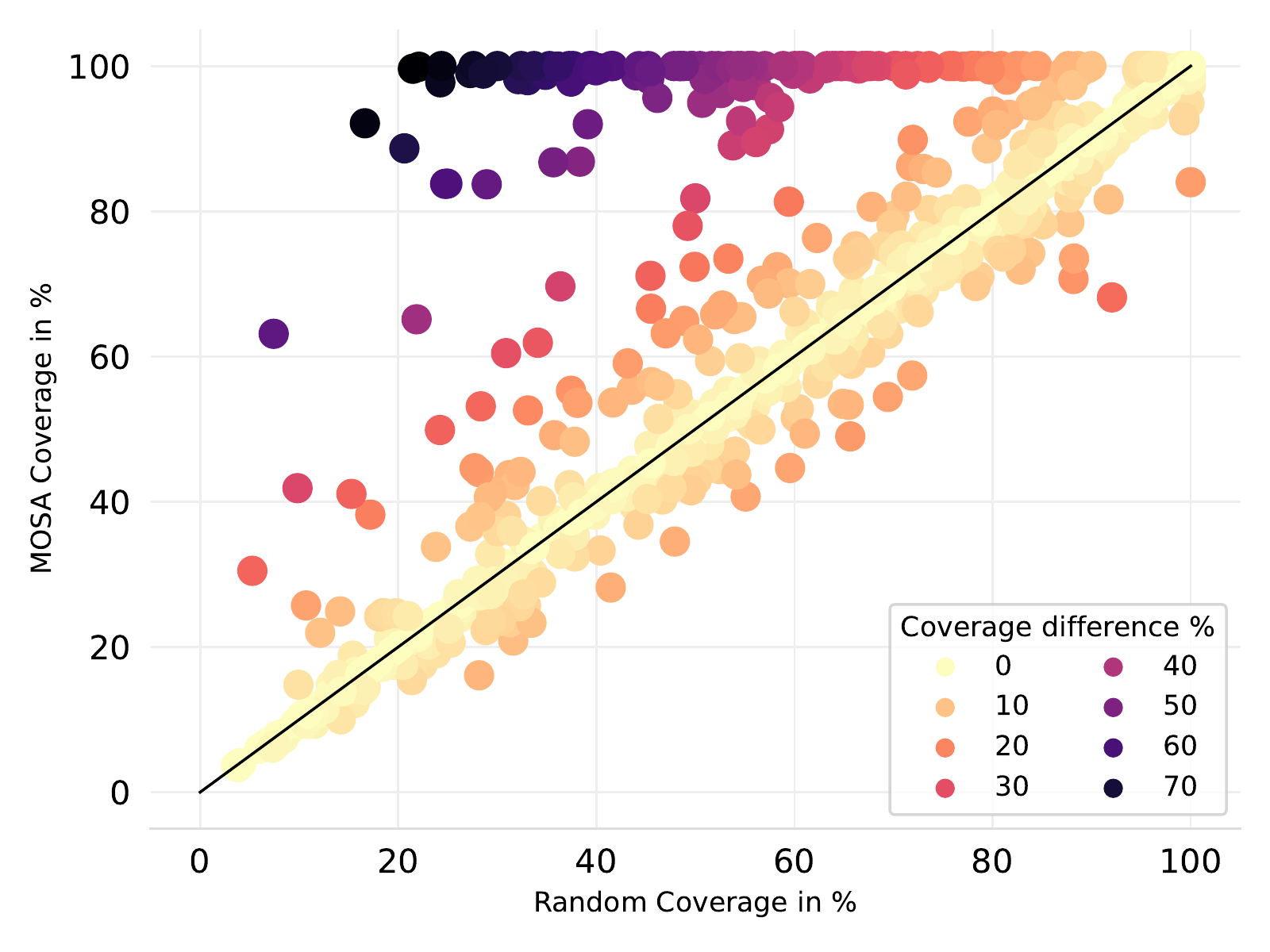}}\\
\caption{Comparison of achieved coverage.}\label{fig:random_vs_search}
\end{figure}

The differences in average coverage can be explained by  \RandomAlgorithmsCovBetterMIOvsRANDOM \ projects in \rand
where MIO performs better than random testing (significant for
\RandomAlgorithmsCovSigBetterMIOvsRANDOM), and
\RandomAlgorithmsCovBetterMOSAvsRANDOM \ for MOSA over random testing
(significant for \RandomAlgorithmsCovSigBetterMOSAvsRANDOM). This is substantially more than
the number of cases where random is better than MIO
(\RandomAlgorithmsCovWorseMIOvsRANDOM, significant for
\RandomAlgorithmsCovSigWorseMIOvsRANDOM) and MOSA
(\RandomAlgorithmsCovWorseMIOvsRANDOM, significant for
\RandomAlgorithmsCovSigWorseMOSAvsRANDOM).
This is also reflected in the average effect size of
\RandomAlgorithmsVDCoverageMOSARandom \ for MOSA vs. random testing, and
\RandomAlgorithmsVDCoverageMIORandom \ for MIO vs. random testing. Consequently,
there are clear benefits to using either of the search algorithms over random
testing on projects similar to those we randomly sampled.

The trade-off between search and random testing is less clear for \topp: MIO
performs better than random testing for \TopRatedAlgorithmsCovBetterMIOvsRANDOM
\ projects (significant for \TopRatedAlgorithmsCovSigBetterMIOvsRANDOM), and
MOSA for \TopRatedAlgorithmsCovBetterMOSAvsRANDOM \
(significant for \TopRatedAlgorithmsCovSigBetterMOSAvsRANDOM); at the same time, however,
random is better than MIO for \TopRatedAlgorithmsCovWorseMIOvsRANDOM \ projects
(significant for \TopRatedAlgorithmsCovSigWorseMIOvsRANDOM) and MOSA for
\TopRatedAlgorithmsCovWorseMOSAvsRANDOM \ (significant for
\TopRatedAlgorithmsCovSigWorseMOSAvsRANDOM). On average the effect size
nevertheless leans towards search
(\TopRatedAlgorithmsVDCoverageMIORandom \ for MIO vs. Random, and
\TopRatedAlgorithmsVDCoverageMOSARandom \ for MOSA vs. Random). To better
understand this result, \cref{fig:random_vs_search} contrasts the coverage per
project between random and the two search algorithms for both datasets. For
all four cases the picture is very similar: A large share of the projects is
clustered around the diagonal with equal coverage, and there is a larger spread
of projects to the left of the diagonal than to the right, meaning that the
search algorithms achieved higher coverage. For \topp
(\cref{fig:random_vs_mio:top1000} and \cref{fig:random_vs_mosa:top1000}) the
spread around the diagonal is notably larger than for \rand
(\cref{fig:random_vs_mio:random1000} and \cref{fig:random_vs_mosa:random1000}),
which shows that even though there are more cases with differences on \topp,
these are often insubstantial. When search is better, it is often better by a
very large margin.

The differences between MOSA and MIO are small, but an \atwelve of
\RandomAlgorithmsVDCoverageMIOMOSA \ for the \rand set and slightly more significantly better results for MIO (\TopRatedAlgorithmsCovSigBetterMIOvsMOSA \ vs. \TopRatedAlgorithmsCovSigWorseMIOvsMOSA) suggest that MIO overall is
the algorithm better suited for the problem at hand.
\Cref{fig:random_vs_search} also shows only small differences between the two
algorithms, confirming that differences are small, though slightly in favour of MIO. We conjecture that
this is influenced by the larger degree of exploration achieved in MIO through
the parameters that emerged from our tuning process.

\begin{figure}[t]
		\subfloat[\label{fig:algorithms_time:random1000}\textsc{Random1000}]{\includegraphics[width=0.49\textwidth]{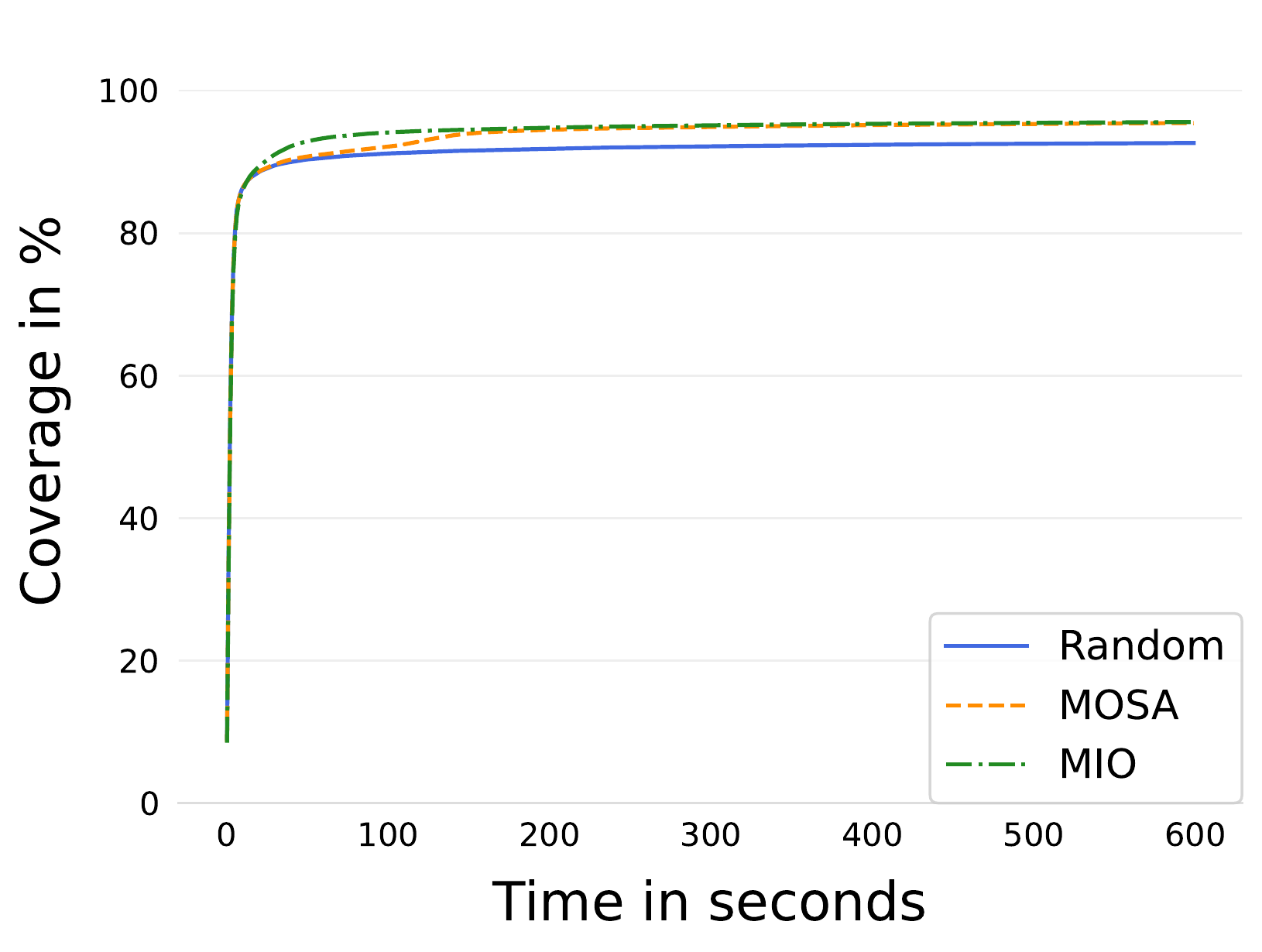}}\hfill
		\subfloat[\label{fig:algorithms_time:top1000}\textsc{Top1000}]{\includegraphics[width=0.49\textwidth]{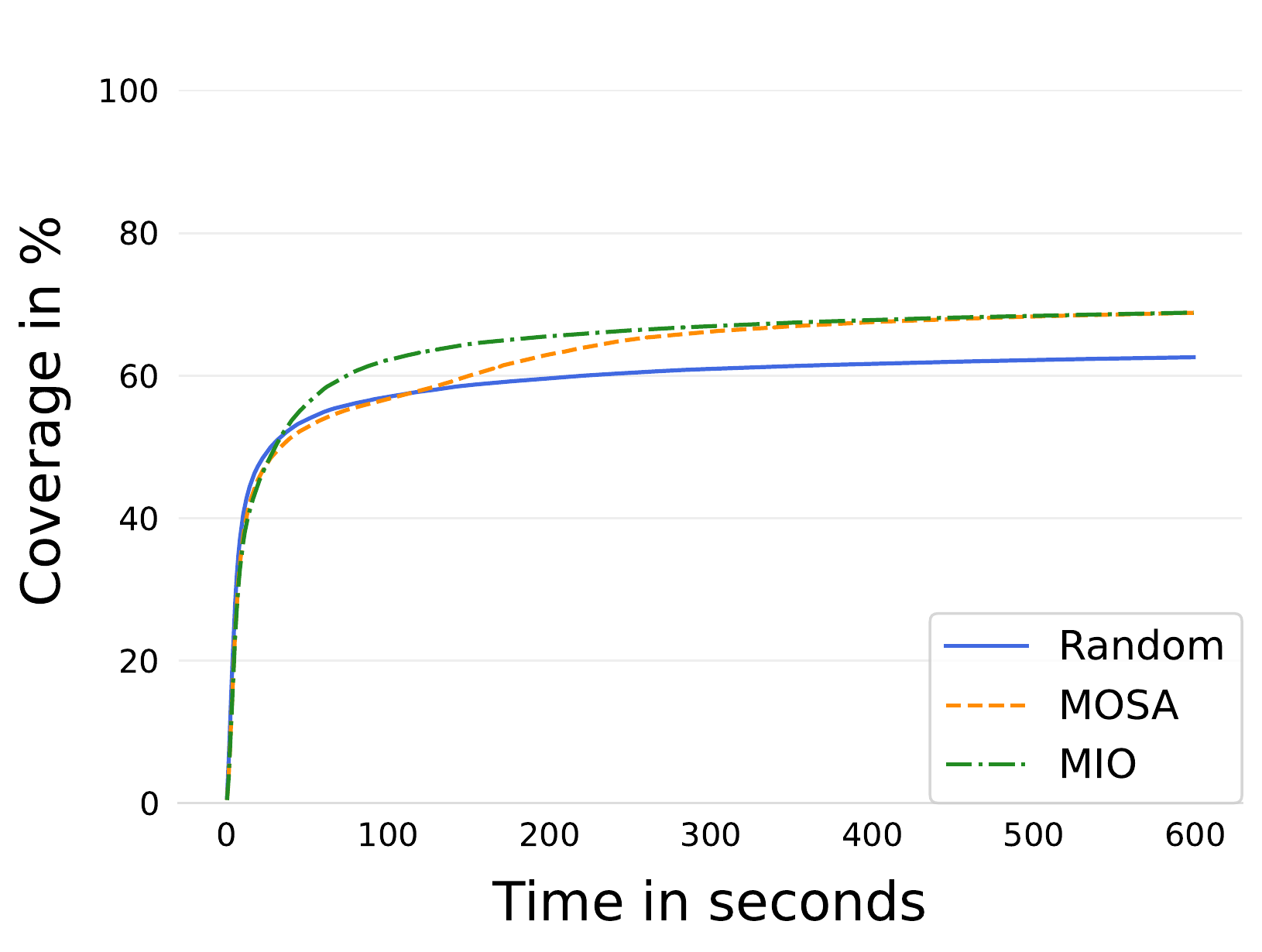}}
\caption{Average coverage over time.}\label{fig:algorithms_time}
\end{figure}

\Cref{fig:algorithms_time} shows how coverage evolves over time for both, \rand
and \topp: Random test generation very quickly converges at a lower coverage
value on both datasets, whereas the two search algorithms successfully evolve
tests to cover more code. The plot also shows a distinct difference between MIO
and MOSA: The MOSA algorithm requires longer to reach a higher level of
coverage, whereas MIO has substantially higher coverage within the first 2--3
minutes of the search. This is due to the population based approach of MOSA,
which applies evolutionary operators to entire generations of the population
size chosen (30 in our experiments). In contrast, MIO produces one test at a
time and directs the search towards promising areas of the search space, which
initially allows it to perform better. Consequently in particular if the time
budget is limited, MIO may be a preferable choice.

\begin{figure}[t]
	\subfloat[Code excerpt of the project]{\includegraphics[scale=0.5]{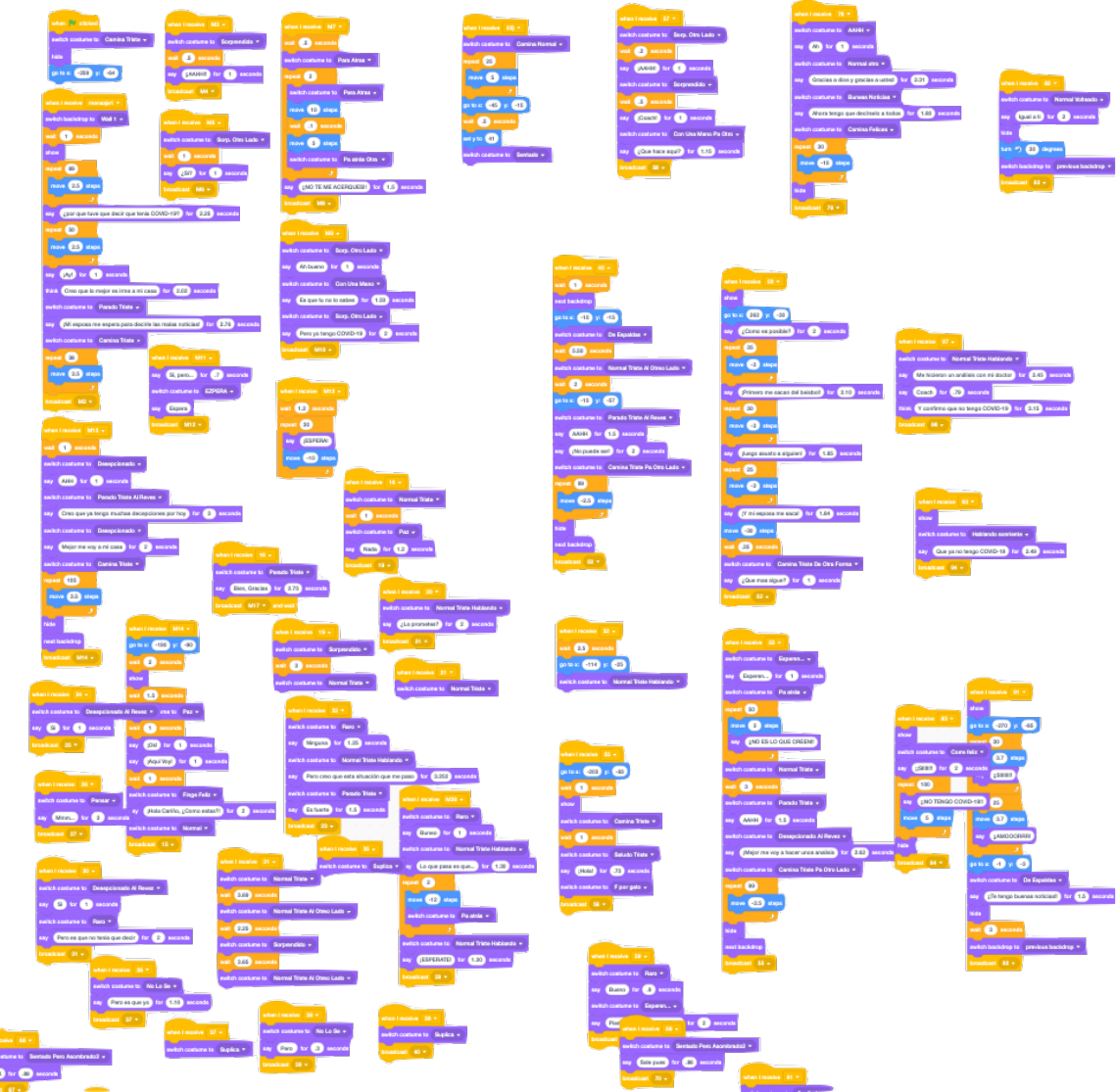}}
	\hfill
	\subfloat[Project stage]{\includegraphics[scale=0.25]{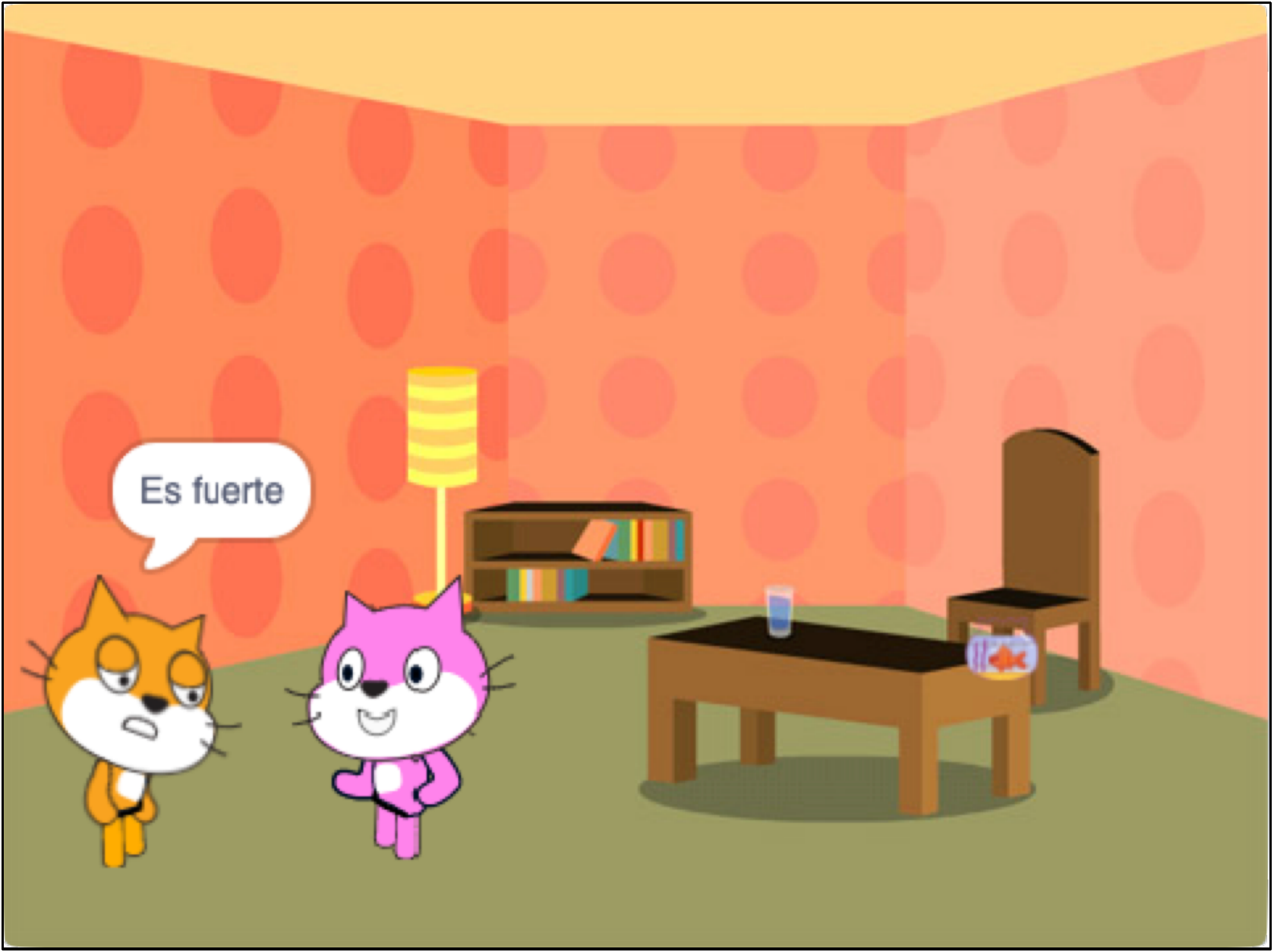}}
\caption{Example project (ID 401050644): The project represents a story with more than 100 scenes, each encoded in an individual script, triggered by a broadcast.\label{fig:example_mio_vs_random}}
\end{figure}

\begin{figure}[t]
		\subfloat[\label{fig:execution_time:random1000}\textsc{Random1000}]{\includegraphics[width=0.49\textwidth]{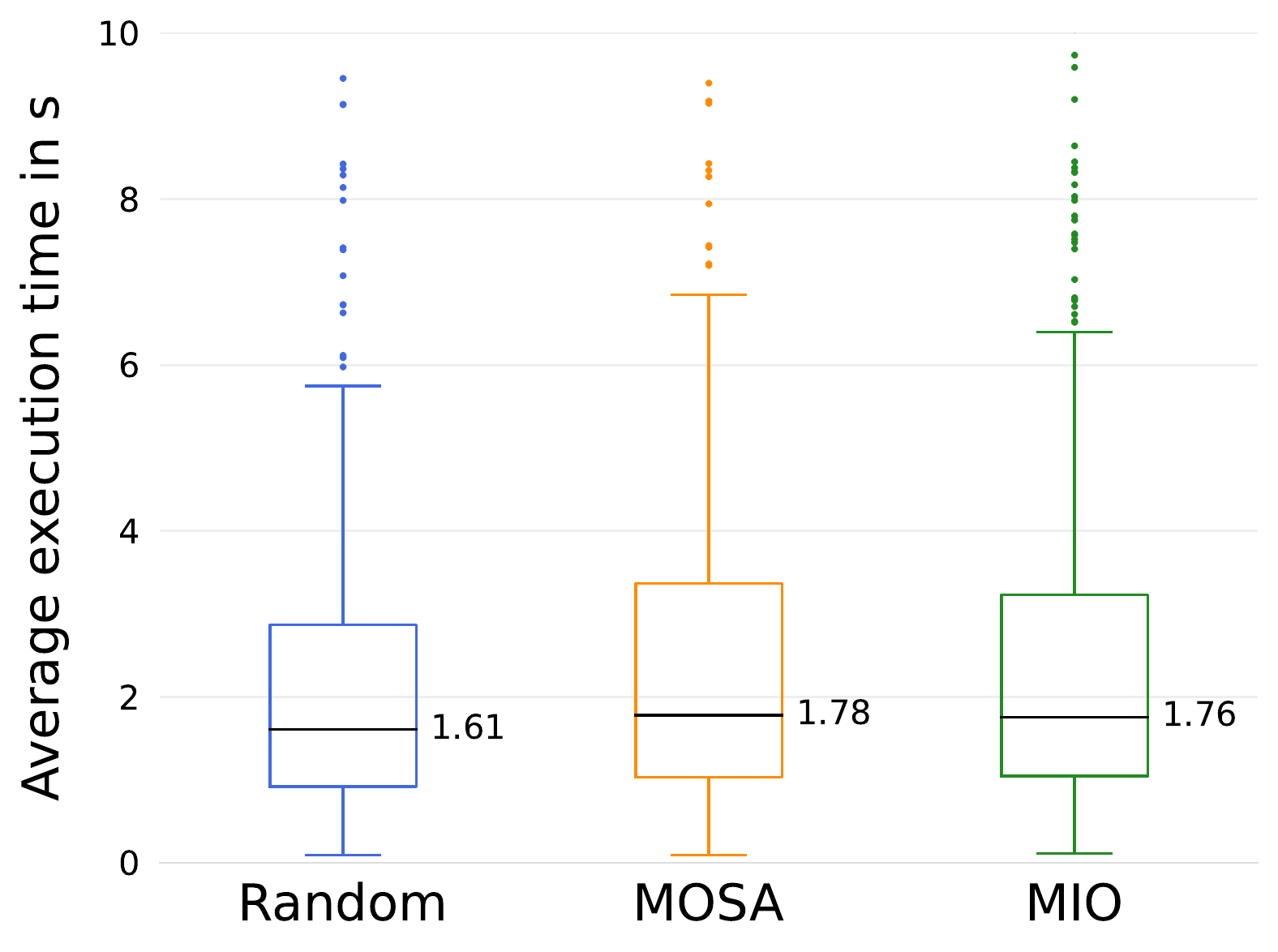}}\hfill
		\subfloat[\label{fig:execution_time:top1000}\textsc{Top1000}]{\includegraphics[width=0.49\textwidth]{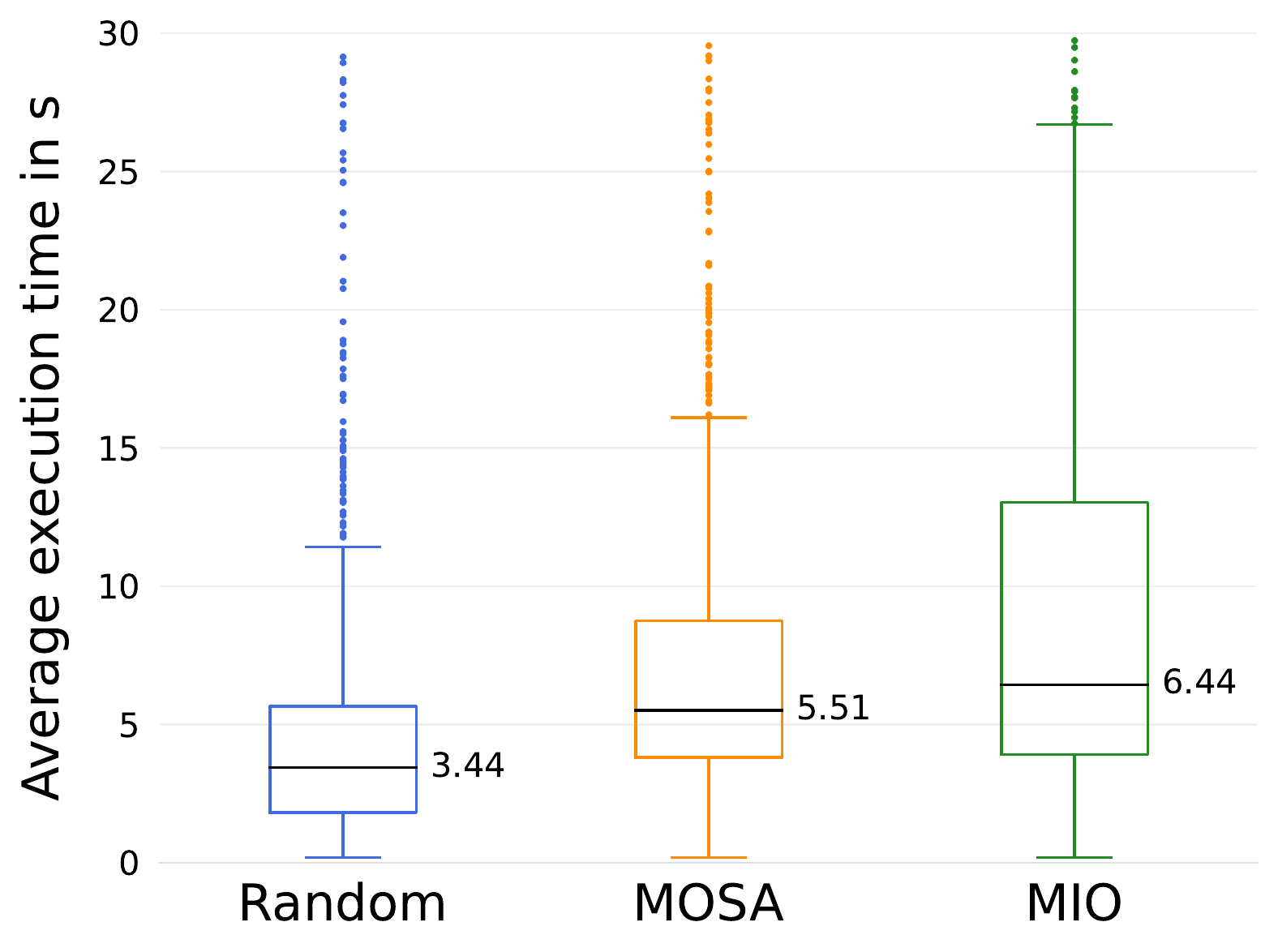}}
\caption{Average test execution time.}\label{fig:execution_time}
\end{figure}

The largest improvement of MOSA and MIO over random testing can be observed for
projects that implement non-trivial story behaviour. For example,
\cref{fig:example_mio_vs_random} shows project ID 401050644, where MOSA and MIO
achieve \SI{67.83}{\percent} and \SI{66.24}{\percent} coverage respectively, while random testing achieves only an average of \SI{10.23}{\percent}. The story consists of more than 100 individual
scenes in which eight different sprites interact. Each scene is encoded as a
script that is triggered by a message with the scene ID, and at the end
broadcasts a message with the next scene ID. Covering the program entirely
requires waiting long, and the fitness function provides a monotonic gradient
to achieve this: The approach level captures the dependencies between the
broadcasts, the control flow distance captures the progress in the scripts, and
the branch distance captures the progression of time-related blocks. In
conjunction with the extension local search, the problem thus becomes easy for
the search. A similar pattern can be observed for many of the projects with
large differences between search and random testing.

Although this type of project results in the largest difference between search
and random testing, \cref{fig:execution_time} suggests that very long execution sequences do not appear to be dominating.
While the execution speed differences for the random set are negligible (Random: \RandomAlgorithmsAverageExecutionTimeRANDOM s, MOSA: \RandomAlgorithmsAverageExecutionTimeMOSA s, MIO: \RandomAlgorithmsAverageExecutionTimeMIO s), we notice more significant differences and overall longer running tests for the \topp projects (Random: \TopRatedAlgorithmsAverageExecutionTimeRANDOM s, MOSA: \TopRatedAlgorithmsAverageExecutionTimeMOSA s, MIO: \TopRatedAlgorithmsAverageExecutionTimeMIO s).
However, a look at the execution times for projects of the \topp dataset in which all algorithms achieve exactly the same amount of coverage reveals that these results correlate with the increased program coverage (Random: \TopRatedAlgorithmsAverageExecutionTimeEqualRANDOM s, MOSA: \TopRatedAlgorithmsAverageExecutionTimeEqualMOSA s, MIO: \TopRatedAlgorithmsAverageExecutionTimeEqualMIO s), as the differences in execution times become negligible again.

Note that these execution times
refer to \emph{accelerated} execution, which means that effectively
(unaccelerated) the tests are running up to an average of more than \emph{two minutes} per test suite! Clearly, test generation without accelerated execution would
be challenging.
Notably these execution times are substantially higher than common values found
in other test generation domains. For example, in search-based unit test
generation~\citep{fraser2012whole} tests tend to execute within a few
milliseconds. Since the computational costs of test execution are the central
bottleneck in search-based test generation, this explains why, even though
\Scratch programs are substantially smaller than other types of software, we
still have to run test generation for 10 minutes for reasonable results.

\begin{figure}[t]
	\subfloat[\label{fig:example_mio_vs_random2:code}Code excerpt]{\includegraphics[scale=0.4]{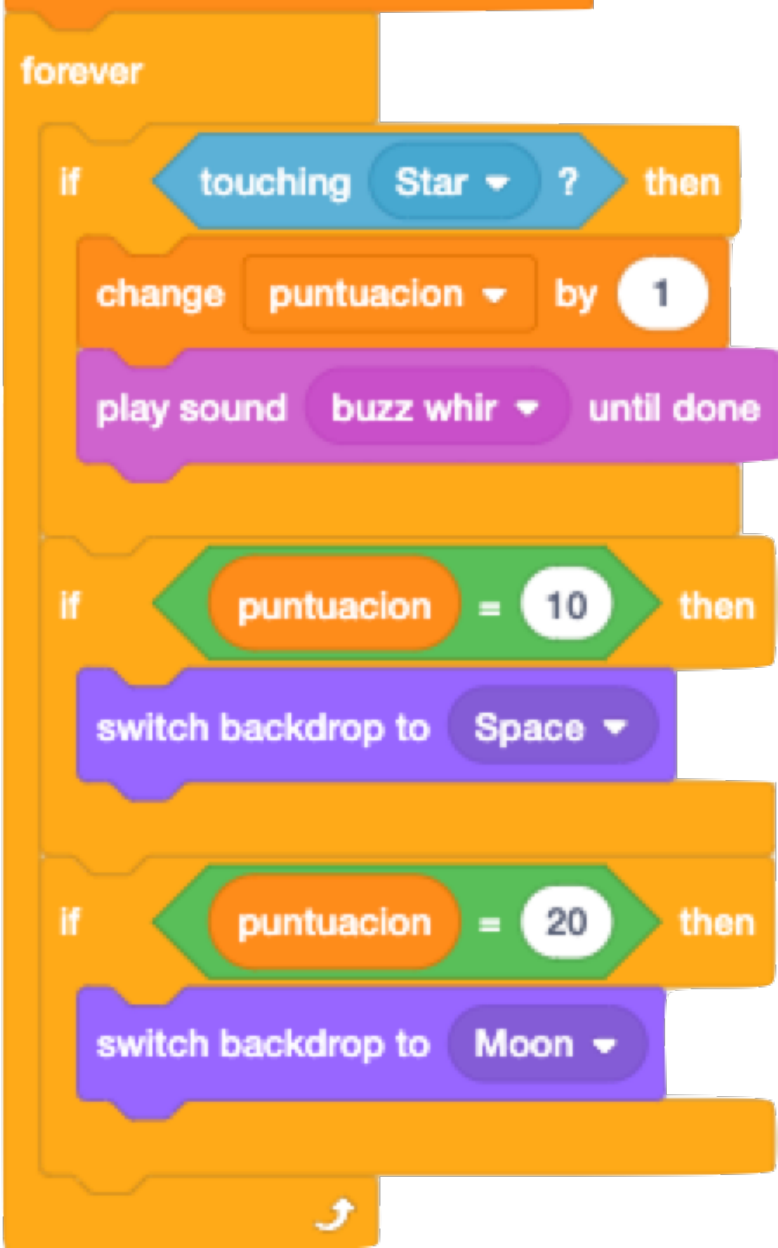}}
	\hfill
	\subfloat[Project stage]{\includegraphics[scale=0.25]{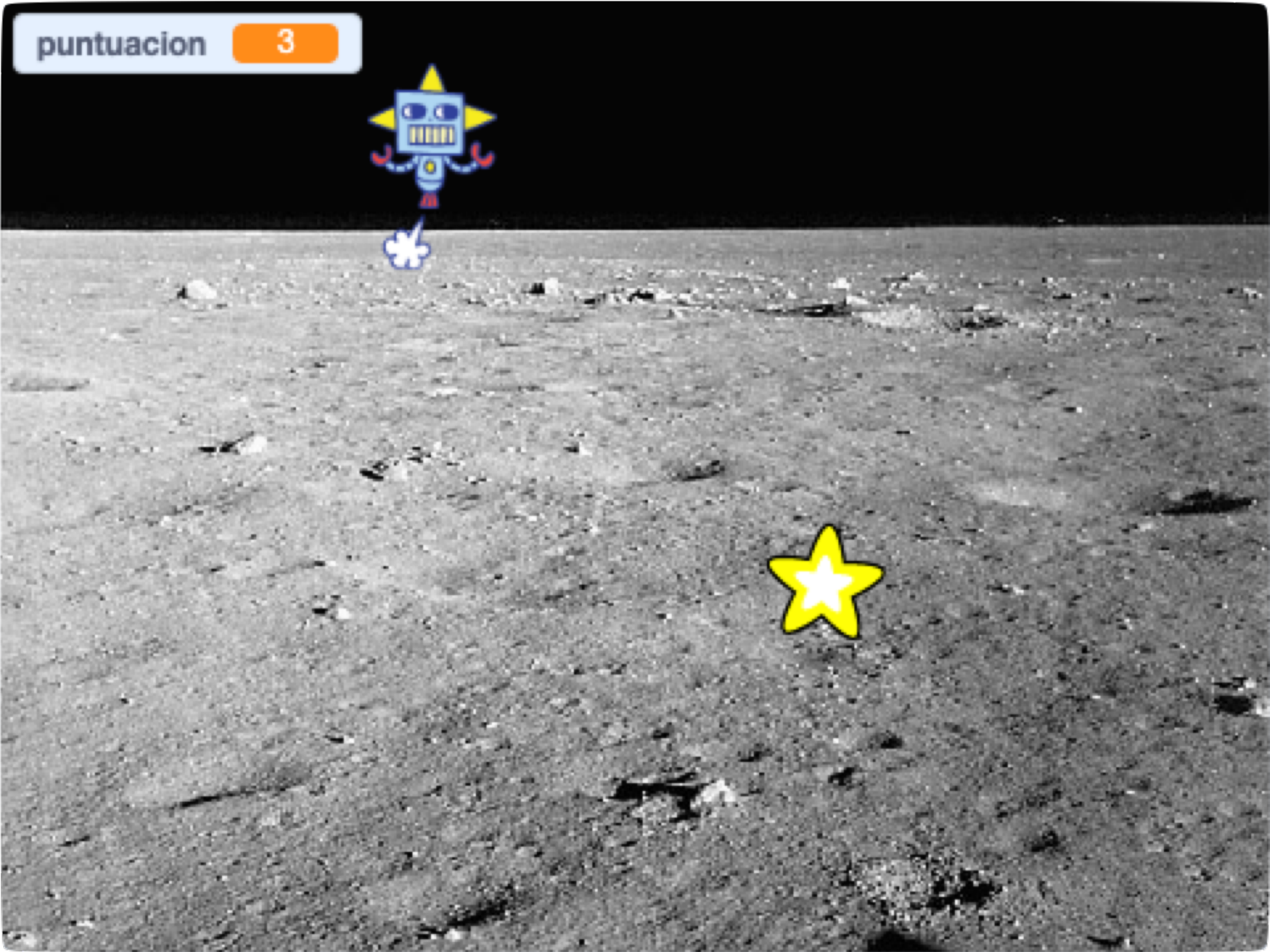}}
\caption{Example project (ID 400148579): The user controls a robot that has to catch the star.\label{fig:example_mio_vs_random2}}
\end{figure}

The search does not only provide advantages when the objective is to
wait long enough. \Cref{fig:example_mio_vs_random2} shows a game (ID 400148579)
where the player controls the robot using the cursor keys, and the aim is to
catch the star, which continuously moves to random positions. The script
controlling the player score (\cref{fig:example_mio_vs_random2:code}) provides
a gradient in the fitness landscape that drives the search towards touching the
star through the \boolsensing{touching \ovalsensing*{Star}} block, and the two
if-conditions checking the score drive the search towards trying to repeat
this. A further script checks for intermediate scores and displays messages.
The search successfully controls the robot, and sometimes drags it, in
order to reach scores that are substantially higher than those achieved by
random testing. Indeed in most cases the search reaches the second (final) level of the game.
%
Consequently, MOSA and MIO achieve an average of \SI{92.57}{\percent} and \SI{92.71}{\percent} coverage, respectively, whereas
random testing only reaches an average of \SI{80.14}{\percent}.


\begin{figure}[t]
	\subfloat[Code excerpt of the project]{\includegraphics[scale=0.35]{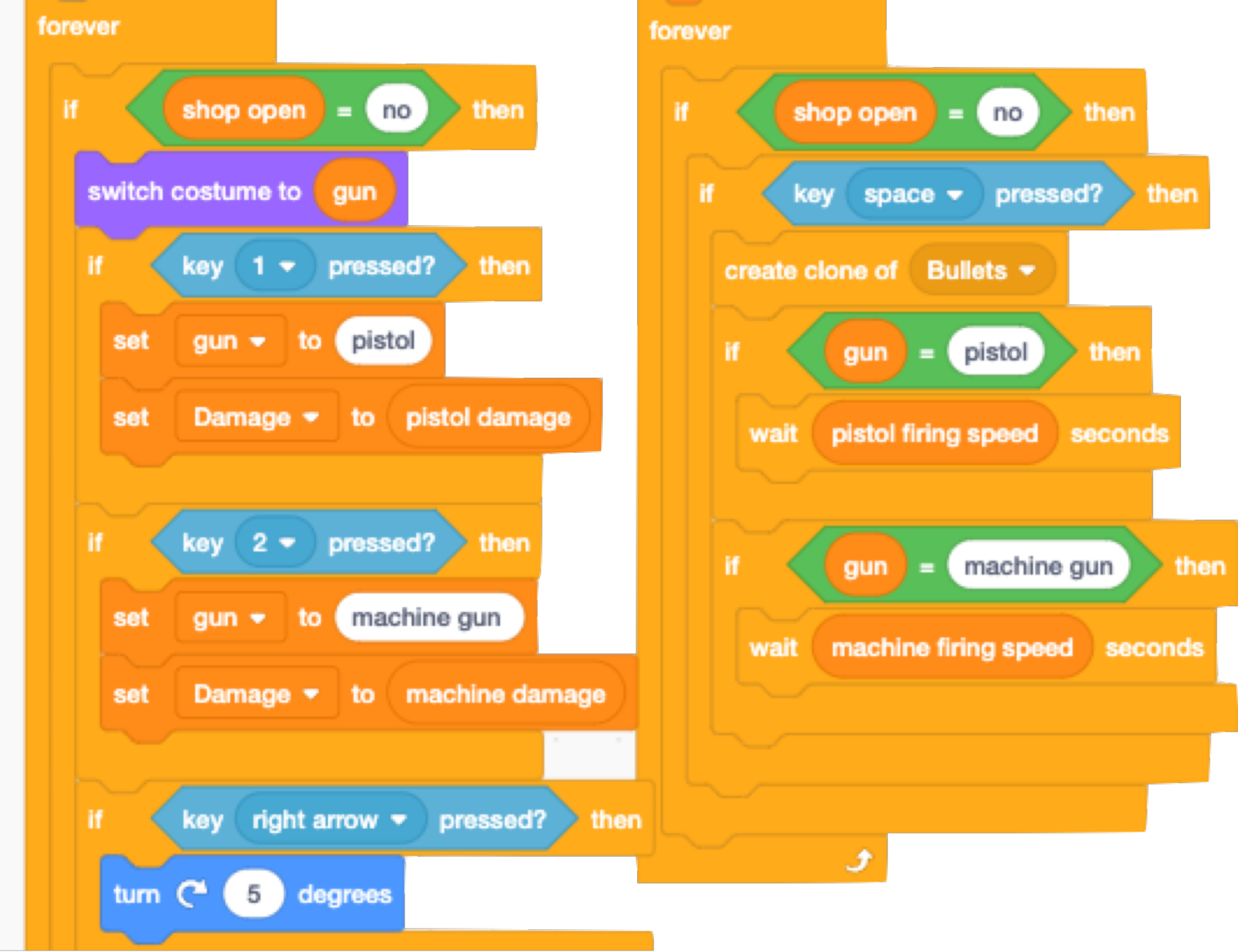}}
	\hfill
	\subfloat[Project stage]{\includegraphics[scale=0.25]{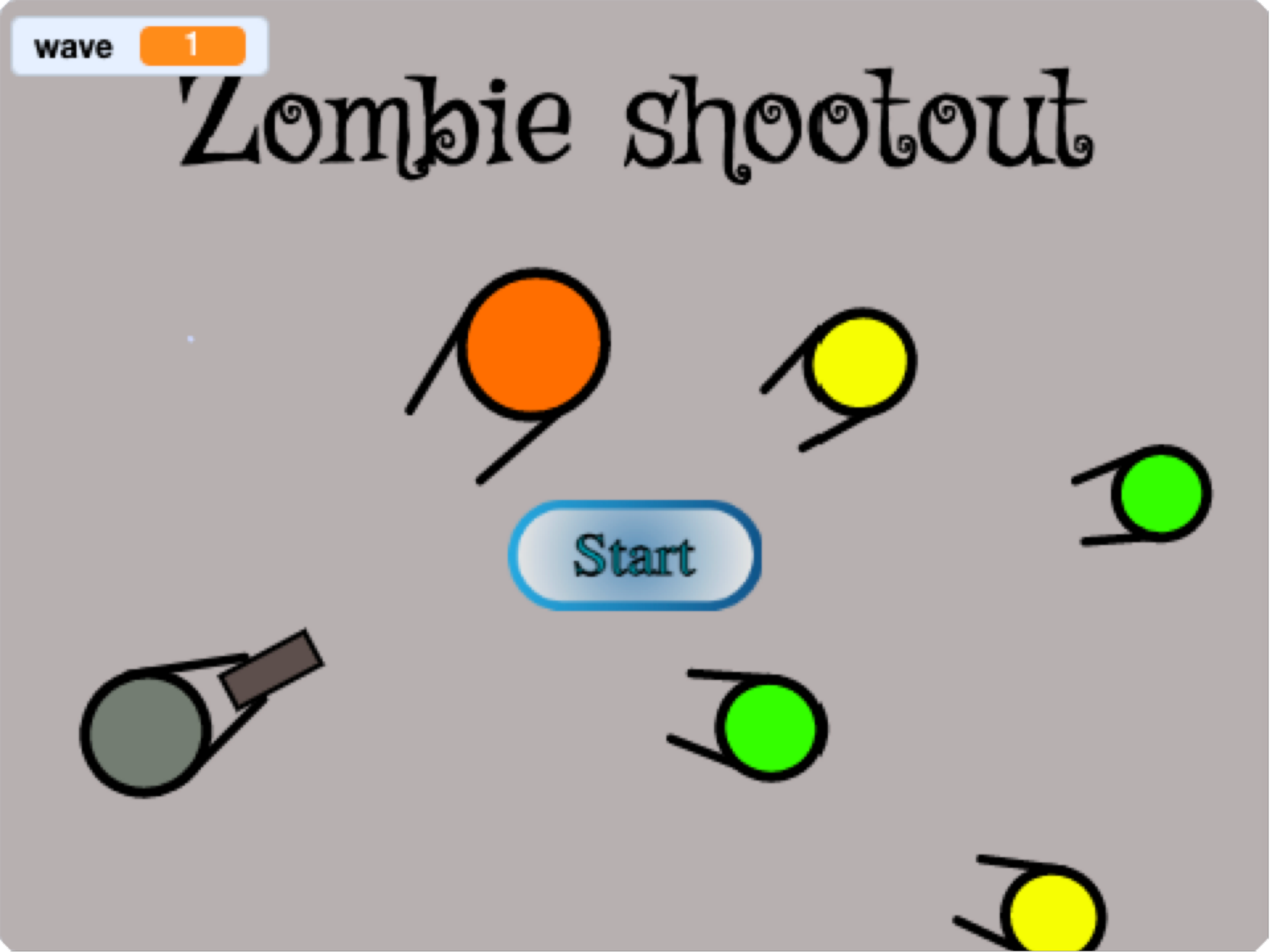}}
\caption{Example project (ID 402089829): Zombie shootout game.\label{fig:example_mosa_vs_mio}}
\end{figure}

\begin{figure}[t]
	\subfloat[Title screen\label{fig:example-minecraft-title}]{\includegraphics[scale=0.3]{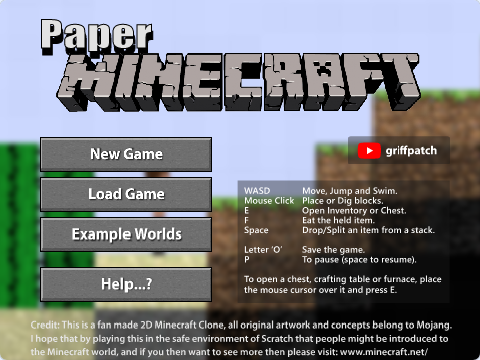}}
	\hfill
	\subfloat[Having started a new game\label{fig:example-minecraft-new-game}]{\includegraphics[scale=0.3]{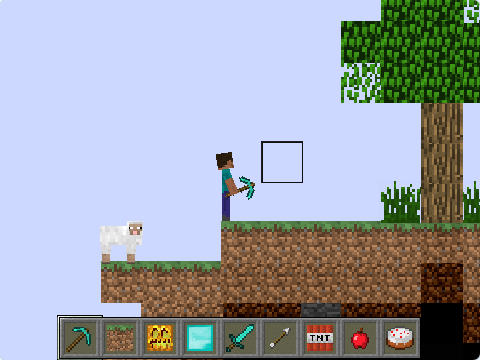}}
\caption{Example project (ID 10128407): ``Paper Minecraft'' sandbox game.\label{fig:example-minecraft}}
\end{figure}

While there are no projects where random testing achieves comparably large
margins in terms of coverage over the search algorithms, there are some
projects where random testing does achieve higher coverage.
%
\Cref{fig:example_mosa_vs_mio} shows a Zombie game where the player has to exterminate zombies without being eaten, using weapons
that can be purchased in a shop. While the
bullet-sprite provides some guidance for the search through a condition that
checks if a zombie is touched, the fitness does not provide guidance towards
shooting \emph{all} zombies, nor to evade them. While random testing appears to
be lucky nevertheless with an average coverage of \SI{53.38}{\percent}, MOSA
only achieves an average of \SI{49.68}{\percent}. MIO benefits from the combination of exploration and exploitation and reaches an average coverage of \SI{60.05}{\percent}.

Many of the projects in the \topp dataset are games with similar challenges. For example, \cref{fig:example-minecraft} shows ``Paper Minecraft'', the most loved
project of the \topp dataset. Containing \num{6715} statements, it is also among
the biggest projects in the dataset (cf.\ \cref{fig:stats_top1000}). Paper
Minecraft implements a so-called sandbox game where players face no pre-determined
objective but are encouraged to be creative by farming resources, creating buildings,
etc. Random testing achieves an average coverage of
\SI{7.73}{\percent}, while
MOSA and MIO achieve \SI{6.77}{\percent} and \SI{4.93}{\percent},
respectively.
To actually play the game (\cref{fig:example-minecraft-new-game}) one has to
select the ``New Game'' option on the title screen (\cref{fig:example-minecraft-title}).
Interestingly, when the button was hovered we observed a click-rate of \num[parse-numbers=false]{17/44}
across all executions for random testing, compared to MOSA (\num[parse-numbers=false]{7/22}) and MIO (\num[parse-numbers=false]{6/10}).
That is, random testing started the game more often than the other algorithms, which explains its
higher coverage, while MIO started the game least often.

In general, for all algorithms the majority of tests for more complex games focus on interacting with the
title screen rather than playing the actual game. Even when a test does
play the game, the maximum length of 20~events per test case prevents it from doing
so long enough, and only simple actions such as walking or switching items in the
inventory can be performed in case of Paper Minecraft. Allowing longer tests would alleviate this problem to
a certain extent, but would increase the computational costs; using variable length might also lead to effects of bloat~\citep{fraser2012whole}.
Like Paper Minecraft, many other games challenge the approach of optimising event-based sequences. Future work might consider reinforcement learning or other related techniques to address this problem.

\begin{figure}[t]
		\subfloat[\label{fig:algorithms_sizes:random1000}\textsc{Random1000}]{\includegraphics[width=0.49\textwidth]{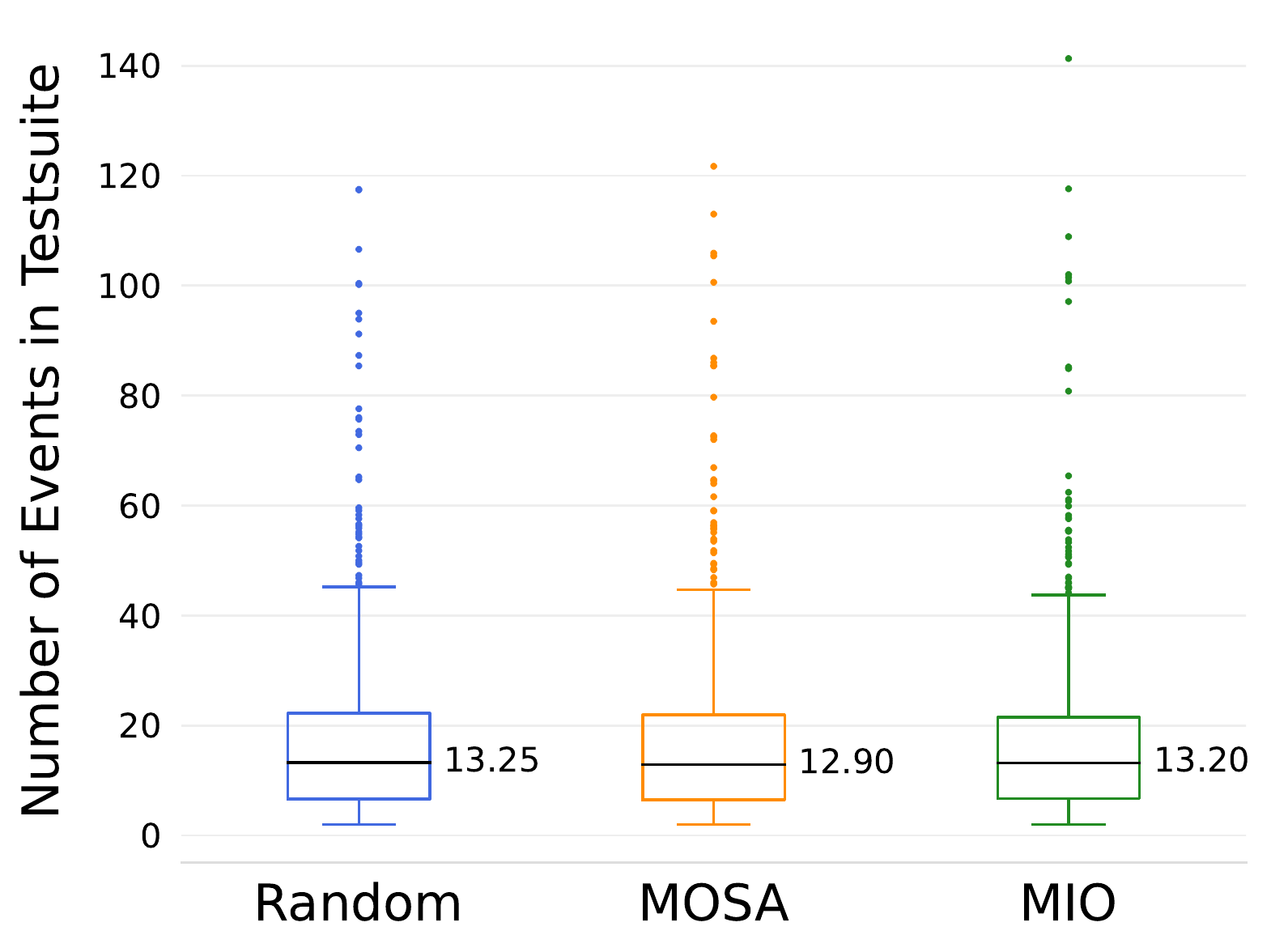}}\hfill
		\subfloat[\label{fig:algorithms_sizes:top1000}\textsc{Top1000}]{\includegraphics[width=0.49\textwidth]{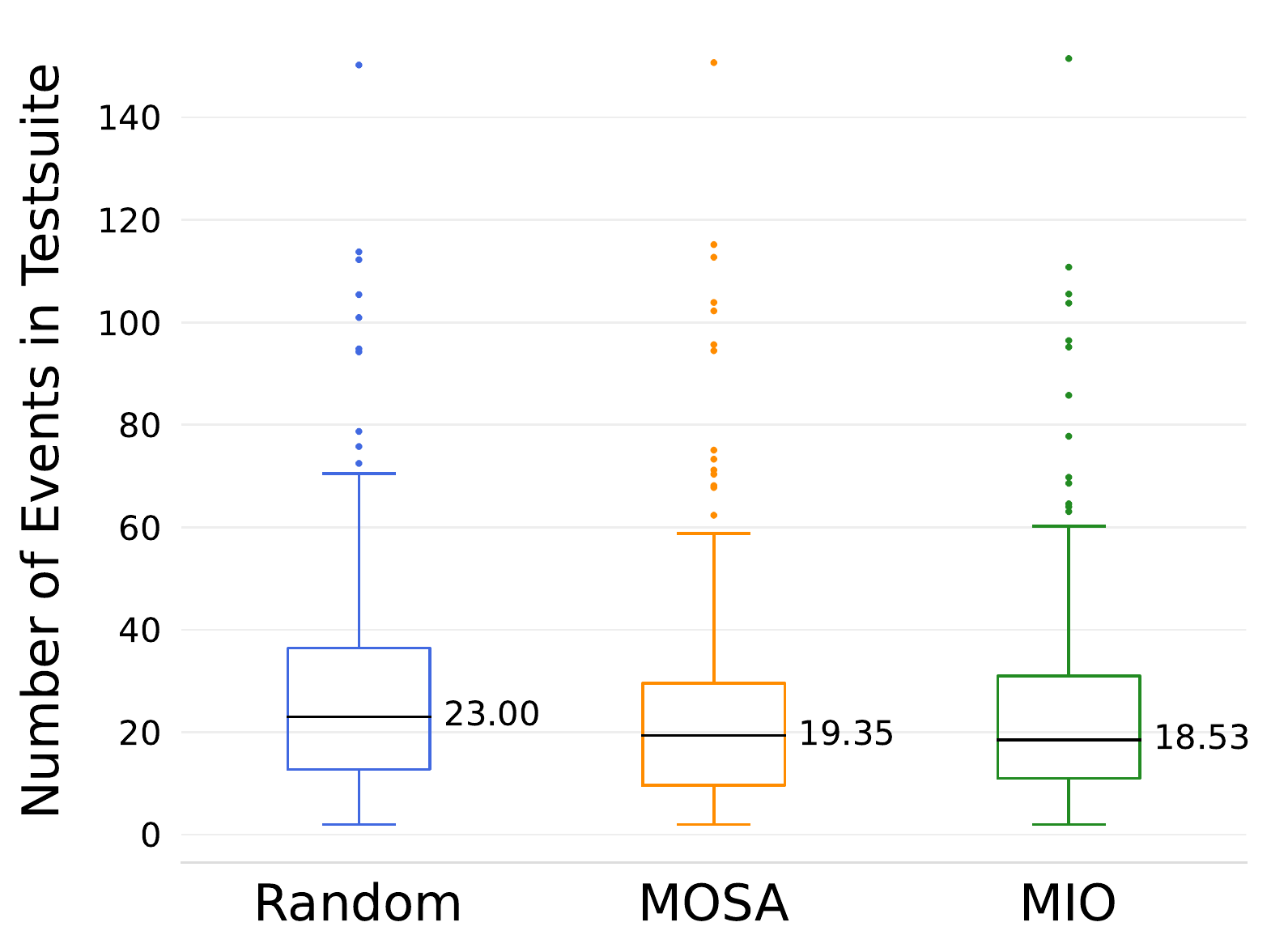}}
\caption{Overall test suite size for projects with equal coverage.}\label{fig:algorithms_sizes}
\end{figure}

\begin{figure}[t]
		\subfloat[\label{fig:algorithms_size_effects:random1000}\textsc{Random1000}]{\includegraphics[width=0.49\textwidth]{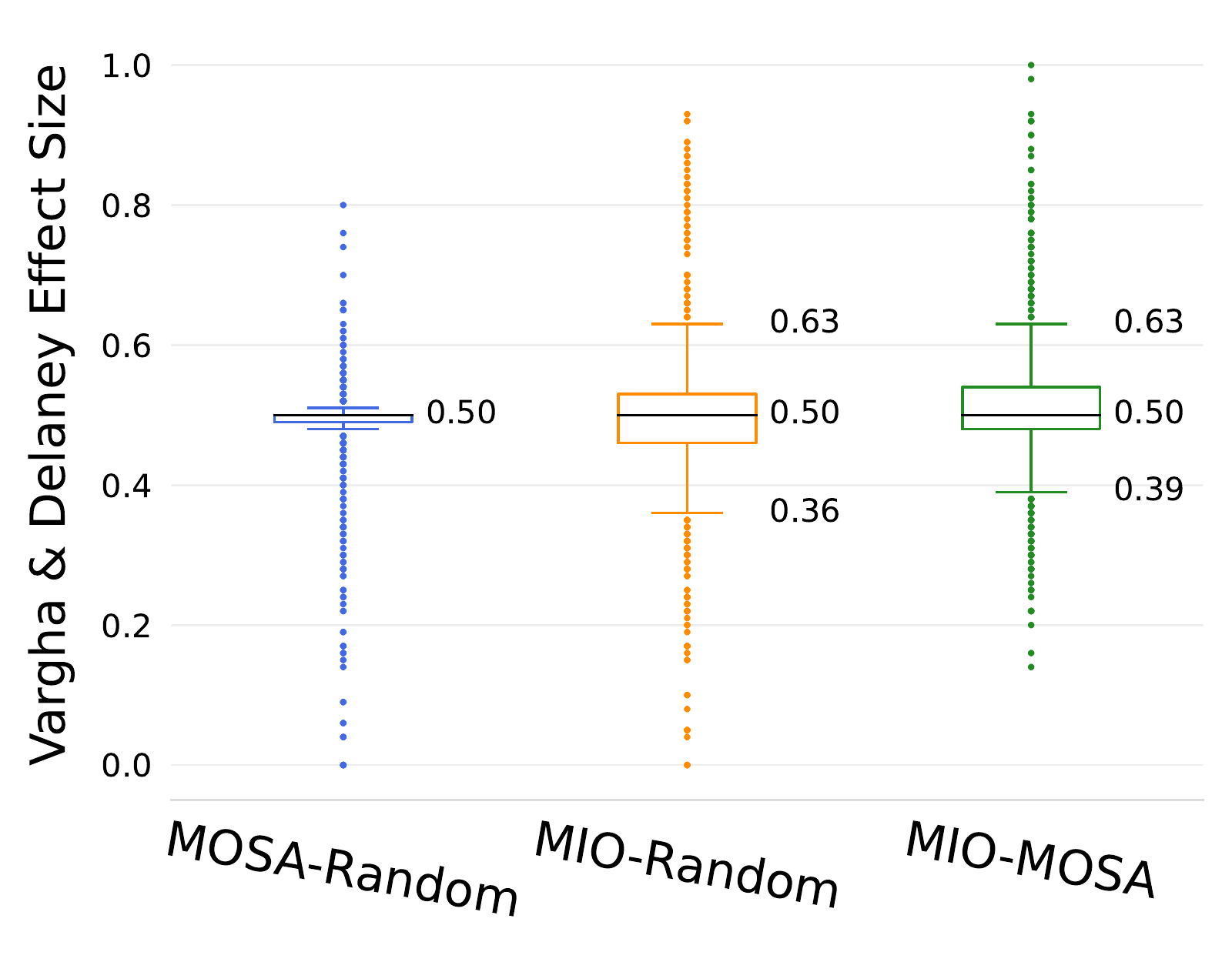}}\hfill
		\subfloat[\label{fig:algorithms_size_effects:top1000}\textsc{Top1000}]{\includegraphics[width=0.49\textwidth]{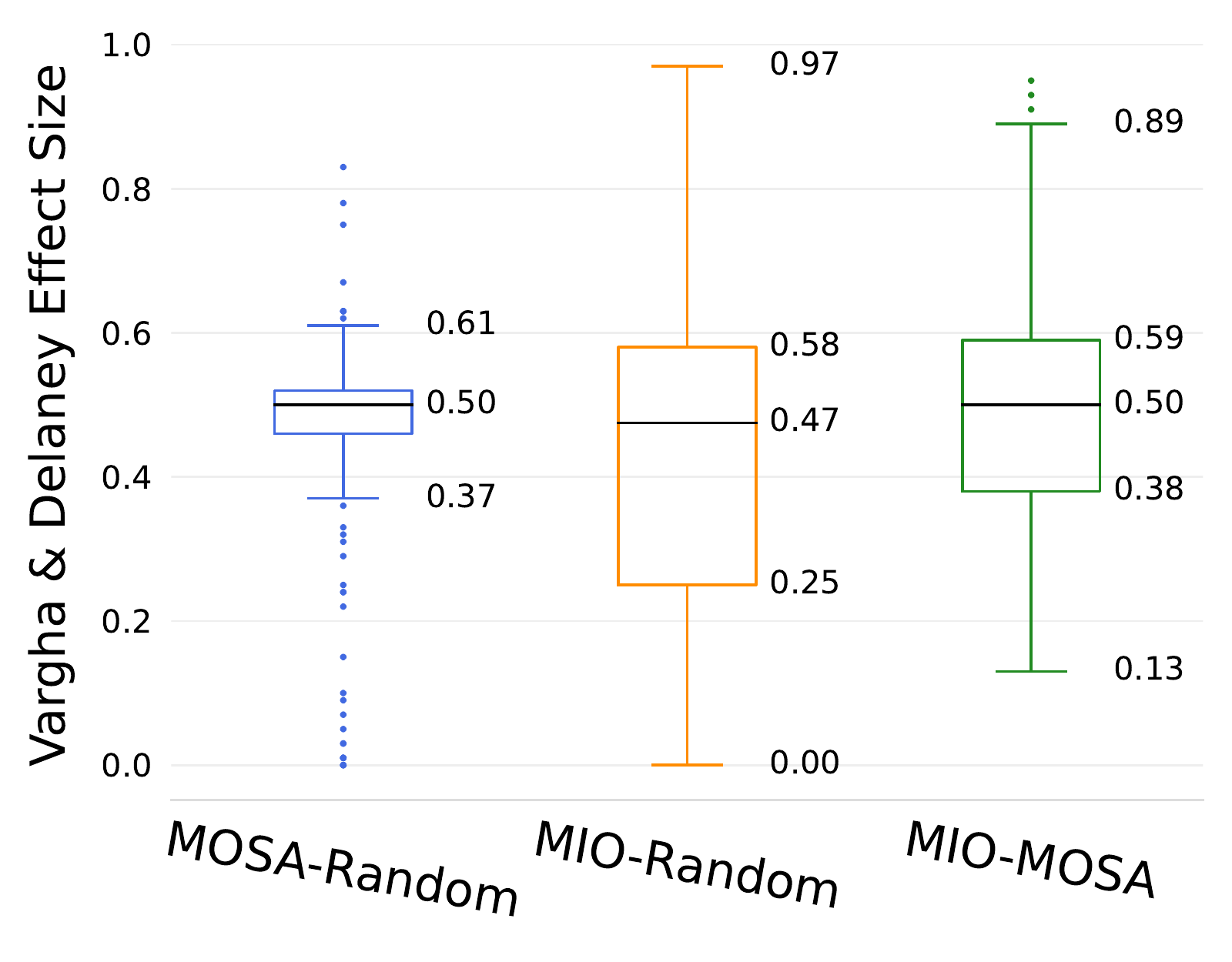}}
\caption{Effect sizes comparing algorithms wrt. test suite size for projects with equal coverage.}\label{fig:algorithms_size_effects}
\end{figure}

Besides the achieved coverage, a further important aspect for consideration is
the size of the resulting test suites, since these may need to be interpreted
by users. As the tests generated by the algorithms vary in length, we quantify
the length in terms of the overall number of events contained in a test suite.
Since the size is influenced by the coverage achieved by a test suite (i.e.,
test suites with higher coverage tend to be larger), we compare the
algorithms only on those projects, where all algorithms achieved the same
coverage. \Cref{fig:algorithms_sizes}
summarises the average number of events in the final test suites for these
projects, and \cref{fig:algorithms_size_effects} shows the distribution of
effect sizes.
The reported results are based on the test lengths \emph{after} conducting the minimisation process, which reduces the average test suite size by an average of \RandomAlgorithmsAverageMinimizedEqualCovRANDOM, \RandomAlgorithmsAverageMinimizedEqualCovMOSA \ and \RandomAlgorithmsAverageMinimizedEqualCovMIO \ events on the \rand set and by \TopRatedAlgorithmsAverageMinimizedEqualCovRANDOM, \TopRatedAlgorithmsAverageMinimizedEqualCovMOSA \ and \TopRatedAlgorithmsAverageMinimizedEqualCovMIO \ events on the \topp set for the Random, MOSA and MIO algorithm, respectively. This comparison shows for both datasets
that MOSA and MIO produce smaller tests than random testing. The fact that this difference is measured \emph{after} the minimisation suggests that the search algorithms succeed in finding targeted tests for more individual coverage goals, rather than accidentaly covering many goals with long execution sequences. However, the minimisation has to remove the most events from MIO's test suites. We conjecture the longer event sequences to be influenced in particular by the extension local search, and how successful longer tests are replicated and mutated in MIO.

\begin{minipage}[t]{\linewidth}
\begin{lstlisting}[label=lst:exampletest,basicstyle=\small\ttfamily,language=ES6,tabsize=2,numbers=none,caption=Test generated for project 400148579 (\cref{fig:example_mio_vs_random2})]
const test0 = async function (t) {
    t.dragSprite('Robot', 143, 126);
    await t.runForSteps(1);
    t.keyPress('right arrow', 3);
    await t.runForSteps(3);
    await t.runForSteps(1);
    t.keyPress('right arrow', 6);
    await t.runForSteps(6);
    await t.runForSteps(1);
    t.keyPress('up arrow', 1);
    await t.runForSteps(1);
	// ...
}
\end{lstlisting}
\end{minipage}

\begin{figure}[t]
		\subfloat[\label{fig:algorithms_lengths:random1000}\textsc{Random1000}]{\includegraphics[width=0.49\textwidth]{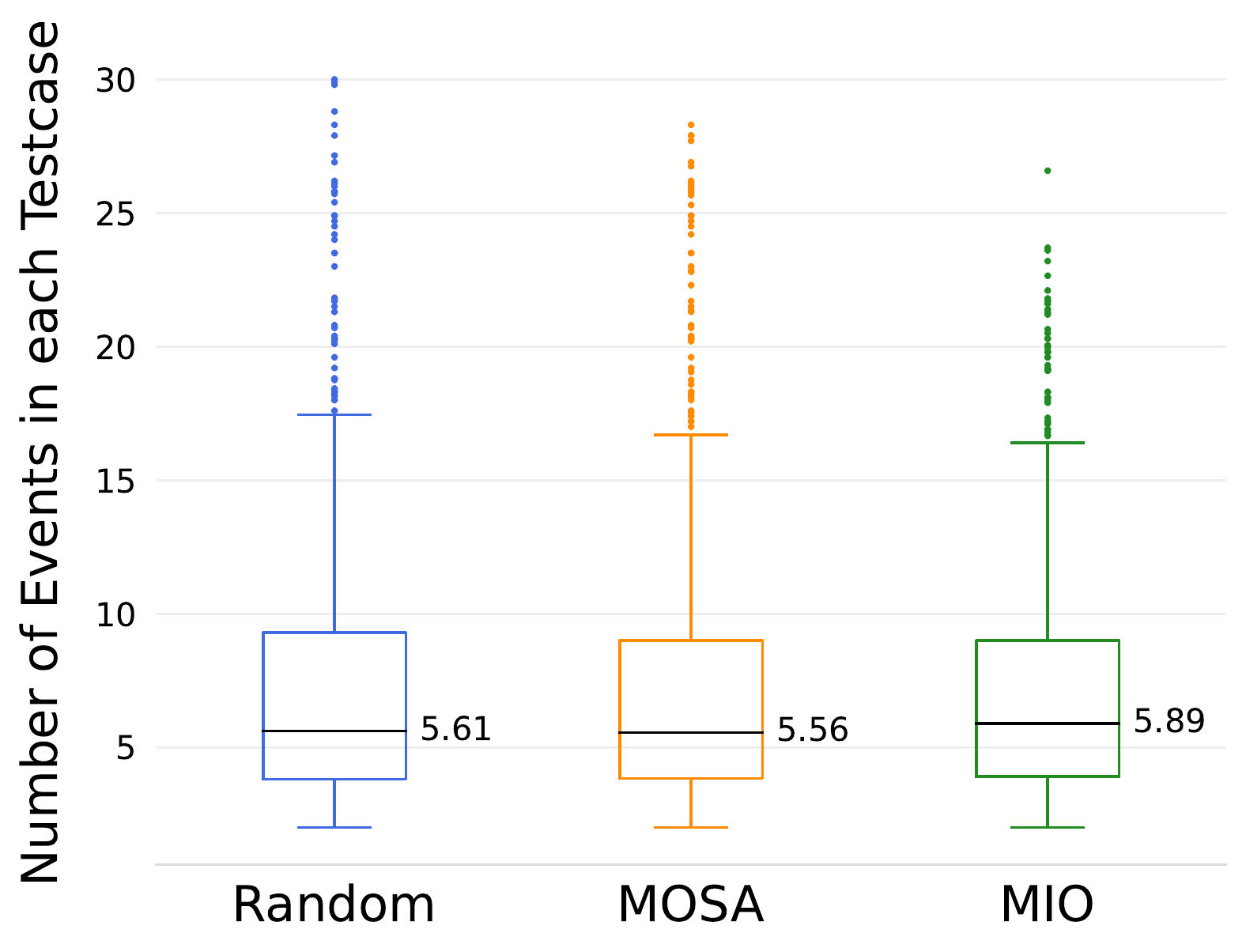}}\hfill
		\subfloat[\label{fig:algorithms_lengths:top1000}\textsc{Top1000}]{\includegraphics[width=0.49\textwidth]{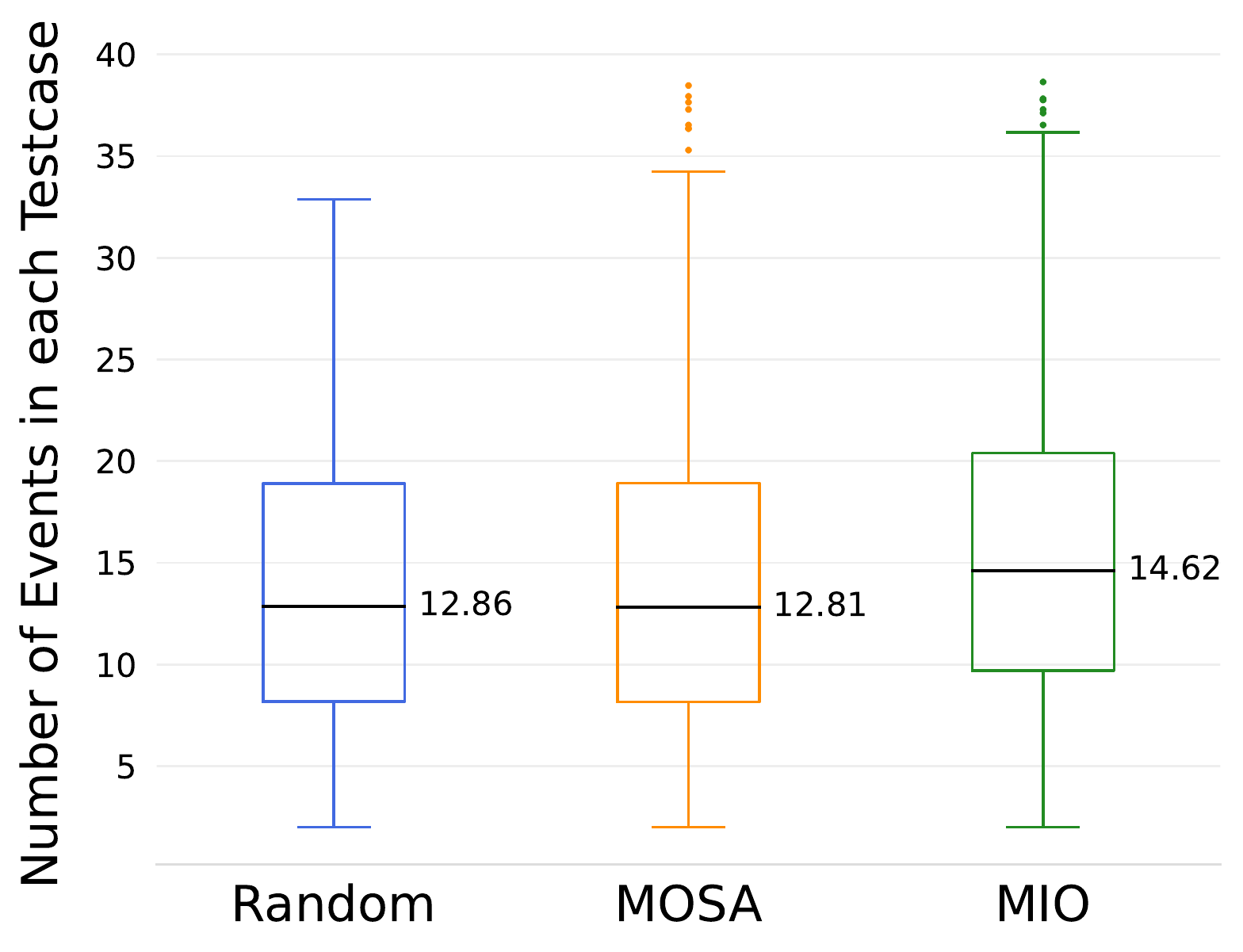}}
\caption{Average length of test cases across all projects.}\label{fig:algorithms_test_length}
\end{figure}

 Note that we count the number of events in the actual \whisker tests in JavaScript format: Since we interleave each event selected by the test generator with a
\event{Wait} event for a single step, the number of events is twice the number of
codons of the internal representation. Listing~\ref{lst:exampletest} shows an
excerpt of a test for the robot-star-game from \cref{fig:example_mio_vs_random2} after
removing automatically generated assertions.
Events resulting from codons are interleaved with steps
(\texttt{t.runForSteps(1)}). \event{KeyPress} events are represented as two statements
in the test code, first the keypress is sent to the \Scratch VM for a certain
number of steps (\texttt{t.keyPress('down arrow', 4)}), and then the test waits
for this number of steps to pass (\texttt{t.runForSteps(4)}).
\Cref{fig:algorithms_test_length}
shows the average length of individual test cases for the different algorithms:
Interestingly, for both datasets the median length of a single test covers between \SI{42.34}{\percent} and \SI{78.90}{\percent} of the entire suite, which indicates that most test suites consist of only 1-3 tests. 
%

\summary{RQ3}{MIO achieves the overall highest code coverage on \Scratch
projects in \rand \ with an average of
\SI{\RandomAlgorithmsAverageCoverageMIO}{\percent}, closely followed by MOSA
with an average of \SI{\RandomAlgorithmsAverageCoverageMOSA}{\percent}, and
random testing with an average of
\SI{\RandomAlgorithmsAverageCoverageRANDOM}{\percent}. On \topp, the ranking is
similar with \SI{\TopRatedAlgorithmsAverageCoverageMIO}{\percent} (MIO),
\SI{\TopRatedAlgorithmsAverageCoverageMOSA}{\percent} (MOSA), and
\SI{\TopRatedAlgorithmsAverageCoverageRANDOM}{\percent} (random testing). }

\subsection{RQ4: How effective are generated tests at detecting faults?}
\begin{figure}[tb]
\includegraphics[width=\textwidth]{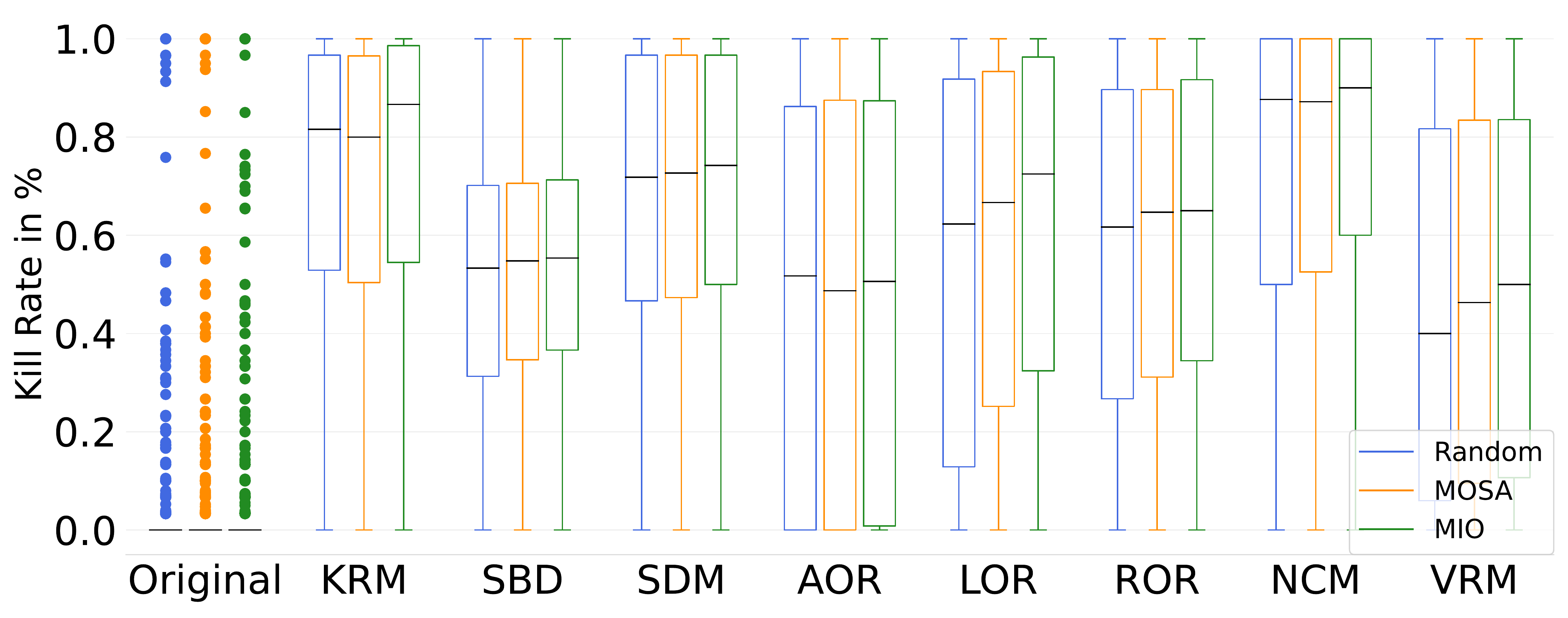}
\caption{Mutant kill rates across the applied mutation operators.}
\label{fig:Mutation-Categories}
\end{figure}

In our last research question, we aim to evaluate the effectiveness of
the tests in detecting faults using assertions generated automatically
using the approach presented in \cref{sec:assertions}. We execute the
test suites produced in \cref{sec:Results-RQ3} on mutated versions of
the program and evaluate whether the synthesised tests detected the
generated mutants. In accordance with common practice in mutation
analysis and in order to ensure a sound evaluation, we exclude test
cases which falsely mark a non-modified program as a mutant.  However,
as shown by \cref{fig:Mutation-Categories}, these false-positives only
occur very rarely in the form of outliers and have a median frequency
of 0 \%. The reason for these rarely occurring false-positives varies
and is very program-specific.  Out of \MutRandomGeneratedMutantsRandom
\ generated mutants, the tests generated by the random search
algorithm were able to reach a \emph{Mutation Score} of
\MutRandomMutationScoreTcSkipRandom \ \%.  In contrast, tests
generated by MOSA and MIO detected \MutRandomMutationScoreTcSkipMOSA \
\% and \MutRandomMutationScoreTcSkipMIO \ \% mutants from a total of
\MutRandomGeneratedMutantsMOSA \ and \MutRandomGeneratedMutantsMIO \
generated mutants, respectively.  Please note that the total number of
generated mutants varies slightly due to memory allocation issues, as
explained in \cref{sec:Methodology-RQ4}.

In addition to the frequency of false-positives,
\cref{fig:Mutation-Categories} also illustrates the distribution of
killed mutants across the applied operators: For all three algorithms,
the results are very similar, with MIO-generated tests detecting
slightly more mutants per projects (\MutRandomMutationScorePerProjectMIO \ \%) than tests synthesised by MOSA (\MutRandomMutationScorePerProjectMOSA \ \%) and random search (\MutRandomMutationScorePerProjectRandom \ \%). This small advantage originates from MIO's ability to
achieve slightly higher coverage than the other algorithms, as
shown in \cref{sec:Results-RQ3}.  All in all, the results demonstrate
that the generated tests are able to detect faulty \Scratch programs
automatically.  Nevertheless, further work should be done to improve the
sensitivity of the generated assertions to reduce the frequency of
false-positives and increase the number of detected program faults.

Finally, \cref{fig:Mutation-Categories} reveals that certain mutants
are harder to detect than others. NCM, KRM and SDM are easiest to
detect since they can fundamentally alter program semantics, e.g.,
diverting control flow to the opposite branch in an if-else, or
preventing the execution of entire scripts. In contrast, SBD, AOR and
VRM show the lowest kill rates. Since they are applicable to many
blocks, and not exclusively to the switching points of control flow,
they have a lower chance of making impactful changes. We hypothesise
that the former operators with their larger changes represent
learners' mistakes well: Previous
work~\citep{fraedrich2020} investigated typical bug
patterns in the \Scratch community, such as broadcast messages that
are never sent or received, or cloning sprites without proper
initialisation. These patterns are among the most common ones, and can
be easily elicited by operators such as SDM or KRM, indicating that
our mutants and tests can produce and detect common real-world faults
in learners' programs. However, a closer analysis of fault coupling
for \Scratch mutation operators is out of scope for this paper and
remains as future work.

\summary{RQ4}{The generated test suites detected more than half of the introduced program faults by achieving \emph{Mutation Scores} of \MutRandomMutationScoreTcSkipRandom\ \%, \MutRandomMutationScoreTcSkipMOSA\ \% and \MutRandomMutationScoreTcSkipMIO\ \% using random search, MIO and MOSA as test generation approaches, respectively. The perceivable advantage of MIO over the other algorithms highlights the importance of generating test suites that manage to cover as many program statements as possible.}

\section{Related Work}
\label{sec:related}

\subsection{Automated Test Generation}

A traditional approach to generate tests automatically is by using symbolic
execution~\citep{baldoni2018survey}, which extracts path conditions from
programs and then generates test inputs by solving the path conditions with
constraint solvers. Symbolic execution is mostly used when testing at unit or
API level, or when inputs can be represented as symbolic variables. In this
paper, however, we consider system testing at the level of a user interface.
While there have been attempts to apply symbolic execution also in this context
(e.g., \citep{ganov2009event,mirzaei2012testing,salvesen2015using}), this is
usually done to generate input values for specific user inputs (e.g., text).
For \Scratch programs, however, the challenge rather lies in finding timed
sequences of simple user interactions. This problem is generally addressed
using random and search-based test generation approaches.

Random testing of GUIs~\citep{miller1995fuzz} consists of sending random user
interactions to a program under test. Search-based testing generalizes this
approach by adding objective functions, such as reaching target points in the
source code~\citep{mcminn2004search}, together with algorithms that optimize
the inputs to reach the objectives. While the bulk of research on search-based
testing considers function inputs or unit tests, the problem of generating
tests for graphical user interfaces (GUIs) has also been successfully addressed
using meta-heuristic search algorithms, for example in the context of Java
Swing applications~\citep{gross2012search} or Android 
apps~\citep{mao2016sapienz,amalfitano2014mobiguitar,mahmood2014evodroid}.

Objective functions in search-based testing are usually based on the notion of
code coverage, and require instrumentation to collect data that allows
calculating fitness values. For some domains, such as Android apps, it is
challenging to provide this instrumentation and to frequently execute
long-running tests, therefore alternative black-box approaches have been
proposed, e.g., aiming to maximise the amount of GUI changes
observed~\citep{mariani2012autoblacktest}. In contrast, the size of \Scratch
programs represents no problems in terms of the scalability of fitness
computations, and we therefore can base our fitness computations on
inter-procedural analysis and instrumentation. However, test executions may
still take a long time due to the time-based behaviour of \Scratch programs.

An issue that is common to search-based GUI testing approaches is the
difficulty of implementing search operators such as crossover, as
crossing two sequences of events is likely to result in invalid sequences,
where the events encoded in the sequences cannot be executed in the actual
program states. Prior approaches to tackle this problem consisted of
restricting crossover to suitable locations and ensuring valid sequences
through repair~\citep{mahmood2014evodroid}, or using set-based representations
where no sequences are modified during crossover~\citep{mao2016sapienz}. A
common alternative is also to resort to heuristics that do not
require such operators but, e.g., rather decide on executions based on
probabilisty distributions~\citep{su2017guided}. To overcome these problems of
representation, we used an integer-based encoding based on grammatical
evolution~\citep{o2001grammatical}, which has not received much attention in the context of test generation yet~\citep{anjum2020seeding}.

Which search algorithm is most effective is highly problem specific. Variants
of random search are often sufficient~\citep{shamshiri2018random}, but more
advanced search algorithms provide clear benefits on more complex test
problems. Our study confirms that this also holds in the domain of \Scratch
programs. At the unit testing level, it has been shown that searching for sets
of tests that aim to cover all code~\citep{fraser2012whole} at once is most
effective~\citep{campos2017empirical,panichella2018large} using many-objective
optimisation algorithms such as MOSA~\citep{MOSA} and MIO~\citep{MIO}, which is
why we chose these many-objective optimisation algorithms also for our study.


\subsection{Automated Testing and Analysis for \Scratch Programs}

Novice programming environments such as \Scratch
\citep{maloney2010scratch} or \textsc{Snap}~\citep{harvey2013snap} 
are widely used in introductory
programming curricula~\citep{franklin2020scratch, garcia2015beauty}. These
environments motivate students by enabling them to create programming artefacts
that they can interact with, and they have
been shown to improve learning gains and long-term interests towards
 programming~\citep{weintrop2017comparing}. Among the available
programming environments, \Scratch is by far the most popular environment, with
the largest online youth programming community~\citep{fields2017youth}.

A core aspect of these programming environments is that they use blocks instead
of text to avoid that learners have to memorise syntax or available
programming commands. While this simplifies initial coding, students have been
shown to still struggle in building logically coherent programs in \Scratch
\citep{meerbaum2011habits}. They have also been reported to create ``smelly''
code~\citep{aivaloglou2016kids,hermans2016smells,techapalokul2017understanding},
 and these code smells have been shown to have a negative impact on
understanding~\citep{hermans2016code}. It is therefore important to provide
tool-based support for learners as well as their teachers.
The majority of prior work focused on statically analysing \Scratch code. For
example, the popular \textsc{Dr. Scratch}~\citep{moreno2015dr} website assesses
evidence of computational thinking in programs and can also point out code smells using the
\textsc{Hairball}~\citep{boe2013hairball} static analysis tool. Similar code
smells are identified by \textsc{Quality hound}~\citep{techapalokul2017quality}
and \textsc{SAT}~\citep{chang2018scratch}.
\textsc{LitterBox}~\citep{fraedrich2020} is an extensible framework that can
identify not only code smells, but also patterns of common bugs as well as
positive aspects such as code perfumes~\citep{obermuller2021code} in \Scratch
programs. A general limitation of these syntax-based approaches which we aim to
address in this paper is that they can only provide limited reasoning about the
actual and intended program behaviour.

Dynamic analysis is required to reason about program behaviour, and automated
testing is a common means to enable dynamic analysis. Automated testing is
commonly applied in the context of programming education for tasks such as
assessing student programs to provide feedback after a task has been completed,
or during its creation~\citep{shute2008formative}. In many text-based
programming environments, automated tests have been shown to enable
various types of feedback, such as by displaying failed test
cases~\citep{edwards2017codeworkout}, suggesting likely
misconceptions~\citep{gusukuma2018misconception}, and highlighting erroneous
code~\citep{edmison2017}. Offering such immediate, automated feedback has been
shown to improve students performance and learning
outcomes~\citep{corbett2001locus}.

However, unlike text-based programming environments, novice programming environments like \Scratch are often centered around custom graphical scenarios that are controlled by input streams of signals from users' input devices such as keyboard and mouse, which causes challenges for automated testing. 
The \textsc{Itch} tool~\citep{johnson2016itch} dynamically tests \Scratch
programs by translating a small subset of \Scratch programs to Python code.
However, such tests are limited to functions that take in static input/outputs,
such as \begin{scratch}\blocksensing{ask}\end{scratch} and
\begin{scratch}\blocklook{say}\end{scratch} blocks. Furthermore, \textsc{Itch} does not automatically generate test cases.
\whisker~\citep{TestingScratch} takes this approach a step further and, besides
execution of automated tests directly in \Scratch, also provides automated
property-based testing. \textsc{SnapCheck}~\citep{wang2021snapcheck} applies
similar concepts in the context of the \textsc{Snap!} programming language.
However, all of these existing testing tools focus on automatically executing
manually written tests. In contrast, the aim of this paper is to automate the test generation process itself.

The work presented in this paper is integrated into
\whisker~\citep{TestingScratch}, but controls the \Scratch VM directly and
represents a separate component which is mainly connected with \whisker through
the result of the test generation, which is saved in \whisker's format and can
be re-executed with \whisker. Our proof-of-concept on \whisker test
generation~\citep{deiner2020search} proposed a codon-based encoding, the use of
interprocedural graphs to calculate fitness values, and accelerated test
execution. This paper extends this initial proof-of-concept by providing an
entirely new execution model, extending the codon encoding and search
operators, providing new search algorithms, adding local search, refining the
fitness function with the concept of control flow distance, adds many
testability transformations to improve the fitness function, adds a new model
for event extraction as well as new events, and adds test minimization as well
as regression assertion generation, and adding many smaller technical
improvements overall. In addition, a central contribution of this paper lies in
the large empirical study.

\section{Conclusions}
\label{sec:conclusions}

The increasing popularity of block-based programming languages leads to a
demand for tools to support programmers. However, even though languages like
\Scratch have millions of users, they lack fundamental analysis frameworks that
are common for other programming languages, which inhibits the development of
tools to support novice programmers. To address this issue, \whisker makes it
possible to run automated tests also on \Scratch, but writing \whisker tests
remains challenging. In this paper we presented a fully automated approach to
generate these tests given a \Scratch program under test. Our experiments on
three different, large datasets have demonstrated that automated test
generation generally achieves very high coverage. This paves the way for
advanced analysis and feedback tools.

Although our experiments suggest that \whisker will fully cover many types of
programs, we also observed two notable patterns of programs where the
search-based test generation approach implemented by \whisker could be
improved: 
\begin{itemize}
\item First, for many of the projects finding the correct sequence of user
inputs is only part of the challenge, while in fact the \emph{timing} is a more
important question, and very often test generation would require waiting long
durations for parts of the animations or sounds playing. While \whisker
accommodates for this through accelerated execution and including timing in the
fitness function, the rather classical search-based testing approach that
\whisker implements nevertheless builds on the assumption that one can run many
short executions. In contrast, many \Scratch programs may be easier to test by
alternative approaches aiming to drive individual, longer executions.
\item Second, many \Scratch programs, in particular the popular ones, represent
games where a traditional test generation approach stands no chance of ever
optimising a sequence that can really succeed at \emph{playing} the
game---which, alas, is a prerequisite to reaching interesting states and parts
of the code. Possible avenues to address this problem will be to record and
integrate user interactions with a program under test, or to apply
reinforcement learning approaches to teach the computer to actually play the
games.
\end{itemize}

\looseness=-1
The techniques described in this paper generalise conceptually in multiple
dimensions: First, there are other block-based languages such as
\textsc{Alice}~\citep{cooper2000alice} or \textsc{Snap}~\citep{harvey2013snap},
which also use a similar concept of stages and sprites. Second, there are also
text-based programming environments such as \textsc{Greenfoot}
\citep{kolling2010greenfoot} that are based on the same concept. Adapting our
approach to these programming environments mainly requires engineering work to
adapt our modified execution model, and to add support for different language
constructs. We also anticipate that our deterministic execution model can influence the design of future programming environments.
More generally, the encoding, search operators, and algorithmic
modifications proposed in this paper are applicable in principle to any
UI-based testing problem, independently of the underlying programming language. 

\looseness=-1
Compared to other testing problems, the code coverage observed in our
experiments is relatively high. Besides the smaller size of \Scratch programs,
one potentially influential factor is the absence of certain types of testing
challenges such as external dependencies; for example, Android apps will
frequently access web services and data storage, which leads to substantially
lower code coverage~\citep{mao2016sapienz}. However, such challenges also exist
in the domain of block-based languages: The \Scratch language provides support
for extensions that can provide arbitrary functionality, ranging from machine
learning functionality to support for controlling external devices.
Furthermore, there are related languages such as
\textsc{mBlock}\footnote{https://mblock.makeblock.com/, last accessed June
2022}, which extends \Scratch with support for a wide range of robots.
Supporting these features will require future work to extend our encoding
as well as the underlying instrumentation.

Given the ability to generate tests for \Scratch programs, we hope to enable  new approaches for automated tutorial systems, automated repair systems, hint generation systems. To support this future work, \whisker is available as open source at:
\begin{center}
	 \url{https://github.com/se2p/whisker}
\end{center}

\begin{acknowledgements}
We thank Christoph Fr\"{a}drich, Sophia Geserer, and Niklas Zantner for their contributions to \whisker. This work is supported by DFG project FR 2955/3-1  ``TENDER-BLOCK: Testing, Debugging, and Repairing
Blocks-based Programs''.
\end{acknowledgements}

%
\section*{Conflict of interests}

The authors declare that they have no conflict of interests.

%
%

\bibliographystyle{plainnat}
\bibliography{related}



\end{document}